\documentclass[12pt, a4paper]{article}
  \usepackage{algpseudocode}
  \usepackage{algorithm}
  \usepackage{amsfonts}
  \usepackage{amsmath}
  \usepackage{amssymb}
  \usepackage{amsthm}
  \usepackage{array}
  \usepackage{bm}
  \usepackage{booktabs}
  \usepackage{caption}
  \usepackage{color}
  \usepackage{float}
  \usepackage[stable, bottom, marginal]{footmisc}
  \usepackage{geometry}
  \usepackage{graphicx}
  \usepackage{algorithm}
  \usepackage{lineno}
  \usepackage{mathrsfs}
  \usepackage{multirow}
  \usepackage{natbib}
  \usepackage[normalem]{ulem}
  \usepackage{subfigure}
  \usepackage{url}
  \usepackage{ulem}
  \usepackage{epstopdf}
  \usepackage{bm}
  \setcitestyle{authoryear, open = {(}, close = {)}}
  \theoremstyle{plain}
  \newtheorem{definition}{Definition}
  
  \theoremstyle{definition}
  
\begin{document}

\begin{center}
  {\bfseries \Large
      Forecasting open-high-low-close data contained in candlestick chart}
\end{center}

\vskip 0.5cm
\begin{center}
 Huiwen Wang$^{1, 2}$, Wenyang Huang$^{1,3}$, and Shanshan Wang$^{1, 2*}$ \\
 \vskip 0.3cm
 \small
  $^1$\emph{School of Economics and Management, Beihang University, Beijing 100191, China} \\
  $^2$\emph{Beijing Advanced Innovation Center for Big Data and Brain Computing, Beijing 100191, China} \\
  $^3$\emph{Shen Yuan Honors College, Beihang University, Beijing 100191, China}
\end{center}

\let \thefootnote \relax \footnotetext{
  * Corresponding author. Correspondence to: School of Economics and Management, Beihang University, Beijing 100191, China.
  E-mail address: sswang@buaa.edu.cn (S.S. Wang).}

\vskip 0.5cm
\begin{center}
  {\bfseries \large
    Abstract}
\end{center}

Forecasting the (open-high-low-close)OHLC data contained in candlestick chart is of great practical importance, as exemplified by applications in the field of finance. Typically, the existence of the inherent constraints in OHLC data poses great challenge to its prediction, e.g., forecasting models may yield unrealistic values if these constraints are ignored. To address it, a novel transformation approach is proposed to relax these constraints along with its explicit inverse transformation, which ensures the forecasting models obtain meaningful open-high-low-close values. A flexible and efficient framework for forecasting the OHLC data is also provided. As an example, the detailed procedure of modelling the OHLC data via the vector auto-regression (VAR) model and vector error correction (VEC) model is given. The new approach has high practical utility on account of its flexibility, simple implementation and straightforward interpretation. Extensive simulation studies are performed to assess the effectiveness and stability of the proposed approach. Three financial data sets of the Kweichow Moutai, CSI $100$ index and $50$ ETF of Chinese stock market are employed to document the empirical effect of the proposed methodology.

\vskip 0.5cm
\noindent
  {\bfseries
    Key Words:}
    OHLC data;
    Candlestick chart;
    Unconstrained transformation;
    VAR;
    VEC

\section{Introduction} \label{Sec 1}

Nowadays, the candlestick chart analysis has become one of the most intuitive and widely used methods for analyzing the price movements of financial products \citep{romeo2015study}.
And there has been vast literature on statistical methods to analyze the (open-high-low-close)OHLC data contained in candlestick chart, which may broadly be categorized into two main aspects: graphical analysis \citep{caginalp1998predictive, goo2007application, Lu2012, Tao2017Further, wan2018hidden} and numerical analysis \citep{pai2005hybrid, Faria2009Predicting, caporin2013predictability, manurung2018algorithm, kumar2019stock} .

From the essence of OHLC data, it represents the prices of certain financial product.
Investors continue to make purchases and sells according to accurate predictions of OHLC data and thus earn profits is the fundamental incentive mechanism to maintain its effective operation \citep{liu2017integrated}.
Therefore, forecasting the OHLC data is one of the most important aspect among the various methods of investigating OHLC data, which is also our concern.

The existing literature on forecasting OHLC data have two shortcomings.
First, the information contained in OHLC data is always under-utilized.
For example, \cite{arroyo2011different} used exponential smoothing method, multi-layer perception, K-nearest neighbor algorithm, autoregressive integrated moving average model, vector auto-regression (VAR) model and vector error correction (VEC) methods to perform regressions based on the Dow Jones Industrial Average index and Euro-Dollar Exchange Rate.
In their study, only the center and range (Center \& Range method, CRM), or the high and low prices (Min \& Max method) of the OHLC data were used, which is the common practice of modelling OHLC data.
However, for OHLC data, in addition to the high and low prices, there are open price and close price in the middle of the two boundaries, which have an explanatory power for the fluctuation of OHLC data and should be carefully considered in the model \citep{cheung2007empirical}.
Regrettably, the open price and close price are beyond the consideration scope of the CRM method and Min \& Max method, and few articles fully considered the information contained in OHLC data.

\begin{figure}[!h]
  \centering
  \subfigure[]{
    \label{fig:subfig:a}
    \resizebox{5.32cm}{4.8cm}{\includegraphics{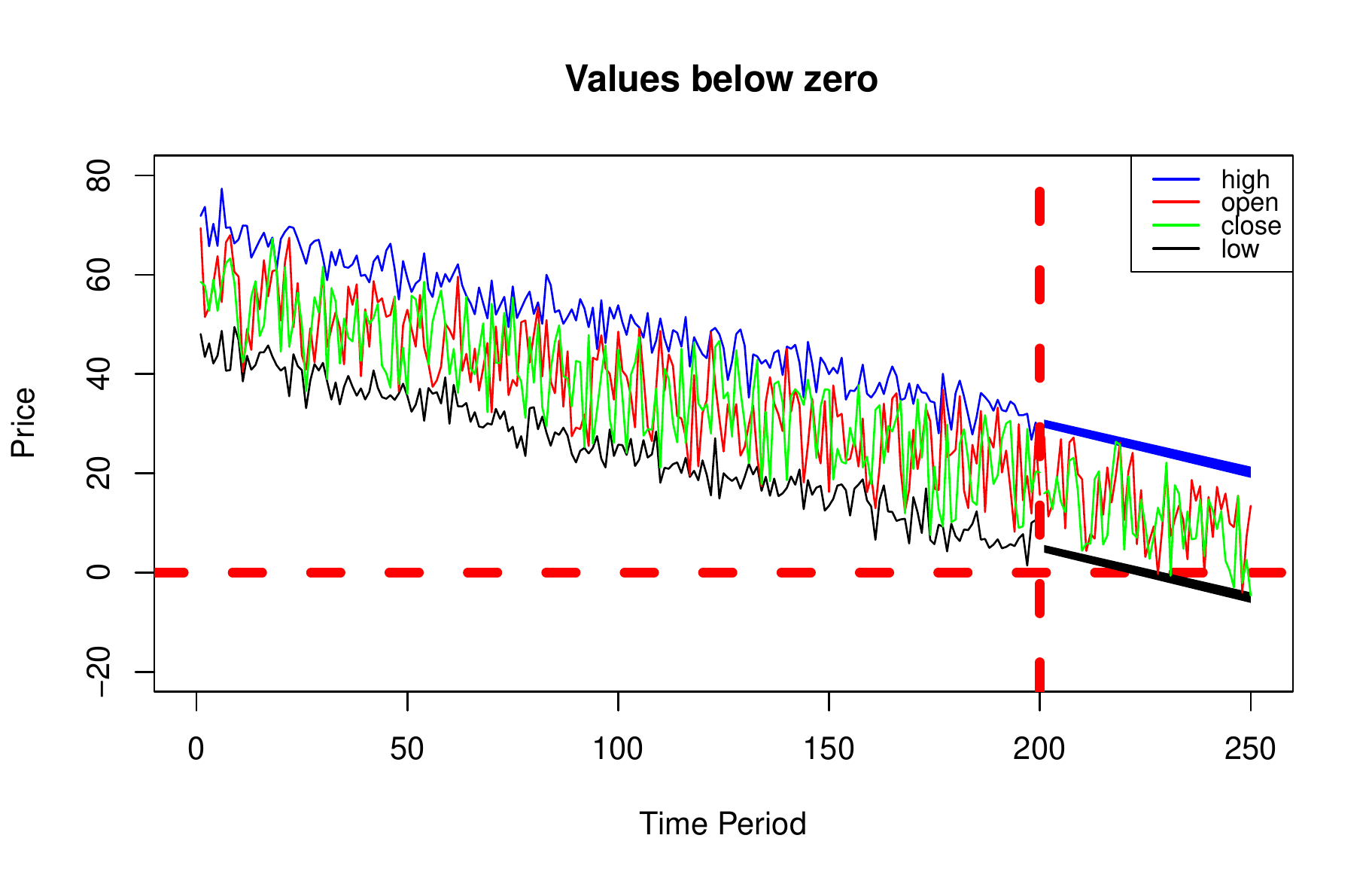}}}
  \hspace{0in}
  \subfigure[]{
    \label{fig:subfig:b}
    \resizebox{5.32cm}{4.8cm}{\includegraphics{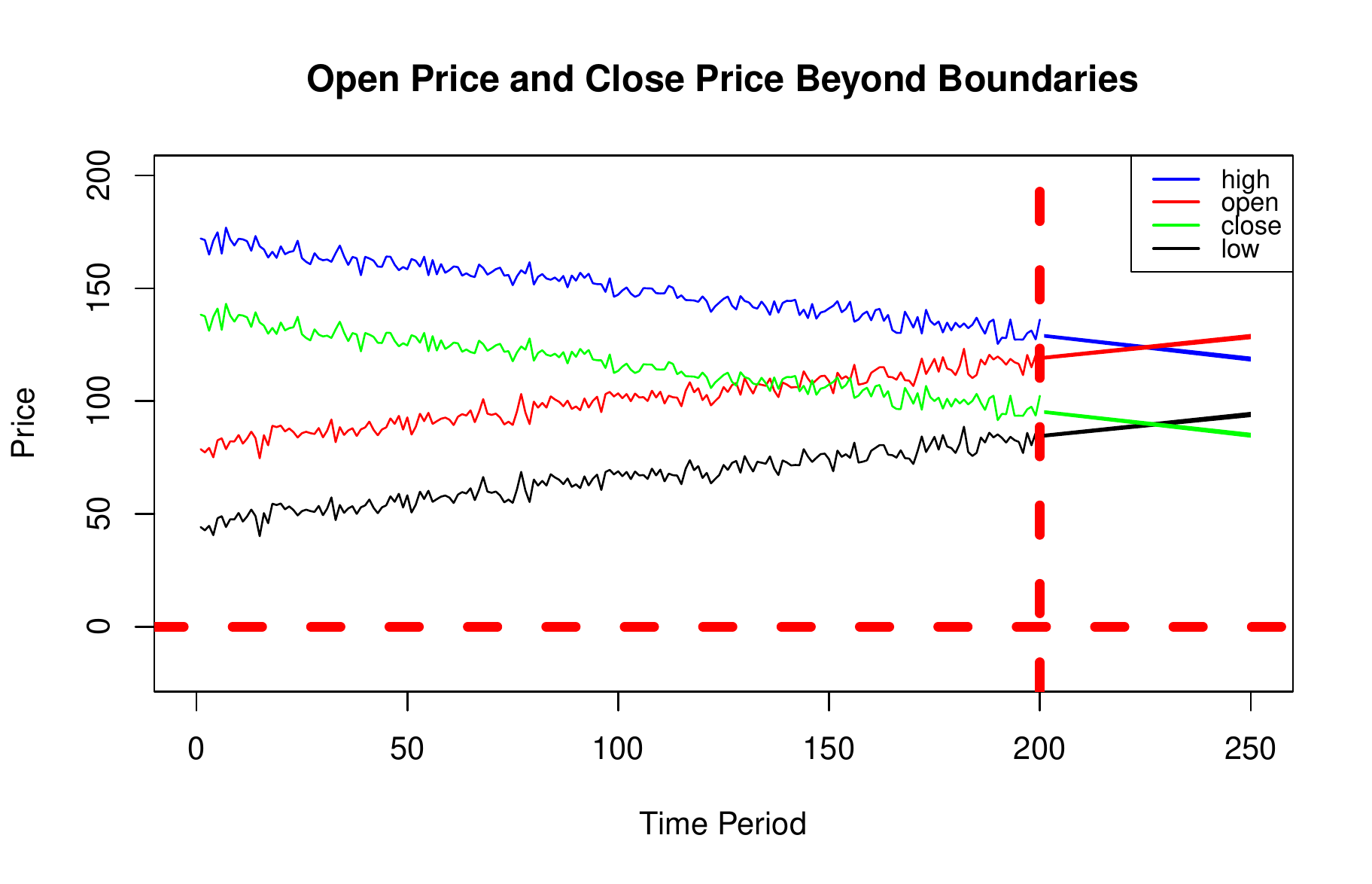}}}
  \hspace{0in}
  \subfigure[]{
    \label{fig:subfig:c}
    \resizebox{5.32cm}{4.8cm}{\includegraphics{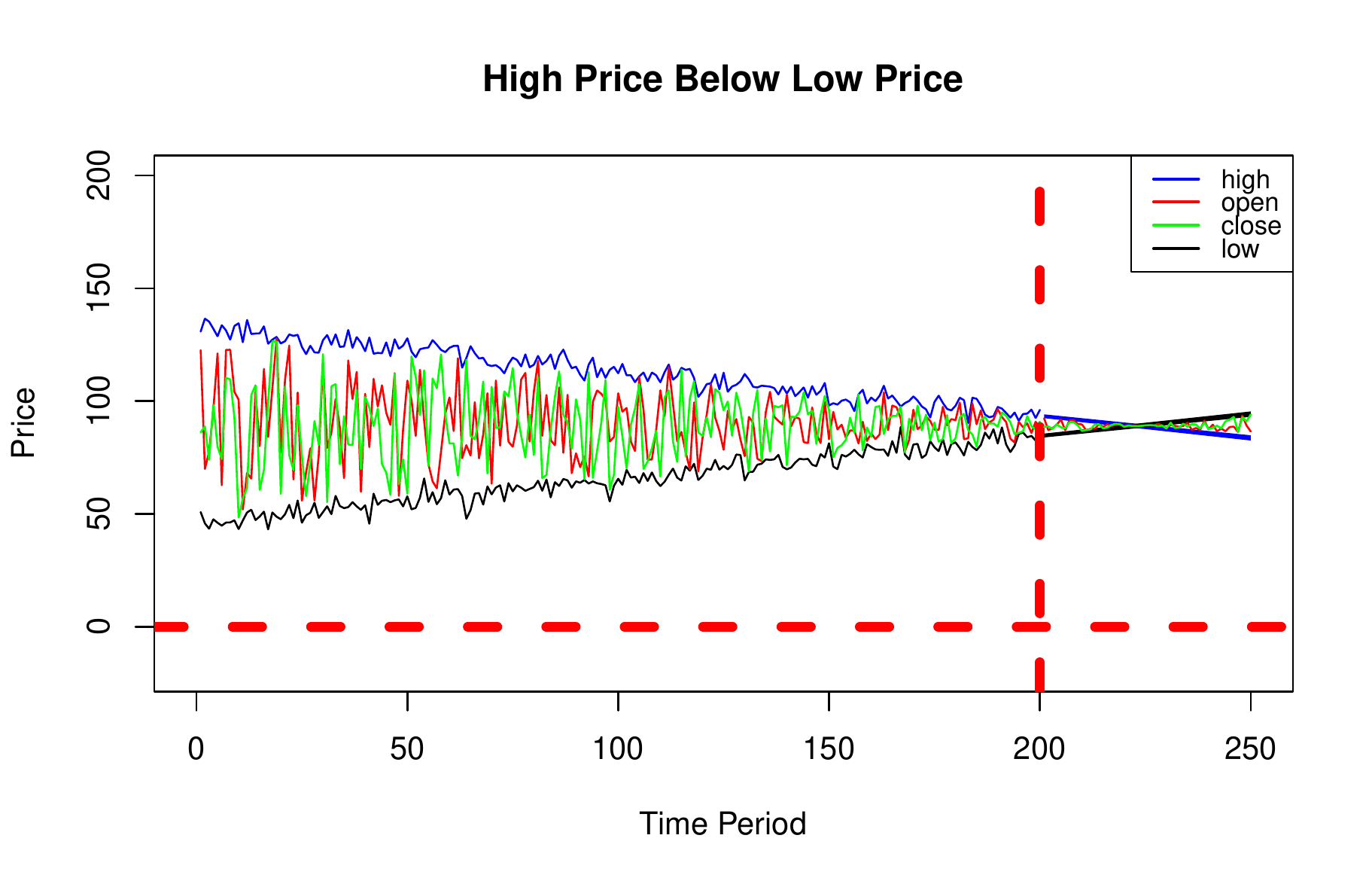}}}
  \caption{Meaningless prediction results caused by ignoring the inherent constraints in OHLC data (The original data contains $200$ periods, and $50$ periods are predicted forward by the linear models. The red dotted line perpendicular to the vertical axis indicates zero value, and the red dotted line perpendicular to the horizontal axis indicates the $200$th period, whose right side are predicted values with a confidence interval of $95\%$ confidence level.)}
  \label{fig:toycase}
\end{figure}

Secondly, existing literature may not guarantee a meaningful forecasting of OHLC data.
On the one hand, most machine learning models can only predict part information of OHLC data, such as high-low range \citep{2009Forecasting, von2012forecasting}, high and low prices \citep{von2014forecasting}, or close price \citep{liu2012fluctuation}.
On the other hand, although some literature attempted to forecast the four variables of OHLC data, the inherent constraints of OHLC data were not well handled, see \cite{manurung2018algorithm} and \cite{kumar2019stock}.
Specifically, the OHLC data requires that all the values are positive, the low price must be smaller than the high price, and the open and close prices must be within the two boundaries.
The direct use of the time series modelling approaches for the four variables of OHLC data without accommodating the constraints may yield unrealistic and meaningless results, i.e., (a) the low price of the forecast period becomes negative without practical significance (shown in Fig.\ref{fig:subfig:a});
(b) the predicted open price (or close price) breaks through the high price (or the low price) boundary (shown in Fig.\ref{fig:subfig:b});
(c) the high price of the forecast period is smaller than the low price (shown in Fig.\ref{fig:subfig:c}).

Apart from the aforementioned  concerns, it is also necessary to propose a novel method to forecast OHLC data as a whole, from which we may benefit compared with the partial information forecasting methods.
First, the prediction of open-high-low-close prices can provide significant help for investors to develop more sophisticated investment plans. Specifically, according to the traditional prediction of the close price, investors can only try to buy in certain financial product on the opening quotation and sell out on the closing quotation if upward trend is forecasted.
While if the OHLC data is predicted, one is able to buy in certain financial product with price near the predicted low price, and sell out with price near the predicted high price to gain excess profits.
Second, candlestick charts can be drew according to OHLC data, whose pattern can reveal the power of demand and supply in financial markets, as well as reflect market conditions and emotion \citep{nison2001japanese, Tsai2014Stock}. Graphical analysis can give further investment advise based on the forecasted candlestick chart.
Third, full information of OHLC data can be reserved. A full information set of the open-high-low-close prices can enhance the reliability and explanatory ability of researches. As pointed out in \cite{cheung2007empirical} and \cite{fiess2002towards}, the open-high-low-close prices as a whole are proved to be of significant power in explaining the price fluctuation.
Finally, opening the possibility of applying a wide range of multivariate modelling techniques to explore the dynamic and structural relationship between the four-dimensional variables of OHLC data \citep{fiess2002towards}.

To this end, we proposed a new transformation method to transform the OHLC data from the original four-dimensional constrained subspace to the four-dimensional real domain full space along with its explicit inverse transformation, which ensures the predicting models obtain meaningful open-high-low-close prices.
As an example combining with time series analysis, we illustrated the detailed procedure of the VAR and VEC modelling of OHLC data.
Ample simulation experiments under different forecast periods, time period basements as well as signal-noise ratios were conducted to validate the effectiveness and stability of the transformation method.
Further, three financial data sets of the Kweichow Moutai, CSI $100$ index and $50$ ETF from their appearance on the Chinese stock market to $14/6/2019$ were used to illustrate the empirical utility of the proposed method.
The results showed a satisfying prediction effect.

Compared with existing literature, the advantages of the proposed transformation method are three folded.
First, it takes full advantage of the information contained in the OHLC data. The open price, high price, low price and close price are all considered, which enables a more efficient analysis.
Second, it can well handle the inherent constraints in the OHLC data. The inherent constraints of OHLC data are satisfied during the whole process of numerical modelling without increasing the complexity of the model, which enables more interpretable results.
Third, the proposed method provides a unified framework for a variety of positive interval data which has minimum and maximum boundaries greater than zero, and multi-valued sequences within the two boundaries, such as daily temperature and profit of companies.
Meanwhile, the method of dealing with the transformed variables can be generalized to other statistical models and machine learning methods.
From this perspective, the proposed method provides a completely new and useful alternative for OHLC data analysis, enriching the existing literature.

The remainder of this paper is organised as follows. In Section \ref{Sec 2}, we introduce the mathematical definition of OHLC data and its inherent constraints, and in Section \ref{Sec 2-methodology}, we propose the transformation and inverse transformation formulas to relax the inherent constraints of OHLC data, and illustrate the VAR and VEC modelling process for OHLC data. Section \ref{Sec 3} presents simulation studies and Section \ref{Sec 4} shows the empirical application of the proposed method in real financial market. Finally, we conclude with a brief discussion in Section \ref{Sec 5}.

\section{Preliminaries}\label{Sec 2}

\begin{figure}[!h]
  \centering
  \includegraphics[scale=0.5]{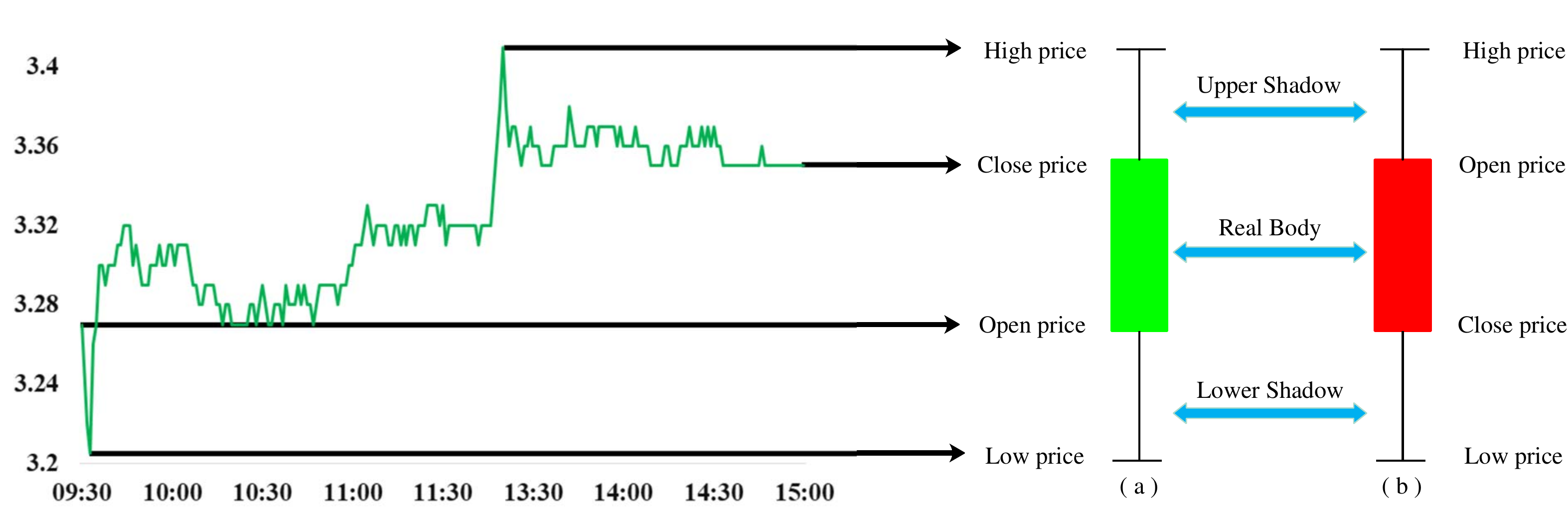}
  \caption{\rm An example of daily candlestick chart}\label{Fig Kline}
\end{figure}

To grasp an intuitive picture of the candlestick chart, here we take the daily candlestick chart as an example (in this article, if there is no special explanation, all candlestick charts refer to daily candlestick chart), as shown in Fig.\ref{Fig Kline}.
Obviously, a daily candlestick chart can not only record the open price, high price, low price and close price of a certain stock on the day, but also can visually reflect the difference between any two prices.

Generally, the candlestick chart is divided into two categories as shown in Fig.\ref{Fig Kline}.
Specifically, Fig.\ref{Fig Kline}(a) indicates that the close price is higher than the open price, which corresponding to a bull market;
while Fig.\ref{Fig Kline}(b) corresponds to a bear market with the close price being lower than the open price. In the stock market of the United States, green and red are habitually used to mark the real body of candlestick chart of bull market and bear market respectively.
If the daily candlestick charts are arranged in chronological order, a sequence reflecting the historical price changes of a certain financial product is formed, called the candlestick chart series, and the corresponding data is called OHLC series.

Actually, the essence of OHLC series is a four-dimensional time series of stock prices with three inherent constraints.
First, all the four prices should be positive due to the limitation that the values of OHLC data in the financial market cannot be less than zero;
Second, the high price must be higher than the low price of the same day;
Third, the open price and close price should fall into the two boundaries.
To represent it in mathematical form, for any time period $t$, we give the following definition of OHLC data.

\begin{definition}\label{k-line}
A four-dimensional vector $\bm{X}_t=(x_t^{(o)}, x_t^{(h)}, x_t^{(l)}, x_t^{(c)})^T$ is a typical OHLC data if it satisfies:
\begin{itemize}
\centering
  \item [1.] $x_t^{(l)} > 0$;
  \item [2.] $x_t^{(l)} < x_t^{(h)}$;
  \item [3.] $x_t^{(o)}, x_t^{(c)} \in [x_t^{(l)},x_t^{(h)}]$.
\end{itemize}
Here $x_t^{(o)}$ is the $t$ period daily open price, $x_t^{(h)}$ is the $t$ period daily high price, $x_t^{(l)}$ is the $t$ period daily low price, and $x_t^{(c)}$ is the $t$ period daily close price.
\end{definition}

For the time period $\mathcal{T}=[1,T]$, the collection of $\bm{X}_t$ for any $t\in\mathcal{T}$ forms the OHLC time series, denoted by
$$\bm{S}=\{\bm{X}_t\}_{t=1}^T.$$
Compared with the ordinary real domain vector, the biggest difference of the vectors in $\bm{S}$ is that there are intrinsic constraint formulas between its four components, which poses great challenge to classical statistical analysis.
Specifically, to establish a time series model of OHLC series, the most difficult problem is how to always ensure that the calculation process and the prediction results are also subject to these constraint formulas.
That is, after obtaining the prediction results at the prediction period $(T+m) \; (m \in \mathbb{R}^+)$ by time series modelling, it must be ensured that:
\begin{eqnarray*}
  x_{T+m}^{(l)}&>&0,\\
  x_{T+m} ^{(l)} &<& x_{T+m} ^{(h)}, \\
  x_{T+m} ^{(o)},x_{T+m}^{(c)} &\in& [x_{T+m}^{(l)},x_{T+m}^{(h)}].
\end{eqnarray*}

Obviously, these constraints are not guaranteed to be valid if we directly apply the time series forecasting methods to the original four time series of OHLC data.
To address this problem, a common practice is to remove these inherent constraints via some proper data transformation, then we can forecast the transformed time series data freely. Finally, we can obtain the forecaster of the original OHLC data using the corresponding inverse transformation.

\section{Methodology} \label{Sec 2-methodology}

In this section, we first proposed a flexible transformation method along with its inverse transformation method for OHLC data, as well as a model-independent framework for modelling OHLC data are proposed in Section \ref{Sec_solve}.
And then we use VAR and VEC models as a case of the framework and present the corresponding forecasting procedure in Section \ref{Sec_VAR}.

\subsection{\textit{Data transformation method}} \label{Sec_solve}

Note that from Definition \ref{k-line}, the first constraint is $x_t^{(l)}>0$, which can be relaxed via the most commonly used logarithm transformation.
That is,
\begin{equation}\label{Eq_1}
  y_t^{(1)} = \ln x_t^{(l)}.
\end{equation}
It is quite clear that the transformed data $y_t^{(1)}$ in Eq.(\ref{Eq_1}) satisfies $-\infty<y_t^{(1)}<+\infty$ with no positive constraint.
Moreover, it not only preserves a positive relative relationship between the original data as logarithm transformation is a monotonous increasing function,
but also compresses the scale of the data, which reduces the absolute values of the original data and makes the data more stable to some extent.

Secondly, in order to guarantee the second constraint $x_t^{(l)}<x_t^{(h)}$, i.e., $x_t^{(h)}-x_t^{(l)}>0$, the same practice as that in Eq.(\ref{Eq_1}) yields
\begin{equation}\label{Eq_2}
  y_t^{(2)} = \ln(x_t^{(h)}-x_t^{(l)}),
\end{equation}
where $y_t^{(2)}$ is also free of any constraint, which can be modelled easily.

Finally, the last constraint is $x_t^{(o)}, x_t^{(c)} \in [x_t^{(l)},x_t^{(h)}]$, implying that both the open price and close price must be greater than the low price and less than the high price.
Obviously, without properly processing the raw data, it is very likely that the predicted open/close price is beyond boundaries.
To remedy this situation, inquired by the idea of convex combination, here we introduce two proxy data $\lambda_t^{(o)}$ and $\lambda_t^{(c)}$, which are formulated as
\begin{equation}\label{Eq_3}
  \lambda_t^{(o)} = \frac{x_t^{(o)}-x_t^{(l)}}{x_t^{(h)}-x_t^{(l)}} \ \ \ \mbox{and}\ \ \ \lambda_t^{(c)} = \frac{x_t^{(c)}-x_t^{(l)}}{x_t^{(h)}-x_t^{(l)}}.
\end{equation}

Obviously, $0\leqslant\lambda_t^{(o)},\lambda_t^{(c)}\leqslant1$, and the original data $x_t^{(o)}$ and $x_t^{(c)}$ can be obtained as follows, respectively.
That is,
\begin{equation}\label{Eq xo}
  x_t^{(o)}=\lambda_t^{(o)}x_t^{(h)}+(1-\lambda_t^{(o)})x_t^{(l)},
\end{equation}
\begin{equation}\label{Eq xc}
  x_t^{(c)}=\lambda_t^{(c)}x_t^{(h)}+(1-\lambda_t^{(c)})x_t^{(l)}.
\end{equation}

Thus, the original constraint $x_t^{(o)}, x_t^{(c)} \in [x_t^{(l)},x_t^{(h)}]$ reduces to $0\leqslant\lambda_t^{(o)},\lambda_t^{(c)}\leqslant1$ if we deal with the proxy data $\lambda_t^{(o)}$ and $\lambda_t^{(c)}$ instead of $x_t^{(o)}$ and $x_t^{(c)}$.
Moreover, $\lambda_t^{(o)}$ and $\lambda_t^{(c)}$ make sense.
Specifically, if $\lambda_t^{(o)}$ has a growing trend, the open price $x_t^{(o)}$ will be closer to the high price $x_t^{(h)}$;
while a downward trend of $\lambda_t^{(o)}$ indicates that the open price $x_t^{(o)}$ will be closer to the low price $x_t^{(l)}$.
Similarly, the explanation for $\lambda_t^{(c)}$ can also be obtained.

To further remove the constraint $0\leqslant\lambda_t^{(o)},\lambda_t^{(c)}\leqslant1$ on $\lambda_t^{(o)}$ and $\lambda_t^{(c)}$, following the idea of logistic regression, we propose the logit transformation to obtain the unconstrained data $y_t^{(3)}$ and $y_t^{(4)}$ as follows:
\begin{equation}\label{Eq 7}
  y_t^{(3)}=\ln\frac{\lambda_t^{(o)}}{1-\lambda_t^{(o)}},
\end{equation}
\begin{equation}\label{Eq 8}
  y_t^{(4)}=\ln\frac{\lambda_t^{(c)}}{1-\lambda_t^{(c)}}.
\end{equation}

Up to now, via the transformation process, the raw OHLC data $\bm{X}_t=(x_t^{(o)}, x_t^{(h)}, x_t^{(l)}, x_t^{(c)})^T$ is transformed to the unconstraint four-dimensional data $\bm{Y}_t=(y_t^{(1)}, y_t^{(2)},y_t^{(3)},y_t^{(4)})^T.$
In summary, the proposed transformation method can be described as
\begin{equation}\label{transform}
  \bm{Y}_t=\left(
         \begin{array}{c}
           y_t^{(1)} \\
           y_t^{(2)} \\
           y_t^{(3)} \\
           y_t^{(4)} \\
         \end{array}
       \right)=\left(
                 \begin{array}{c}
                   \ln x_t^{(l)} \\
                   \ln(x_t^{(h)}-x_t^{(l)}) \\
                   \ln\Big(\frac{\lambda_t^{(o)}}{1-\lambda_t^{(o)}}\Big) \\
                   \ln\Big(\frac{\lambda_t^{(c)}}{1-\lambda_t^{(c)}}\Big) \\
                 \end{array}
               \right),
\end{equation}
where $\lambda_t^{(o)}$ and $\lambda_t^{(c)}$ are defined in Eq.(\ref{Eq_3}). Not only does the transformation in Eq.(\ref{transform}) have ranges from $-\infty$ to $+\infty$ and an explicit inverse for the values in its range, but it also shares the flexibility of the well known log- and logit- transformation.

Therefore, the predictive modelling of the OHLC series $\{\bm{X}_t\}_{t=1}^T$ can be transformed into the predictions for the unconstrained series $\{\bm{Y}_t\}_{t=1}^T$ with the whole real number domain and variance stability, which will provide significant convenience for subsequent statistical modelling.
That is, we can apply the classical forecasting models (ARMA, ARIMA, VAR, VEC etc) or machine learning models to $\{\bm{Y}_t\}_{t=1}^T$.
After obtaining the forecaster of $\bm{Y}_t$, we can obtain the corresponding forecaster of $\bm{X}_t$ via the inverse transformation as follows:
\begin{equation}\label{inversetransform}
  \bm{X}_t=\left(
         \begin{array}{c}
           x_t^{(o)} \\
           x_t^{(h)} \\
           x_t^{(l)} \\
           x_t^{(c)} \\
         \end{array}
       \right)=\left(
                 \begin{array}{c}
                   \lambda_t^{(o)}(\exp\{y_t^{(1)}\}+\exp\{y_t^{(2)}\})+(1-\lambda_t^{(o)})\exp\{y_t^{(1)}\} \\
                   \exp\{y_t^{(1)}\}+\exp\{y_t^{(2)}\} \\
                   \exp\{y_t^{(1)}\} \\
                   \lambda_t^{(c)}(\exp\{y_t^{(1)}\}+\exp\{y_t^{(2)}\})+(1-\lambda_t^{(c)})\exp\{y_t^{(1)}\}\\
                 \end{array}
               \right),
\end{equation}
where
\begin{equation}\label{lambda}
  \lambda_t^{(o)}=\frac{\exp\{y_t^{(3)}\}}{1+\exp\{y_t^{(3)}\}}  \ \ \ \mbox{and}\ \ \ \lambda_t^{(c)}=\frac{\exp\{y_t^{(4)}\}}{1+\exp\{y_t^{(4)}\}}.
\end{equation}

The above transformation Eq.(\ref{transform}) and inverse transformation Eq.(\ref{inversetransform}) methods provide a new perspective for forecasting the OHLC data, which will make the prediction results obey its three inherent constraints listed in Definition \ref{k-line}.
Furthermore, the proposed method has great feasibility, which can be easily generalized to any type of positive interval data that owns the minimum and maximum boundaries greater than zero, and multi-valued sequences between the two boundaries.

It should be noticed that in the transformation  process, we  assume that $x^{(o)}, x^{(h)}, x^{(l)}$, and $x^{(c)}$ are not equal to each other $(\text{except for} \; x^{(o)} \equiv x^{(c)})$. In other words, $x^{(h)} \neq x^{(l)} \neq 0$ and $\lambda^{(o)}, \lambda^{(c)} \notin \left\{ 0,1 \right\}$. However, such assumptions are inevitably spoiled sometimes in the real financial market. Here we list the circumstances that makes these assumptions invalid and give the measurement to deal with them accordingly.
(1) When the subject is on trade suspension and all prices equal to $0$, namely, $x^{(o)}=x^{(h)}=x^{(l)}=x^{(c)}=0$, we exclude these extreme cases in the raw data.
(2) When $\lambda^{(o)}$ or $\lambda^{(c)}$ is equal to $0$, it corresponds to $x^{(o)}=x^{(l)}$ or $x^{(c)}=x^{(l)}$, respectively.
We add a random term to $x^{(o)}$ or $x^{(c)}$ and make $\lambda^{(o)}$ or $\lambda^{(c)}$ slightly greater than 0.
(3) When $\lambda^{(o)}$ or $\lambda^{(c)}$ is equal to $1$, it indicates that $x^{(o)}=x^{(h)}$ or $x^{(c)}=x^{(h)}$, respectively. We subtract a random term from $x^{(o)}$ or $x^{(c)}$ to make $\lambda^{(o)}$ or $\lambda^{(c)}$ slightly less than $1$.
(4) When certain subject reaches limit-up or limit-down as soon as the opening quotation, that is, $x^{(o)}=x^{(h)}=x^{(l)}=x^{(c)}\neq0$.
If limit-up(limit-down) happens, we firstly multiply $x^{(c)}(x^{(o)})$ and $x^{(h)}$ by 1.1 to make a relatively large interval. And then conduct measurements given in circumstances (2) and (3).

In summary, the  general forecasting framework for OHLC data with $T$ periods is described in Algorithm \ref{algorithm1}.
\begin{algorithm}[!h]
  \caption{General forecasting framework for OHLC data}\label{algorithm1}
  \begin{algorithmic}[1]
\State Get the raw candlestick charts with $T$ periods from the capital market;
\State Extract the four-dimensional time series data of the candlestick charts, record as $\{\bm{X}_t\}_{t=1}^T$;
\State Conduct transformation method to $\{\bm{X}_t\}_{t=1}^T$ and obtain $\{\bm{Y}_t\}_{t=1}^T$ according to Eq.(\ref{transform});
\State Model $\{\bm{Y}_t\}_{t=1}^T$ via prediction models to get forecasted future values $\widehat{\bm{Y}_t}$;
\State Conduct inverse transformation method to $\widehat{\bm{Y}_t}$ according to Eq.(\ref{inversetransform}) and the original forecaster of OHLC data $\widehat{\bm{X}_t}$ can be derived.
    \end{algorithmic}
\end{algorithm}

\subsection{\textit{The VAR and VEC modelling process for OHLC data}} \label{Sec_VAR}

Here we employ the VAR and VEC models as an example of the framework proposed in Algorithm \ref{algorithm1} and present the corresponding procedure for forecasting OHLC data.

\subsubsection{\textit{VAR model for OHLC data}} \label{VAR}

As one of the most widely used multiple time series analysis methods, the VAR model proposed by  \cite{sims1980macroeconomics} has become an important research tool in economic studies, with advantages of capturing the linear interdependencies among multiple time series \citep{pesaran1998autoregressive}.
According to Algorithm \ref{algorithm1}, we should build the forecast model for the unconstrained four-dimensional time series data $\{\bm{Y}_t\}_{t=1}^T$.
Without loss of generality, we first assume that all time series in $\bm{Y}_t$ are stationary, then a $p$-order ($p\geqslant1$) VAR model, denoted by VAR($p$), can be formulated as:
\begin{equation}\label{VAR}
  \bm{Y}_t=\bm{\alpha}+\bm{A}_1\bm{Y}_{t-1}+...+\bm{A}_p\bm{Y}_{t-p}+\bm{w}_t= \bm{\alpha}+\sum\limits_{j = 1}^{p}\bm{A}_j\bm{Y}_{t-j}+\bm{w}_t, \ \ \  t=(p+1),...,T
\end {equation}
where $\bm{Y}_{t-j}$ is the $j$-th lag of $\bm{Y}_t$; $\bm{\alpha}=(\alpha_1,\alpha_2,\alpha_3,\alpha_4)^T$ is a four-dementional vector of intercepts;
$\bm{A}_j$ stands for the time-invariant $4\times4$ coefficient matrix; and $\bm{w}_t=(w_t^{(1)},w_t^{(2)},w_t^{(3)},w_t^{(4)})^T$ is a four-dementional error term vector satisfying:
\begin{itemize}
\item[(1)] Mean zero: $E(\bm{w}_t)=\bm{0}$;
\item[(2)] No correlation across time: $E(\bm{w}_{t-k}^T\bm{w}_t)=0$, for any non-zero $k$.
\end{itemize}

Writing Eq.(\ref{VAR}) in the concise matrix form yields
\begin{equation}\label{VAR-matrix}
  \bm{Y}=\bm{B}\bm{Z}+\bm{U},
\end{equation}
where $\bm{Y}=[\bm{Y}_{p+1}, \bm{Y}_{p+2},\cdots,\bm{Y}_T]$ is a $4\times(T-p)$ matrix; $\bm{B}=[\bm{\alpha},\bm{A}_1,\cdots,\bm{A}_p]$ is a $4\times (4p+1)$ coefficient matrix; $\bm{U}=[\bm{w}_{p+1},\bm{w}_{p+2},\cdots,\bm{w}_T]$ is a $4\times (T-p)$ error term matrix; and
\begin{equation*}
  \bm{Z}=\begin{pmatrix}
  1&1&\cdots&1\\
  \bm{Y}_p&\bm{Y}_{p+1}&\cdots&\bm{Y}_{T-1}\\
  \vdots&\vdots&\ddots&\vdots\\
  \bm{Y}_1&\bm{Y}_2&\cdots&\bm{Y}_{T-p}
  \end{pmatrix}
\end{equation*}
is a $(4p+1)\times (T-p)$ matrix.
Then we can solve for the coefficient matrix $\bm{B}$ using the least squares estimation \citep{lutkepohl2005new}, that is:
\begin{equation}\label{solution}
  \widehat{\bm{B}}=\big(\bm{Z}^T\bm{Z}\big)^{-1}\bm{Z}^T\bm{Y}.
\end{equation}

\subsubsection{\textit{VEC model for OHLC data}} \label{VEC}

Note that the reliability of the VAR model estimation is closely related to the stationarity of the variable sequences.
If this assumption does not hold, we may use a restricted VAR model, i.e., the VEC model, in the presence of the co-integration among variables;
otherwise the variables have to be differenced by $d$ times firstly until they can be modelled by VAR or VEC model.
As evidenced in \cite{cheung2007empirical}, for the US stock markets, stock prices are typically characterized by $I(1)$ processes and the daily highs and lows follow a co-integration relationship.
This implies that perhaps the VEC model is a more practically relevant model than the VAR model in the context of forecasting the OHLC series.
Here we use the augmented Dickey-Fuller (ADF) unit root test to examine the stationary of each variable, and the Johansen test \citep{johansen1988statistical} is employed to determine the presence or absence of the co-integration relationship.

Assume that $\bm{Y}_t$ is integrated of order one, then the corresponding VEC model takes the following form:
\begin{equation}\label{vecm}
\Delta\bm{Y}_t=\sum_{j=1}^{p-1}\bm{\Gamma}_j\Delta\bm{Y}_{t-j}+\bm{\gamma}\bm{\beta}^T\bm{Y}_{t-p}+\bm{\alpha}+\bm{w}_t,
\end{equation}
where $\Delta$ denotes the first difference, $\sum_{j=1}^{p-1}\bm{\Gamma}_j\Delta\bm{Y}_{t-j}$ and $\bm{\gamma}\bm{\beta}^T\bm{Y}_{t-p}$ are the VAR component of the first difference and error-correction component, respectively.
Here $\bm{\Gamma}_j$ is a $4\times 4$ matrix that represents short-term adjustments among variables across four equations at the $j$-th lag;
two matrices, $\bm{\gamma}$ and $\bm{\beta}$, are of dimension $4\times r$ with $r$ being the order of co-integration, where $\bm{\gamma}$ denotes the speed of adjustment (loading) and $\bm{\beta}$ represents the co-integrating vector, which can be obtained by the Johansen test \citep{johansen1988statistical, cuthbertson1992applied};
$\bm{\alpha}$ is a $4\times 1$ constant vector representing a linear trend; $p$ is a lag structure; and $\bm{w}_t$ is the $4\times 1$ vector of white noise error term.

For the VEC model in Eq.(\ref{vecm}), \cite{johansen1991estimation} employed the full information maximum likelihood method to implement its estimation.
Specifically, the main procedure consists of (i) testing whether all variables are integrated of order one by applying a unit root test \citep{Lai1991A},
(ii) determining the lag order $p$ such that the residuals from each equation of the VEC model are uncorrelated,
(iii) regressing $\Delta\bm{Y}_t$ against the lagged differences of $\Delta\bm{Y}_t$ ($\bm{Y}_{t-p}$) and estimating the eigenvectors (co-integrating vectors) from the Canonical correlations of the set of residuals from these regression equations, and finally (iv) determining the order of co-integration $r$.

\subsubsection{\textit{Discussion of parameter selection in VAR and VEC models}} \label{parameters}

Finally, we discuss the determination of the lag order $p$ in the VAR model and the order of co-integration $r$ in the VEC model for modelling OHLC data.
First, for $p$, on the one hand, it should be made large enough to fully reflect the dynamic characteristics of the constructed model; while on the other hand, an increase of $p$ will cause an increase of the parameters to be  estimated, thus the freedom degree of the model decreases. A trade-off must be evaluated to choose $p$, the common used criterions in practice are AIC, BIC and HQ (Hannan-Quinn).
In this paper, we prefer AIC because of its conciseness, which is formulated as
\begin{equation}\label{aic}
  \mbox{AIC}(p)=\ln\frac{\sum\limits_{i = 1}^{4}\sum\limits_{j = 1}^{T}\hat{u}_{ij}^2}{{T}}+\frac{2pK^2}{T},
\end{equation}
where $T$ stands for the total period number of OHLC series, $p$ is VAR lag order, $K$ is the VAR dimension, and $\hat{u}_{ij}=\hat{Y}_{j}^{(i)}-Y_{j}^{(i)}(1\leq i\leq4,1\leq j\leq T)$ represents for the residuals of the VAR model.
The optimal $p$ is obtained via minimizing $\mbox{AIC}(p)$.

Second, as the order of co-integration, $r$ indicates the dimension of the co-integrating space, and (i) if the rank of $\bm{\gamma}\bm{\beta}^T$ is $4$, i.e., $r=4$, $\bm{Y}_t$ is already stationary;
(ii) if $\bm{\gamma}\bm{\beta}^T$ is a null matrix, i.e., $r=0$, the proper specification of Eq.(\ref{vecm}) is one without the error correction term and degenerate to a VAR model;
and (iii) if the rank of $\bm{\gamma}\bm{\beta}^T$ is between $0$ and $4$, i.e., $0<r<4$, there exist $r$ linearly independent columns in the matrix and $r$ co-integrating relations in the system of equations.
Along the line of \cite{johansen1991estimation}, $r$ is determined by constructing the $``$Trace" or $``$Eigen" test statistics, which are two widely used methods in Johansen test. For more details, please refer to the monograph \cite{johansen1995likelihood} and \cite{lutkepohl2005new}.

\subsubsection{\textit{Unified modelling framework for OHLC data}} \label{framework}

In summary, as one of the popular econometric forecasting models in Step $4$ of Algorithm \ref{algorithm1}, the main implementation of VAR and VEC modelling can be summarized as Algorithm \ref{algorithm2}.

\begin{algorithm}[!h]
  \caption{VAR and VEC modelling framework}\label{algorithm2}
  \begin{algorithmic}[1]
\State Divide $\{\bm{Y}_t\}_{t=1}^T$ into segments $\bm{Y}_i^{(q)}$, each segment involves $q$ periods, and each segment rolls forward by one period;
\State Conduct ADF test on the four-dimensional time series data in $\bm{Y}_i^{(q)}$. If they are all stationary processes, proceed to Step $5$, otherwise proceed to Step 3;
\State Use Johansen method to perform the co-integration test on $\bm{Y}_i^{(q)}$, and determine $r$.
If there is a co-integration relationship, proceed to Step $6$, otherwise go to Step 4;
\State  Take one-order difference on non-stationary time series in $\bm{Y}_i^{(q)}$, and then proceed back to Step 2. The optimal situation is iterating until $\bm{Y}_i^{(q)}$ can be modelled by VAR in Eq.(\ref{VAR}) or VEC model in Eq.(\ref{vecm}).
After establishing the VAR or VEC model, the data is then restored back to the corresponding values before the difference.
\State Model $\bm{Y}_i^{(q)}$ with VAR model and predict $p$ periods forwards to obtain $\widehat{\bm{Y}_i}^{(q)}$. Conduct inverse transformation method on $\widehat{\bm{Y}_i}^{(q)}$ to obtain $\widehat{\bm{X}_i}^{(q)}$, evaluate the prediction accuracy and end.
\State Model $\bm{Y}_i^{(q)}$ with VEC model and predict $p$ periods forwards to obtain $\widehat{\bm{Y}_i}^{(q)}$. Conduct inverse transformation method on $\widehat{\bm{Y}_i}^{(q)}$ to obtain $\widehat{\bm{X}_i}^{(q)}$, evaluate the prediction accuracy and end.
    \end{algorithmic}
\end{algorithm}

\begin{figure}[!h]
  \centering
    \resizebox{17cm}{5.5cm}{\includegraphics{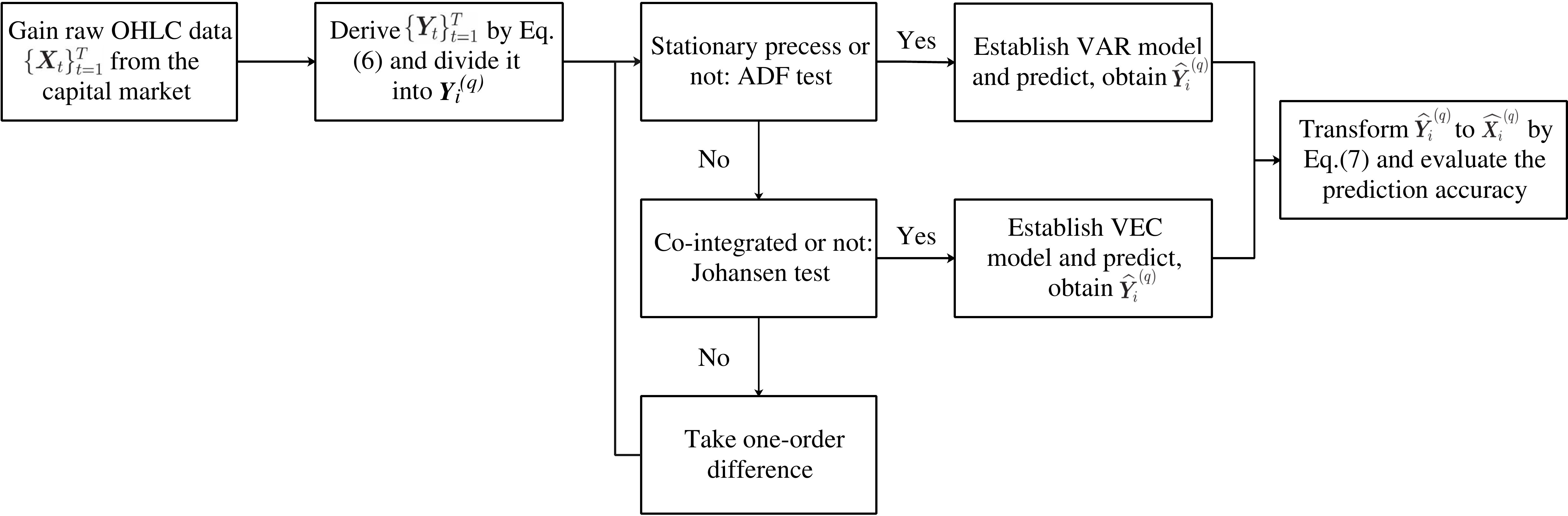}}
  \caption{\rm An unified framework for modelling OHLC data based on VAR and VEC models}\label{Fig process}
\end{figure}

Incorporating Algorithm \ref{algorithm2} into Algorithm \ref{algorithm1}, we can obtain the unified framework for statistical modelling of the OHLC data, shown in Fig.\ref{Fig process}.

\vspace{5mm}

\section{Simulations} \label{Sec 3}

We assess the performance of our proposed method via finite sample simulation studies. We firstly describe the data construction in Section \ref{sec3-sub1}; then give five indicators to measure the difference between predicted values and observed ones in Section \ref{sec3-sub2}; and finally report the simulation results in Section \ref{sec3-sub3}.

\subsection{\textit{Data construction}}\label{sec3-sub1}

We generate simulation data under the VAR model structure in Eq.(\ref{VAR}) as follows: (i) Assign the lag period $p$, and the coefficient matrices $\bm{A}_1, \bm{A}_2, \cdots, \bm{A}_p$; (ii) Generate an original four-dimensional vector $\bm{Y}_1=[y^{(1)}_{1}, y^{(2)}_{1}, y^{(3)}_{1}, y^{(4)}_{1}]^T$; (iii) Generates $\{\bm{Y}_t\}_{t=2}^T$ in a sequence via the VAR($p$) model
$$\bm{Y}_t=\bm{A}_1\bm{Y}_{t-1}+...+\bm{A}_p\bm{Y}_{t-p}+\bm{w}_t,$$ where $\bm{w}_t$ follows the multivariate normal distribution with zero mean and covariance matrix $\Sigma_{\bm{w}}$.
Finally, the simulated OHLC data $\{\bm{X}_t\}_{t=1}^T$ are generated by applying the inverse transformation formula in Eq.(\ref{inversetransform}).

In order to evaluate the performance of the proposed method with different variance component levels, we consider the following scenarios:
\begin{itemize}
  \item [] \textbf{Scenario 1}: $p=1, T=220, \bm{Y}_1=[4,0.7,-0.85,0]^T$ and
  \begin{equation*}
  \bm{A}_1=\begin{pmatrix}
  0.55 & 0.12 & 0.12 & 0.12 \\
  0.12 & 0.55 & 0.12 & 0.12 \\
  0.12 & 0.12 & 0.55 & 0.12 \\
  0.12 & 0.12 & 0.12 & 0.55
  \end{pmatrix},
  \end{equation*}
  and $\Sigma_{\bm{w}}$ is a $4\times 4$ diagonal matrix with diagonal element being $0.05^2$, i.e.,  $$\Sigma_{\bm{w}}=\mbox{diag}\{0.05^2,0.05^2,0.05^2,0.05^2\}.$$
  \item [] \textbf{Scenario 2}:  $p, T, \bm{Y}_1$ and $\bm{A}_1$ follows Scenario $1$ except that $$\Sigma_{\bm{w}}=\mbox{diag}\{0.07^2,0.07^2,0.07^2,0.07^2\}.$$
  \item [] \textbf{Scenario 3}: $p, T, \bm{Y}_1$ and $\bm{A}_1$ follows Scenario $1$ except that $$\Sigma_{\bm{w}}=\mbox{diag}\{0.03^2,0.03^2,0.03^2,0.03^2\}.$$
\end{itemize}

All these scenarios present the transformed unconstrained time series data $\{\bm{Y}_t\}_{t=1}^T$ follows VAR$(1)$, only with different variance component levels according to median (Scenario $1$), low (Scenario $2$) and strong (Scenario $3$) signal-to-noise ratios, respectively.
A higher signal-noise ratio means the information contained in the data comes more from the signal rather than the noise, indicating a better quality of data.
On the contrary, a lower signal-noise ratio means that the noise carries more interference, indicating a worse quality of data.

\begin{figure}[!h]
  \centering
  \resizebox{9cm}{6cm}{\includegraphics{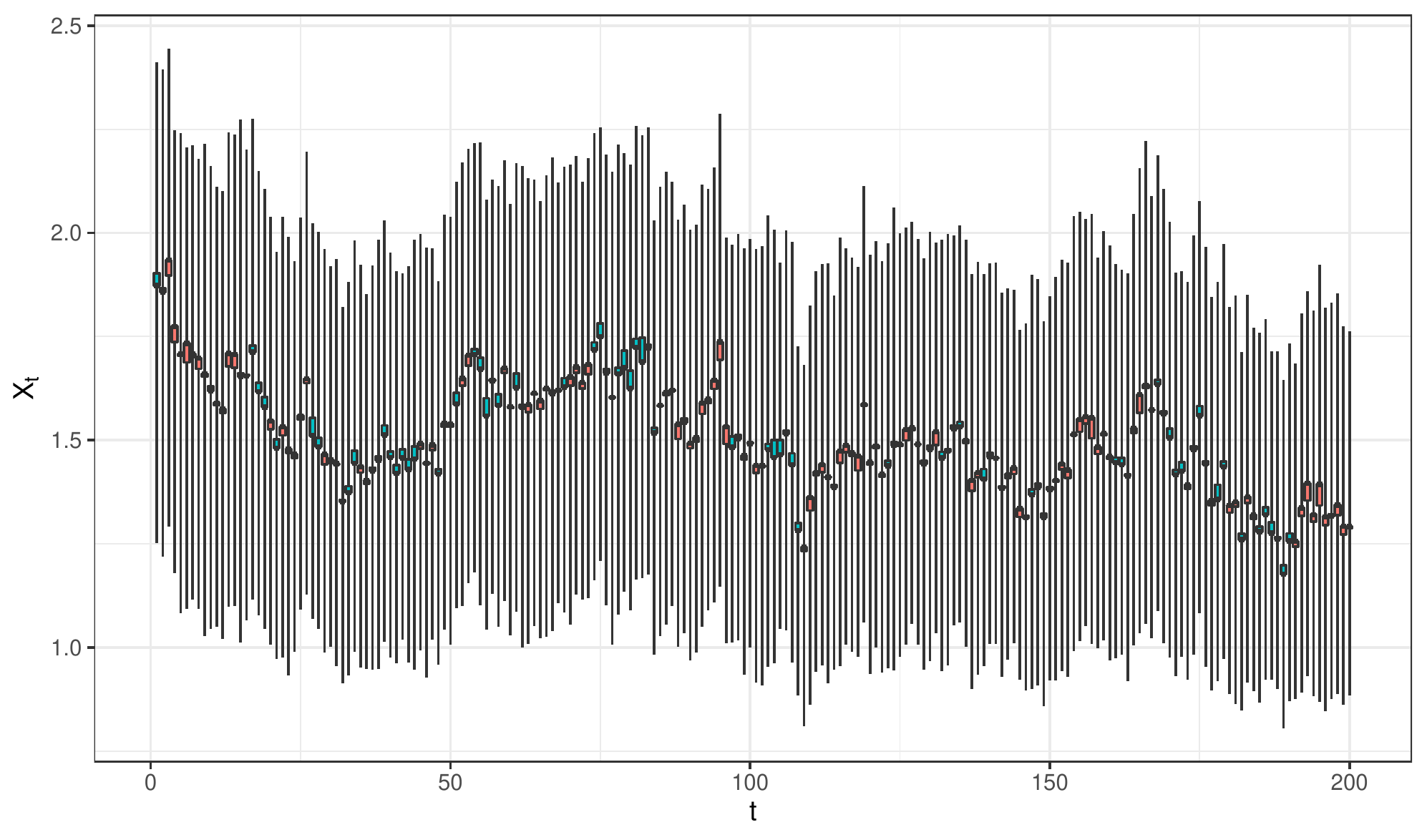}}
  \caption{\rm Simulation OHLC data under Scenario 1}\label{Fig simulation}
\end{figure}

Note that the raw simulation data has $220$ periods, we only take $21$ to $220$ periods as the final simulated data set as the data generated initially may be highly volatile. Take Scenario $1$ as an illustration, Fig.\ref{Fig simulation} shows the simulated OHLC series $\{\bm{X}_t\}_{t=1}^T$.

\subsection{\textit{Measurements}}\label{sec3-sub2}

\begin{figure}[!h]
  \centering
   \resizebox{12cm}{2cm}{\includegraphics{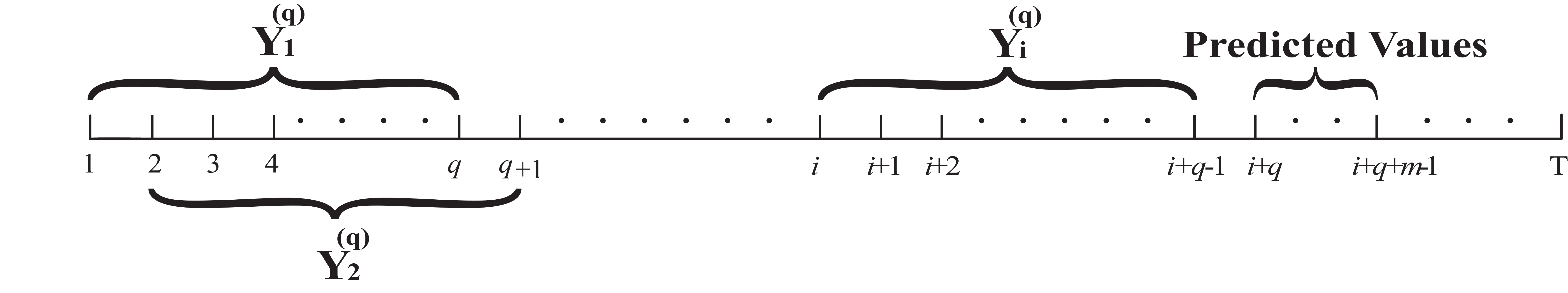}}
  \caption{\rm Specific segment method}\label{Fig introduction}
\end{figure}

According to the process illustrated in Fig.\ref{Fig process}, the VAR and VEC models are used to measure the statistical relationships between $\bm{X}_t$ and $\bm{Y}_t$.
As \cite{corrado1992filter} and \cite{marshall2008candlestick} pointed out, short-term technical analysis can be more helpful to investors than long-term technical analysis. Thus, we focus on relatively short-time analysis here. Specifically, $q$ periods of the simulated data, namely the time period basement, are used to train the model, and make prediction ahead of $m$ periods.
And we set $q$ ranges from $30$ to $70$, and $m=1,2,3$.
For each setting $(q,m)$, $\bm{Y}_i^{(q)}$ scroll forward one period, and predict $(T-q-m+1)$ times in total, as indicated in Fig.\ref{Fig introduction}.

The predicted $\widehat{\bm{Y}_i}^{(q)}$ are firstly derived, then the predicted $\widehat{\bm{X}_i}^{(q)}$ are obtained based on the inverse transformation formulas Eq.(\ref{inversetransform}).
We evaluated the effectiveness of prediction in term of five measurements, which are defined as follows:
\begin{itemize}
  \item The mean absolute percentage error (MAPE)
  \begin{equation*}
  \mbox{MAPE} = \frac{100\%}{k}\sum\limits_{i = 1}^{k}\left|\frac{x^{(*)}_i-\widehat{x}^{(*)}_i}{x^{(*)}_i}\right|,
  \end{equation*}
  where $x^{(*)}_i$ and $\widehat{x}^{(*)}_i$ are the actual and forecasted values with $x^{(*)}_i$ indicating $x^{(o)}_i$, $x^{(h)}_i$, $x^{(l)}_i$, or $x^{(c)}_i$, respectively; $k$ is the number of forecasted points;
  \item The standard deviation (SD) is defined as the empirical standard derivation of the forecasted values $\{\widehat{x}^{(*)}_i\}_{i=1}^k$, i.e.,
  \begin{equation*}
  \mbox{SD} =\sqrt{ \frac{1}{k-1}\sum\limits_{i = 1}^{k}\Big(\widehat{x}^{(*)}_i-\bar{\widehat{x}}^{(*)}\Big)^2 },
  \end{equation*}
  where $\bar{\widehat{x}}^{(*)}=\sum_{i=1}^k\widehat{x}^{(*)}_i/k$;
  \item The root mean squared error (RMSE) as defined in \citep{neto2008centre}
  \begin{equation*}
  \mbox{RMSE}=\sqrt{\frac{1}{k}\sum\limits_{i = 1}^{k}\Big(x^{(*)}_i-\widehat{x}^{(*)}_i\Big)^2}
  \end{equation*}
  \item The RMSE based on the Hausdorff distance (RMSEH) defined in \citep{de2006adaptive}
  \begin{equation*}
  \mbox{RMSEH}=\sqrt{\frac{1}{k}\sum\limits_{i = 1}^{k}\Big(\left|\frac{x_i^{(h)}+x_i^{(l)}}{2}-\frac{\widehat{x}_i^{(h)}+\widehat{x}_i^{(l)}}{2}\right|
  +\left| \frac{x_i^{(h)}-x_i^{(l)}}{2}-\frac{\widehat{x}_i^{(h)}-\widehat{x}_i^{(l)}}{2}\right|\Big)^2}
  \end{equation*}
  \item The accuracy ratio (AR) adopted in \citep{hu2007application}
  \begin{equation*}
  \mbox{AR}=\left\{
             \begin{array}{lcl}
             \frac{1}{k}\sum\limits_{i = 1}^{k}\frac{w(SP_i \cap \widehat{SP_i})}{w(SP_i \cup \widehat{SP_i})}, \qquad \qquad \mbox{if}\ \ \ \ (w(SP_i \cap \widehat{SP_i})\not=0)\\
             0, \qquad \qquad \qquad \qquad \qquad  \mbox{if}\ \ \ (w(SP_i \cap \widehat{SP_i})=0)\\
             \end{array}
        \right.,
  \end{equation*}
  where $w(SP_i \cap \widehat{SP_i})$ and $w(SP_i \cup \widehat{SP_i})$ represent the length of the intersection and union between the observation interval $[x_i^{(l)}, x_i^{(h)}]$ and the prediction interval $[\widehat{x}_i^{(l)}, \widehat{x}_i^{(h)}]$, respectively.
\end{itemize}
Smaller values of MAPE, RMSE, RMSEH and a larger AR indicate a better result; and a smaller SD indicates a more stable result.

\subsection{\textit{Results of simulations}}\label{sec3-sub3}

For Scenario $1$, we summarize the results with $q=40, 50, 70$ and $m=1,2,3$ in Table \ref{Tab_Results}.
From which we can see that: (i) the overall performance of the proposed method in terms of these five measurements are satisfying and stable;
(ii) for fixed $q$, a smaller forecast period $m$ will make the forecasted results more accurate with smaller values of MAPE, RMSE, RMSEH and larger AR.
And there is no obvious pattern in terms of SD.

\begin{table}[!h]
  \caption {Simulation results for Scenario 1 when $q=40, 50, 70$ and $m=1,2,3$}
  \label{Tab_Results}
  \vspace{-0.5cm}
  \setlength\tabcolsep{3 pt}
\renewcommand{\arraystretch}{0.8}
\begin{center}
  \begin{tabular}{ccccccccccc}
  \hline
 \multirow{2}{*}{Criterion}         &    & \multicolumn{3}{c}{$q=40$} & \multicolumn{3}{c}{$q=50$} & \multicolumn{3}{c}{$q=70$} \\
                      \cline{3-11}
                      &    & $m=1$    & $m=2$    & $m=3$    & $m=1$    & $m=2$    & $m=3$   & $m=1$    & $m=2$    & $m=3$   \\ \hline
  \multirow{4}{*}{MAPE} & $x^{(o)}_t$ & 3.73\% & 4.95\% & 5.41\% & 4.19\% & 4.82\% & 5.91\% & 3.58\% & 4.67\% & 5.80\% \\
                      & $x^{(h)}_t$ & 3.75\% & 4.71\% & 4.92\% & 3.93\% & 4.65\% & 5.59\% & 3.43\% & 4.37\% & 5.23\% \\
                      & $x^{(l)}_t$ & 4.30\% & 5.28\% & 6.28\% & 4.80\% & 5.17\% & 6.21\% & 4.60\% & 5.64\% & 6.81\% \\
                      & $x^{(c)}_t$ & 3.69\% & 4.94\% & 5.53\% & 4.31\% & 4.87\% & 5.96\% & 3.58\% & 4.67\% & 5.70\% \\\hline
  \multirow{4}{*}{SD}   & $x^{(o)}_t$ & 0.115 & 0.122  & 0.115  & 0.127  & 0.127  & 0.122  & 0.104  & 0.105  & 0.104  \\
                      & $x^{(h)}_t$ & 0.126  & 0.138  & 0.130  & 0.150  & 0.147  & 0.143  & 0.123  & 0.123  & 0.122  \\
                      & $x^{(l)}_t$ & 0.073  & 0.079  & 0.076  & 0.072  & 0.074  & 0.072  & 0.070  & 0.072  & 0.072  \\
                      & $x^{(c)}_t$ & 0.112  & 0.123  & 0.116  & 0.131  & 0.129  & 0.126  & 0.103  & 0.104  & 0.104  \\\hline
  \multirow{4}{*}{RMSE} & $x^{(o)}_t$ & 0.073 & 0.096  & 0.108  & 0.081  & 0.093  & 0.110  & 0.067  & 0.088  & 0.106  \\
                      & $x^{(h)}_t$ & 0.094  & 0.123  & 0.134  & 0.101  & 0.118  & 0.139  & 0.084  & 0.110  & 0.131  \\
                      & $x^{(l)}_t$ & 0.052  & 0.067  & 0.075  & 0.059  & 0.065  & 0.076  & 0.055  & 0.068  & 0.081  \\
                      & $x^{(c)}_t$ & 0.071  & 0.099  & 0.109  & 0.083  & 0.096  & 0.114  & 0.066  & 0.088  & 0.106  \\\hline
  \multicolumn{2}{c}{RMSEH}  & 0.098  & 0.127  & 0.137  & 0.099  & 0.122  & 0.142  & 0.088  & 0.114  & 0.135  \\\hline
  \multicolumn{2}{c}{AR}     & 0.891  & 0.868  & 0.858  & 0.886  & 0.872  & 0.849  & 0.895  & 0.871  & 0.848  \\ \hline
\end{tabular}
\end{center}
\end{table}

\begin{figure}[!h]
  \centering
  \subfigure[$x^{(o)}_t$]{
    \label{0.05_mape_o}
    \resizebox{5cm}{4cm}{\includegraphics{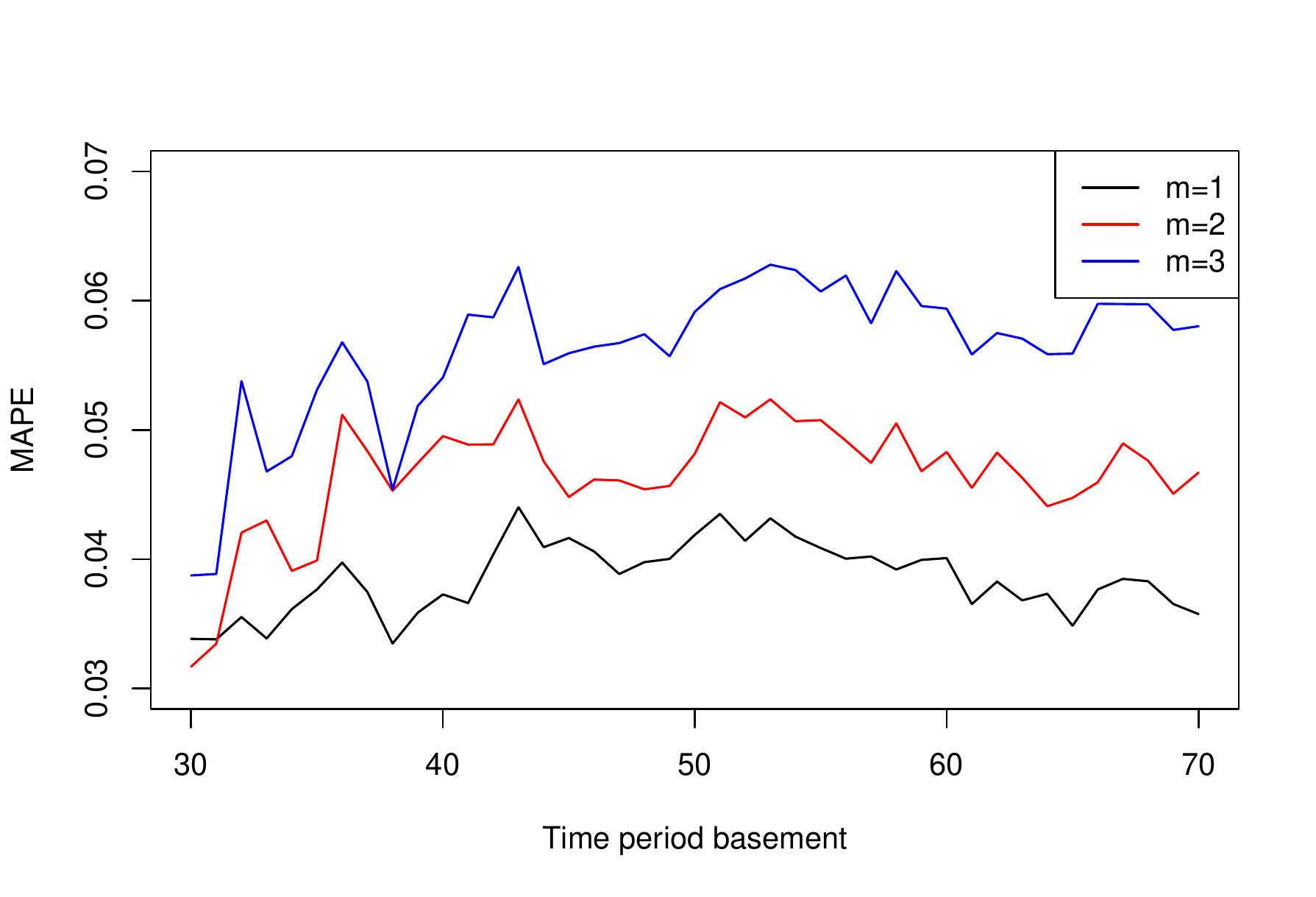}}}
  \hspace{0in}
   \subfigure[$x^{(o)}_t$]{
    \label{0.05_SD_o}
    \resizebox{5cm}{4cm}{\includegraphics{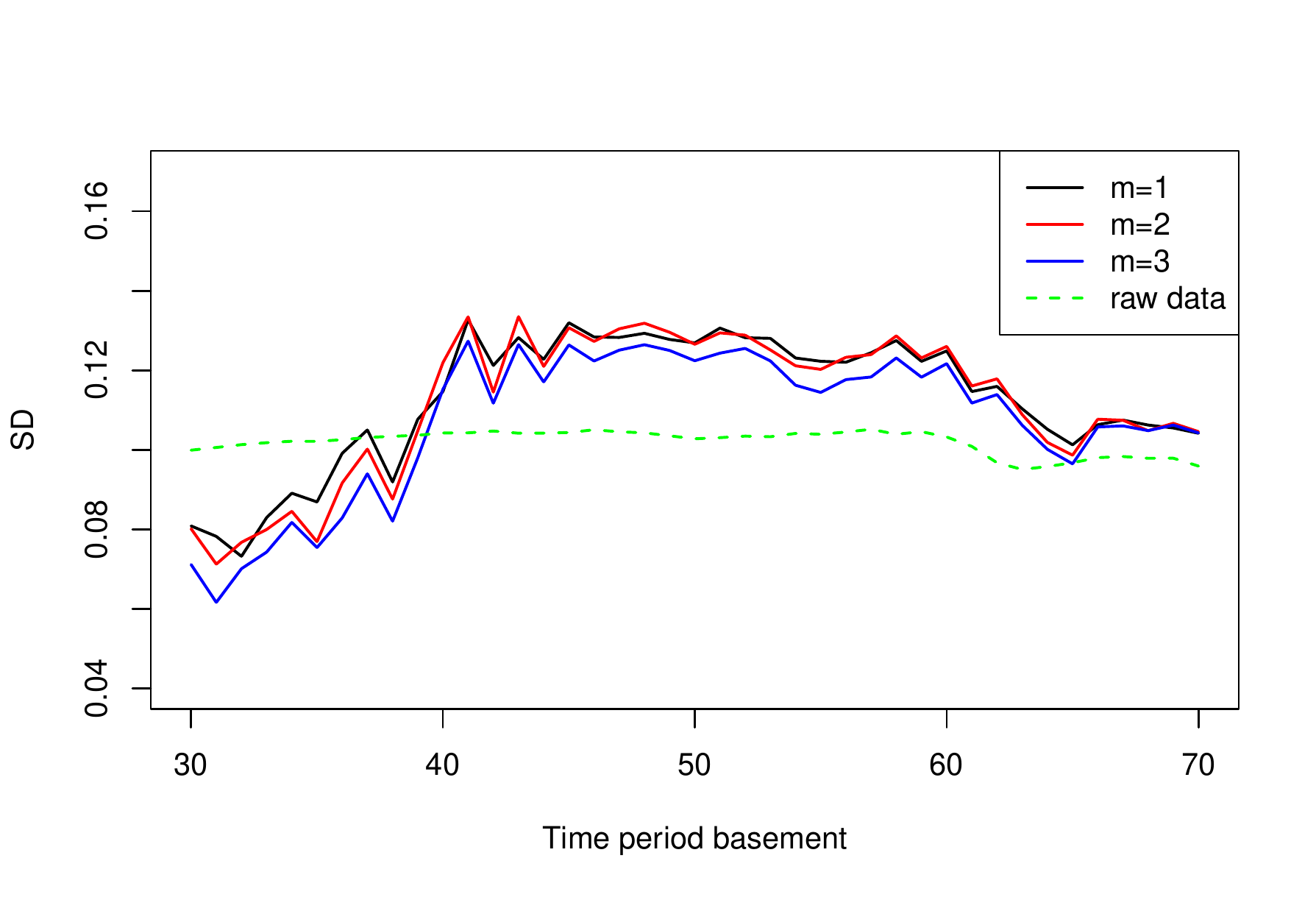}}}
  \hspace{0in}
   \subfigure[$x^{(o)}_t$]{
    \label{0.05_RMSE_o}
    \resizebox{5cm}{4cm}{\includegraphics{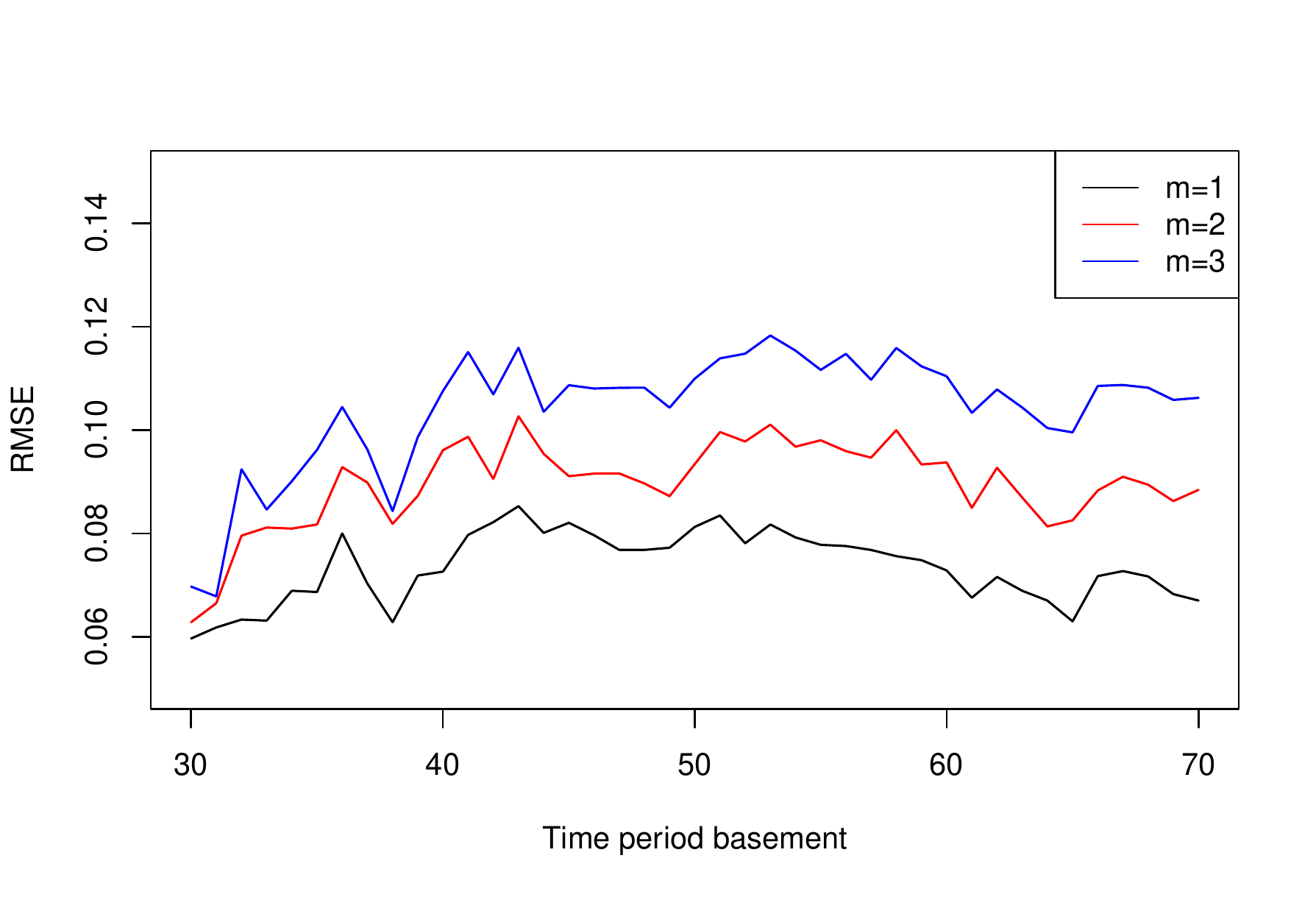}}}
  \hspace{0in}
  \subfigure[$x^{(h)}_t$]{
    \label{0.05_mape_h}
    \resizebox{5cm}{4cm}{\includegraphics{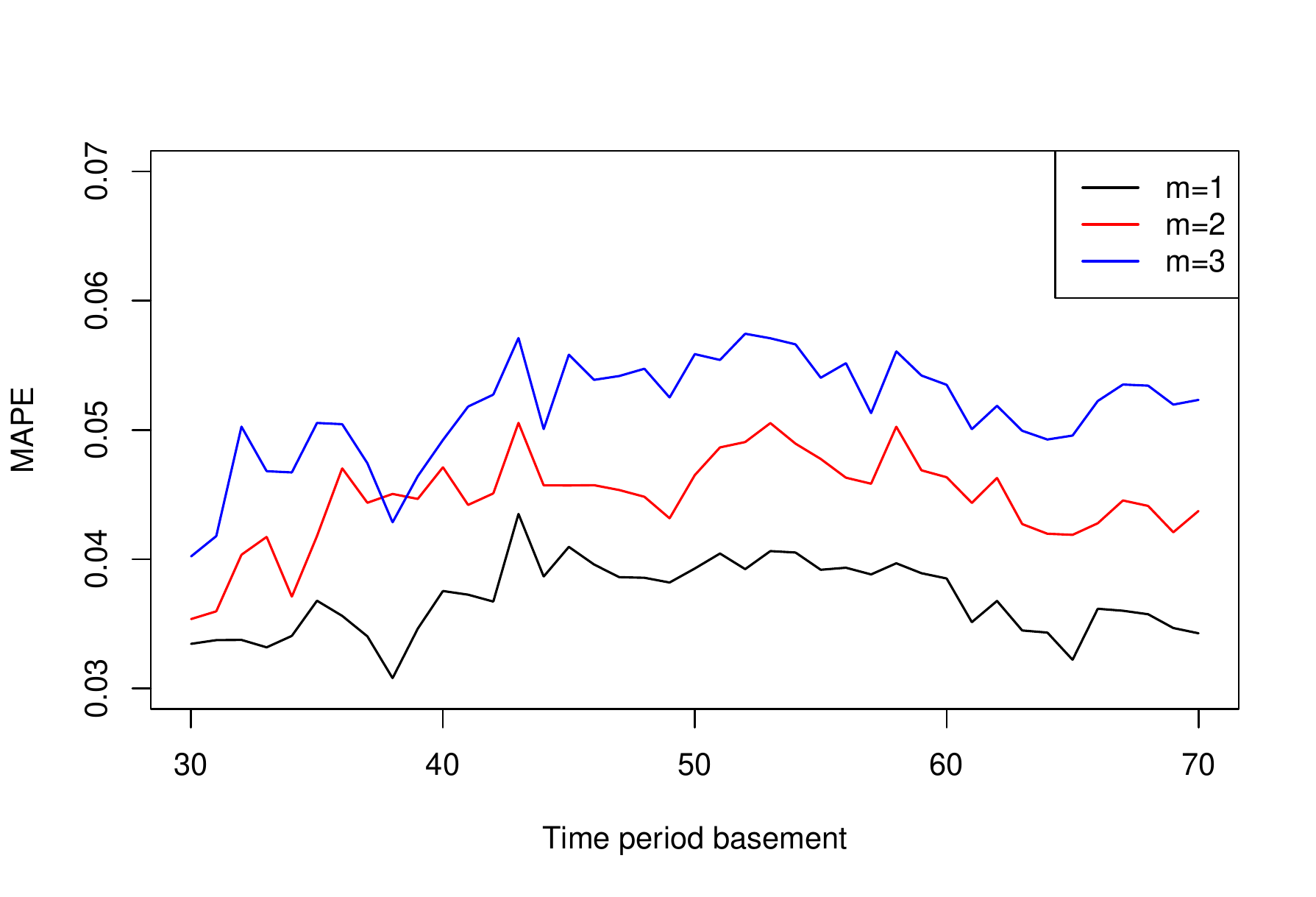}}}
  \hspace{0in}
   \subfigure[$x^{(h)}_t$]{
    \label{0.05_SD_h}
    \resizebox{5cm}{4cm}{\includegraphics{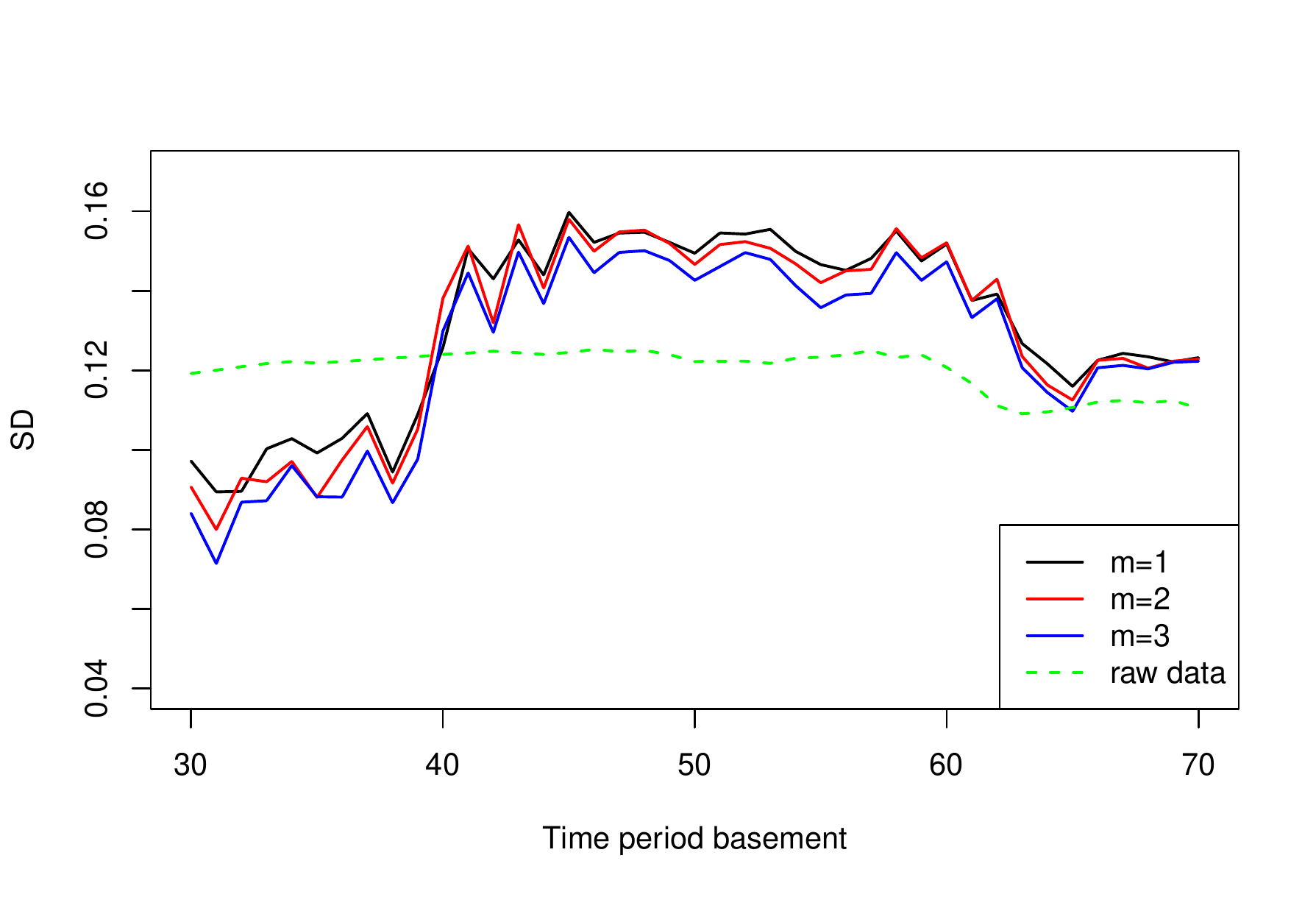}}}
  \hspace{0in}
    \subfigure[$x^{(h)}_t$]{
    \label{0.05_RMSE_h}
    \resizebox{5cm}{4cm}{\includegraphics{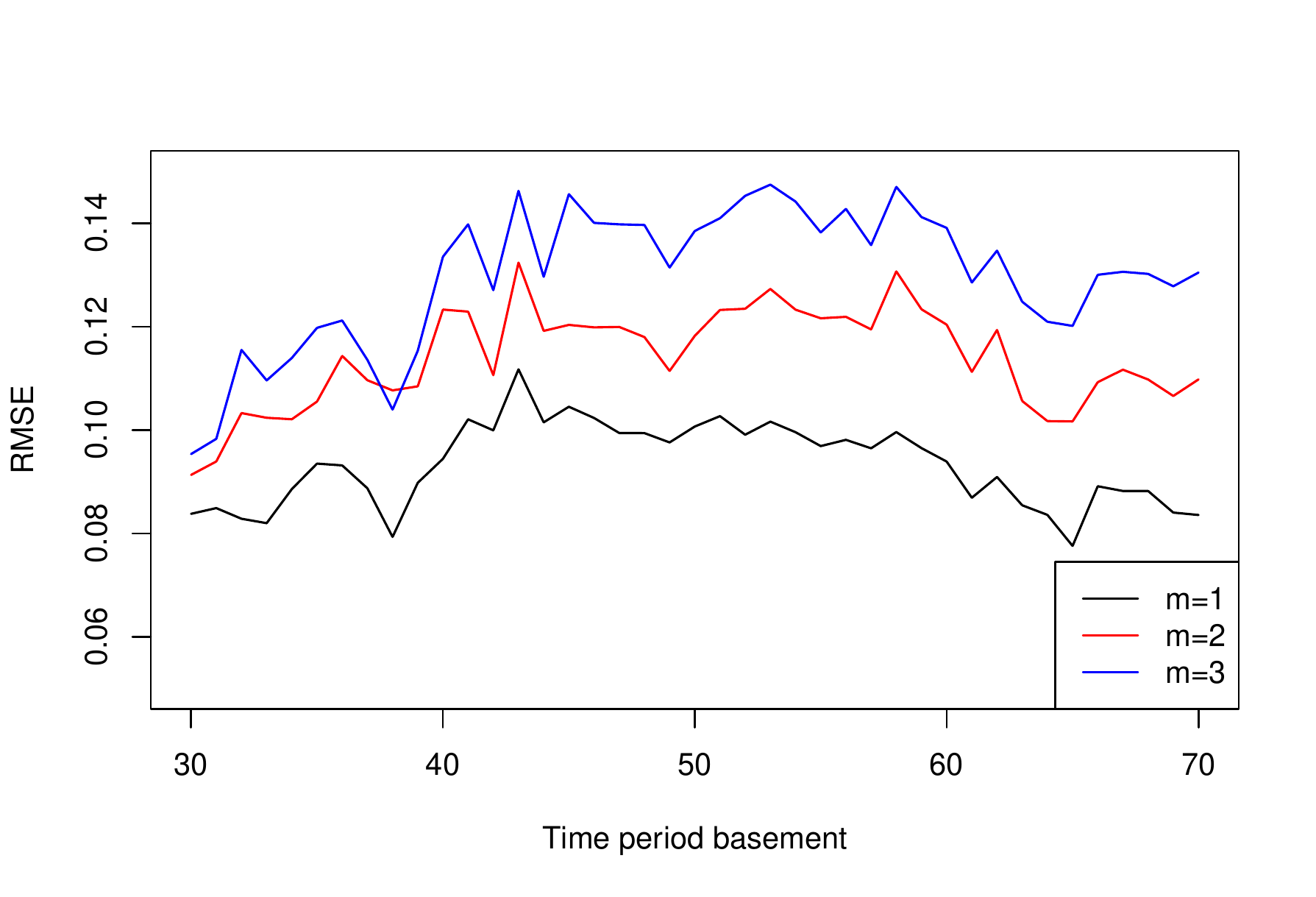}}}
  \hspace{0in}
   \subfigure[$x^{(l)}_t$]{
    \label{0.05_SD_l}
    \resizebox{5cm}{4cm}{\includegraphics{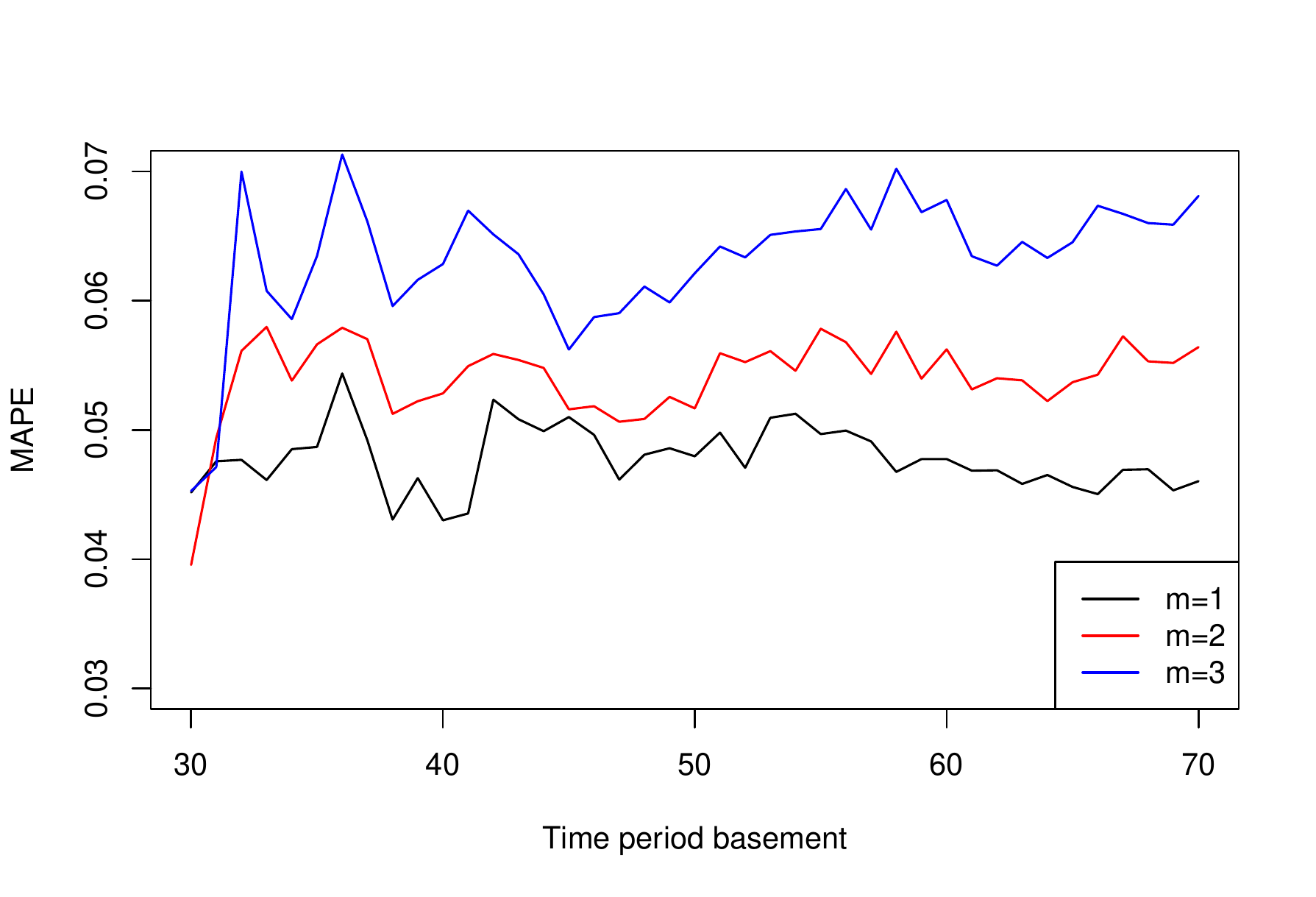}}}
    \hspace{0in}
   \subfigure[$x^{(l)}_t$]{
    \label{0.05_RMSE_c}
    \resizebox{5cm}{4cm}{\includegraphics{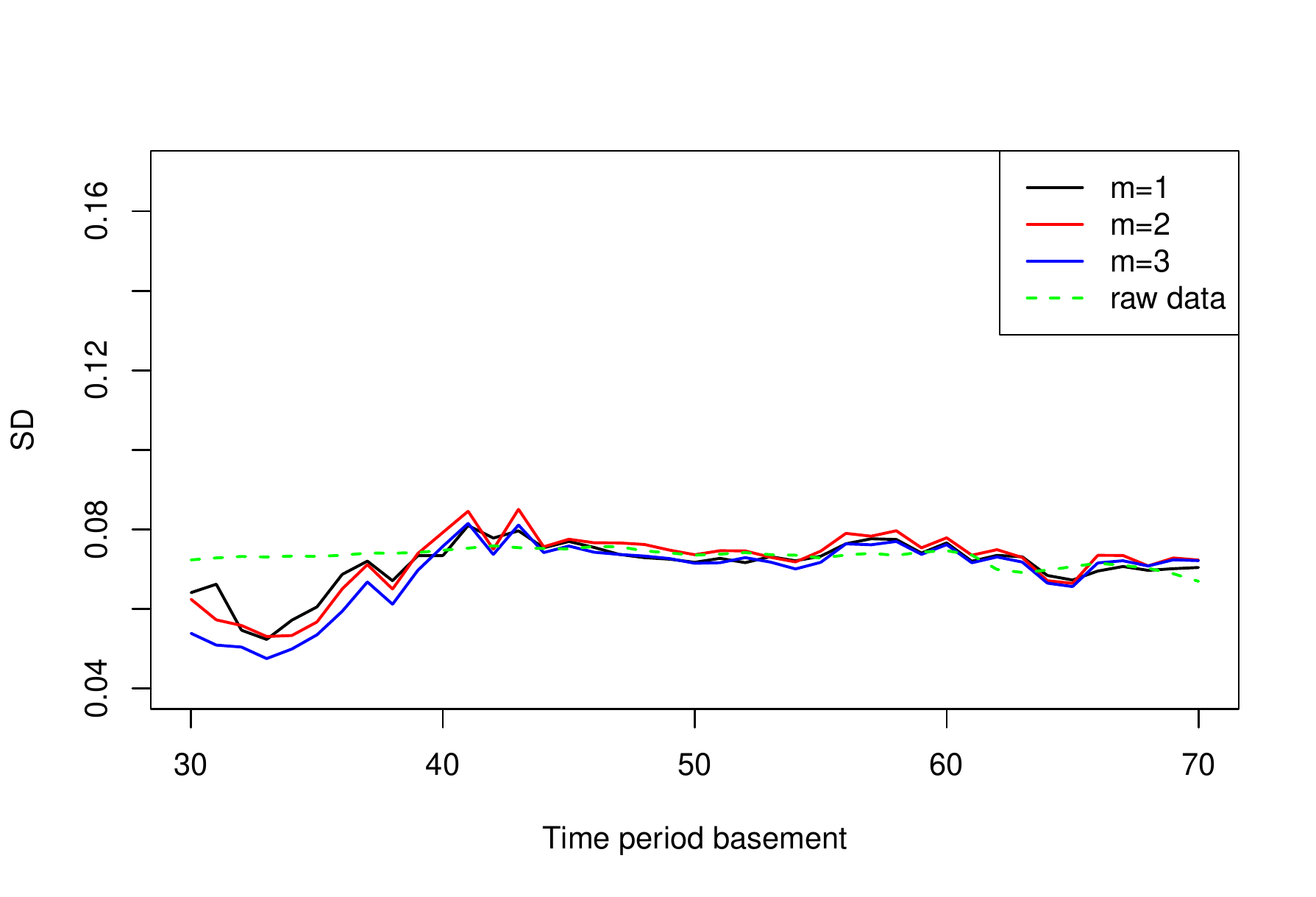}}}
  \hspace{0in}
     \subfigure[$x^{(l)}_t$]{
    \label{0.05_RMSE_l}
    \resizebox{5cm}{4cm}{\includegraphics{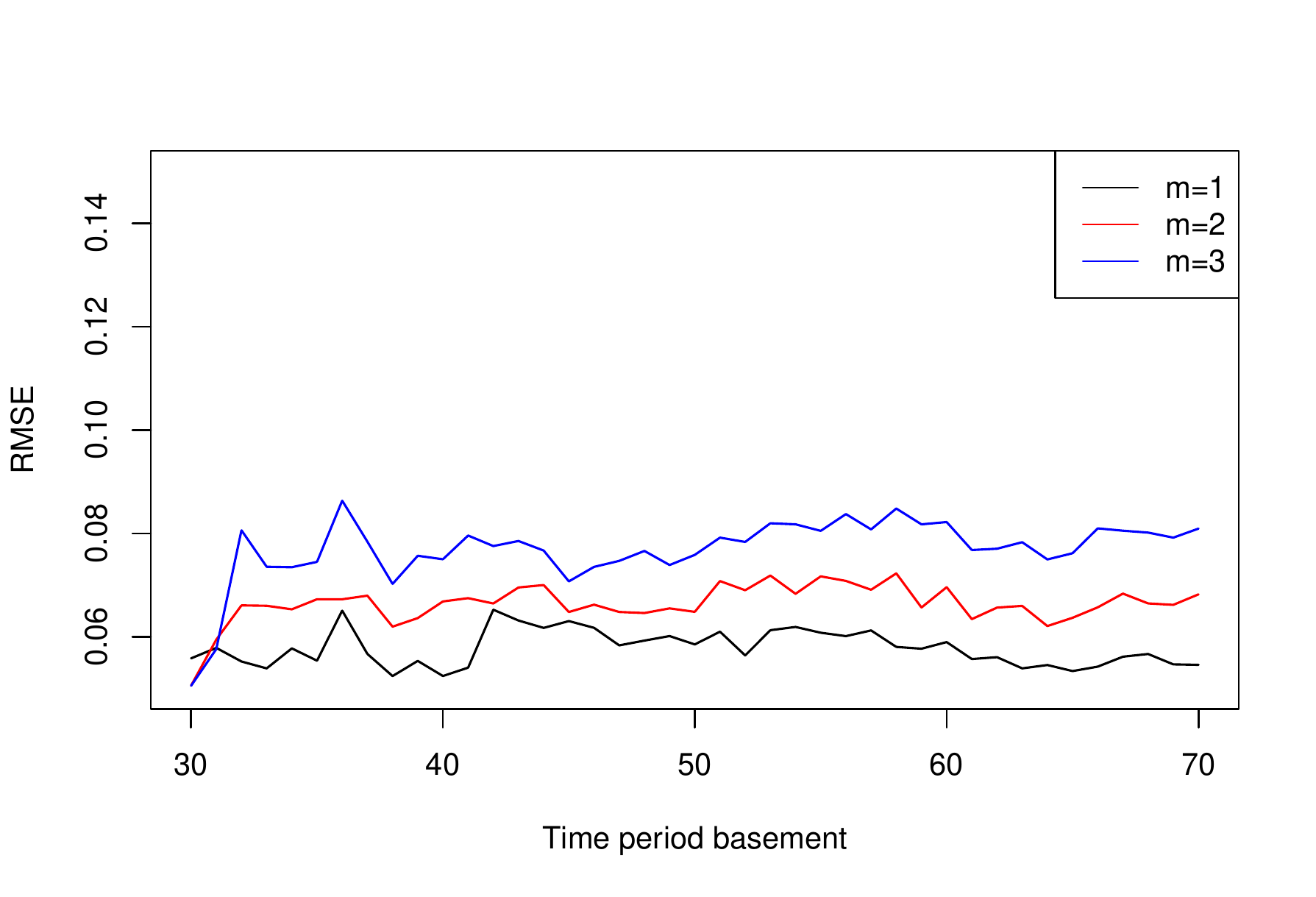}}}
  \hspace{0in}
     \subfigure[$x^{(c)}_t$]{
    \label{0.05_mape_c}
    \resizebox{5cm}{4cm}{\includegraphics{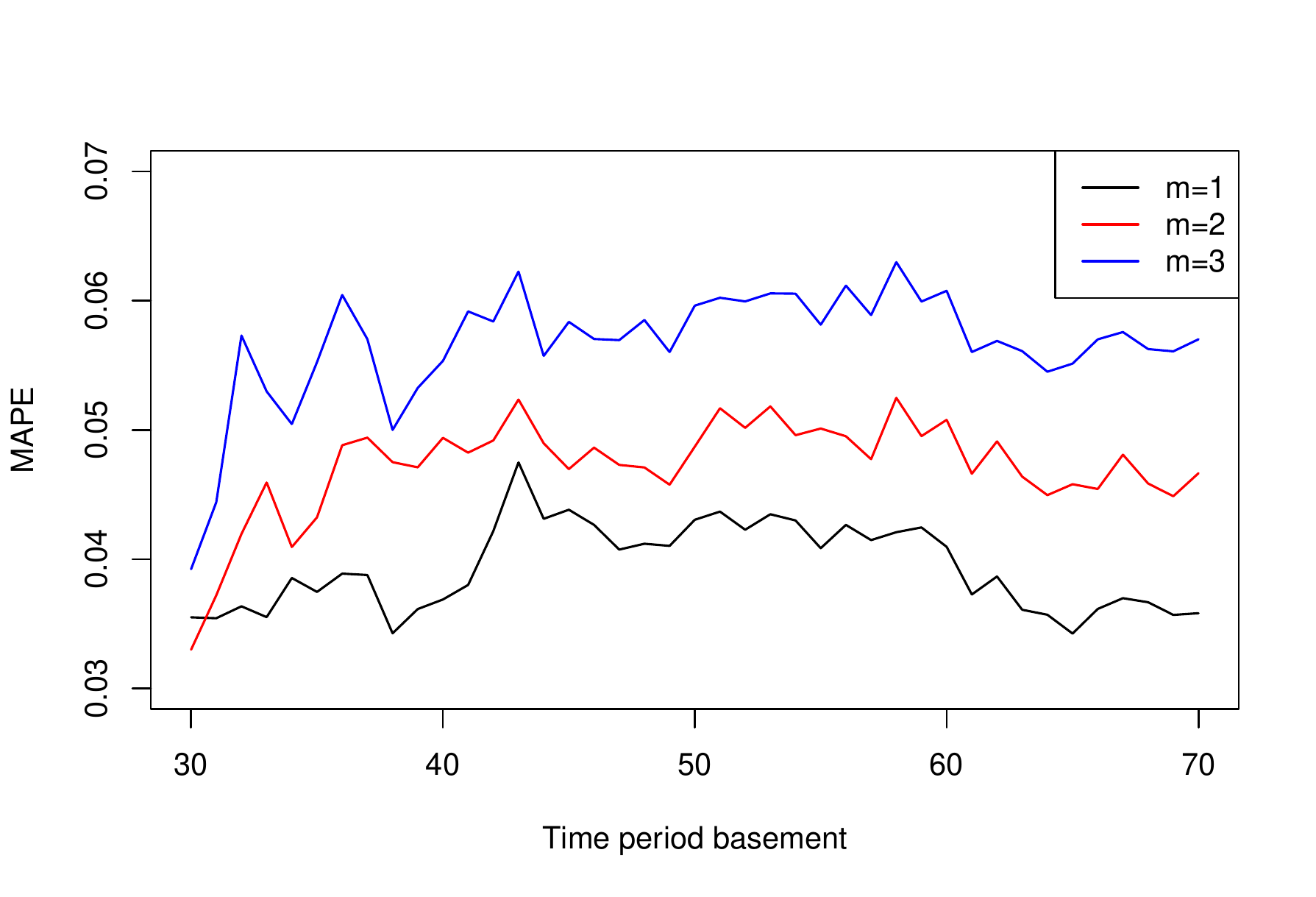}}}
  \hspace{0in}
    \subfigure[$x^{(c)}_t$]{
    \label{0.05_SD_c}
    \resizebox{5cm}{4cm}{\includegraphics{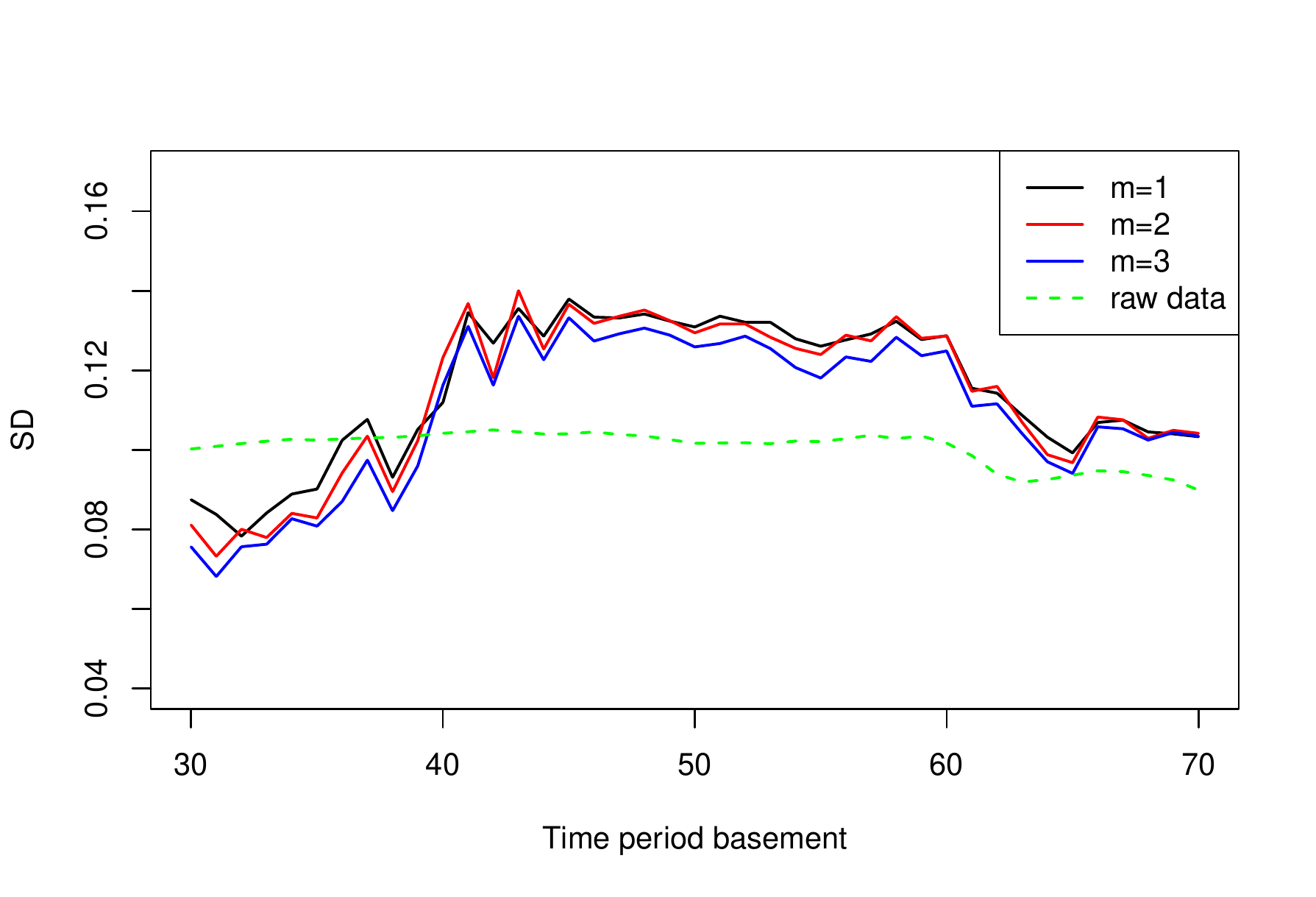}}}
  \hspace{0in}
  \subfigure[$x^{(c)}_t$]{
    \label{0.05_mape_l}
    \resizebox{5cm}{4cm}{\includegraphics{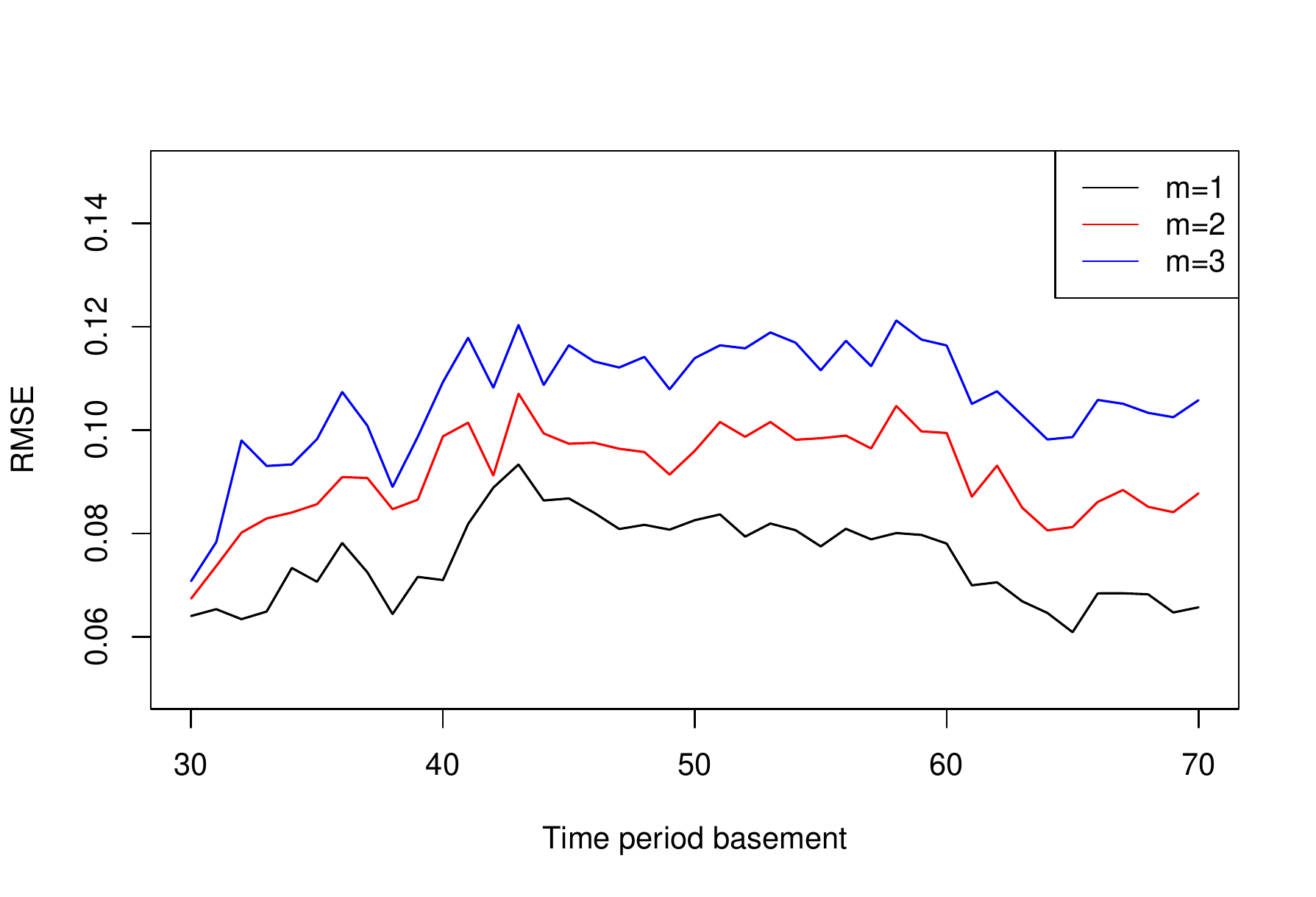}}}
  \caption{\rm MAPE (Left panel), SD (Middle panel) and RMSE (Right panel) of forecasted values for $x^{(o)}_t$ (The first row), $x^{(h)}_t$ (The second row), $x^{(l)}_t$ (The third row) and $x^{(c)}_t$ (The fourth row) with different $q$ and $m$ for Scenario 1 respectively}\label{Scenario1}
\end{figure}

\begin{figure}[!h]
  \centering
  \subfigure[RMSEH]{
    \label{0.05_RMSEH}
    \resizebox{6cm}{5cm}{\includegraphics{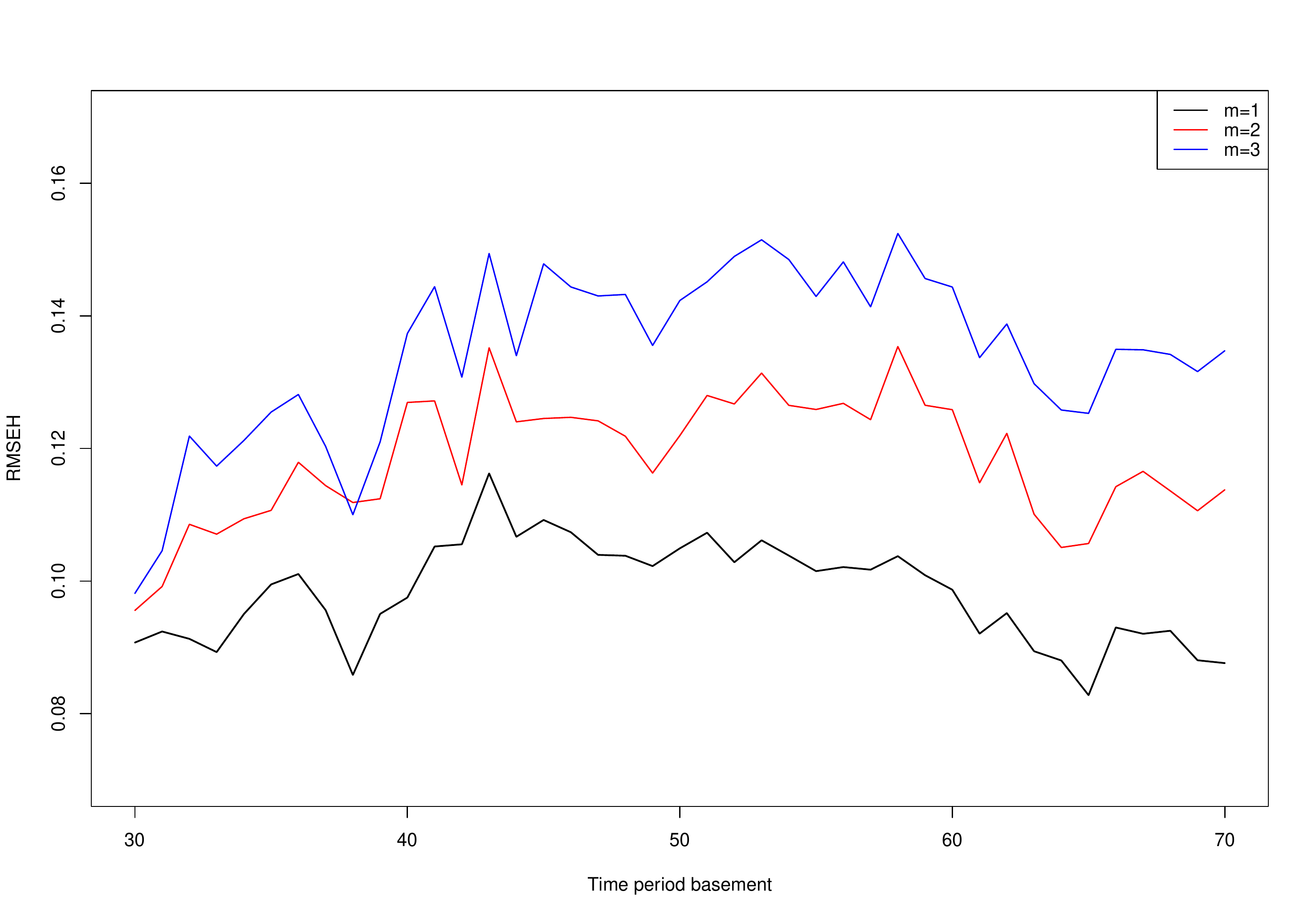}}}
  \hspace{0in}
  \subfigure[AR]{
    \label{0.05_AR}
    \resizebox{6cm}{5cm}{\includegraphics{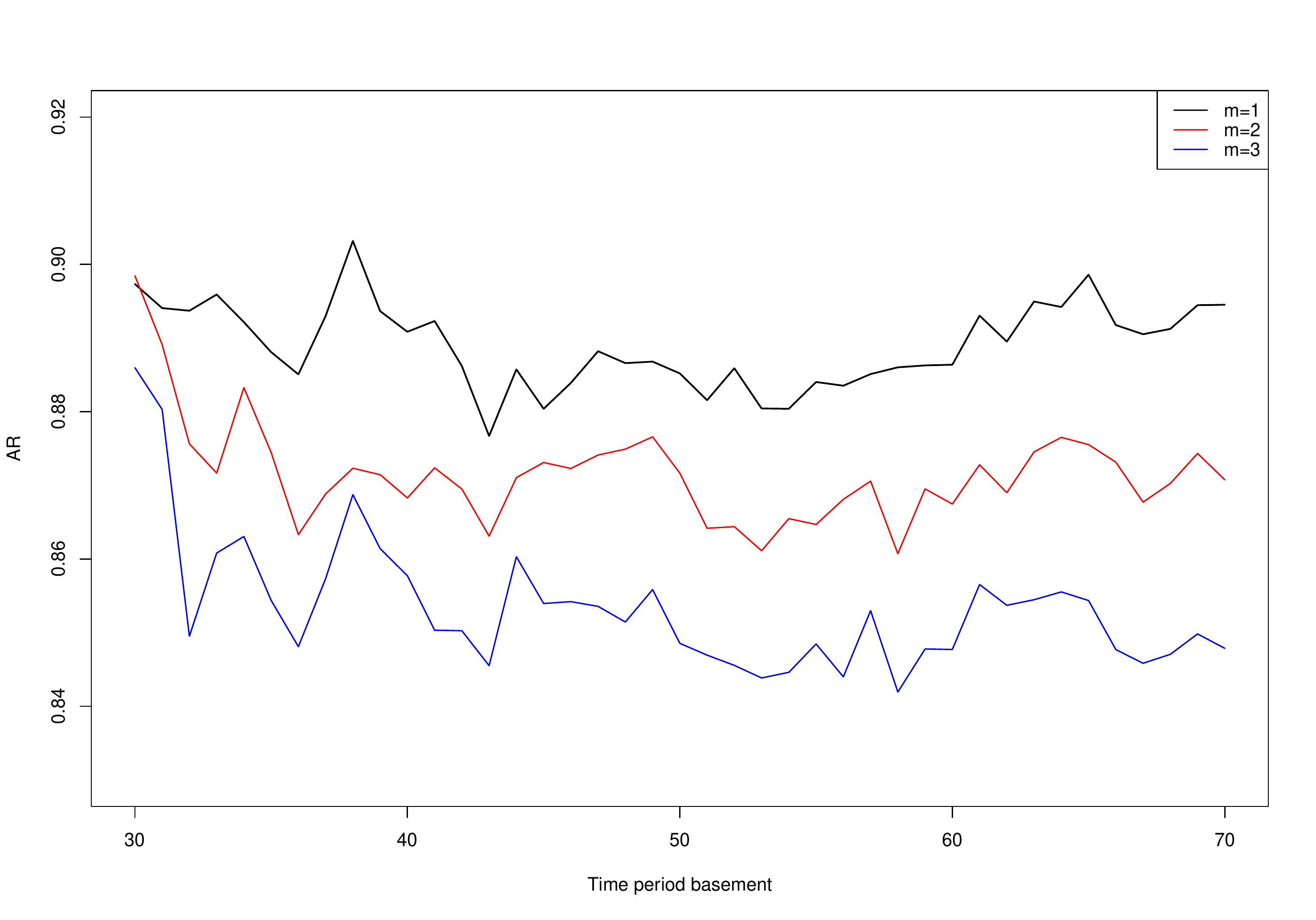}}}
  \caption{\rm RMSEH (Left) and AR (Right) of forecasted values for different $q$ and $m$ in Scenario 1}\label{Fig RMSEH+AR5}
\end{figure}

Moreover, we show more results with $q$ ranging from $30$ to $70$ and $m=1,2,3$ to further demonstrate the performance of the proposed method.
Specifically, Fig.\ref{Scenario1} summarize the results in terms of MAPE (the left panel), SD (the middle panel), and RMSE (the right panel), respectively; while Fig.\ref{Fig RMSEH+AR5} shows the RMSEH and AR of the forecasted results.

Basically from Fig.\ref{Scenario1}, under different $q$ and $m$, the MAPE is between $3.08\%$ and $7.13\%$, the standard deviation is between $0.048$ and $0.158$ and the RMSE is between $0.051$ and $0.148$, indicating a good prediction accuracy and stability. As the forecast period $m$ increases, these three indicators will increase synchronously, decreasing the prediction accuracy. While the prediction performance shows a trend of getting better first and then getting worse as $q$ increases. From Fig.\ref{Fig RMSEH+AR5}, the RMSEH maintains a small value between $0.083$ and $0.152$. Meanwhile, the AR is relatively high, varies from $0.842$ to $0.903$, which illustrates the prediction interval is closely coincided with the observation interval, indicating a satisfying prediction effect.

\begin{figure}[!h]
  \centering
  \subfigure[$x^{(o)}_t$]{
    \label{0.07_mape_o}
    \resizebox{3.5cm}{3.5cm}{\includegraphics{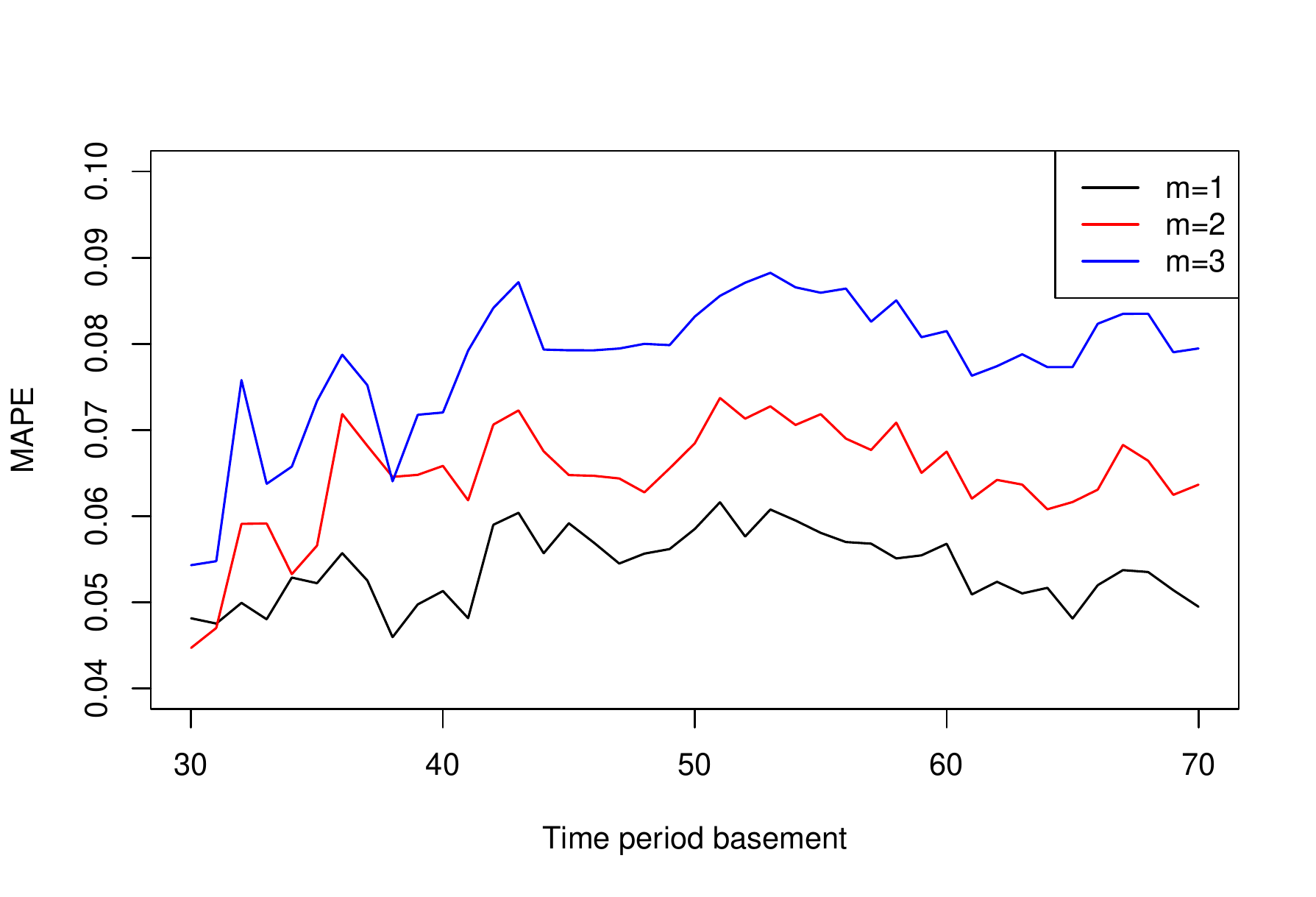}}}
  \hspace{0in}
  \subfigure[$x^{(h)}_t$]{
    \label{0.07_mape_h}
    \resizebox{3.5cm}{3.5cm}{\includegraphics{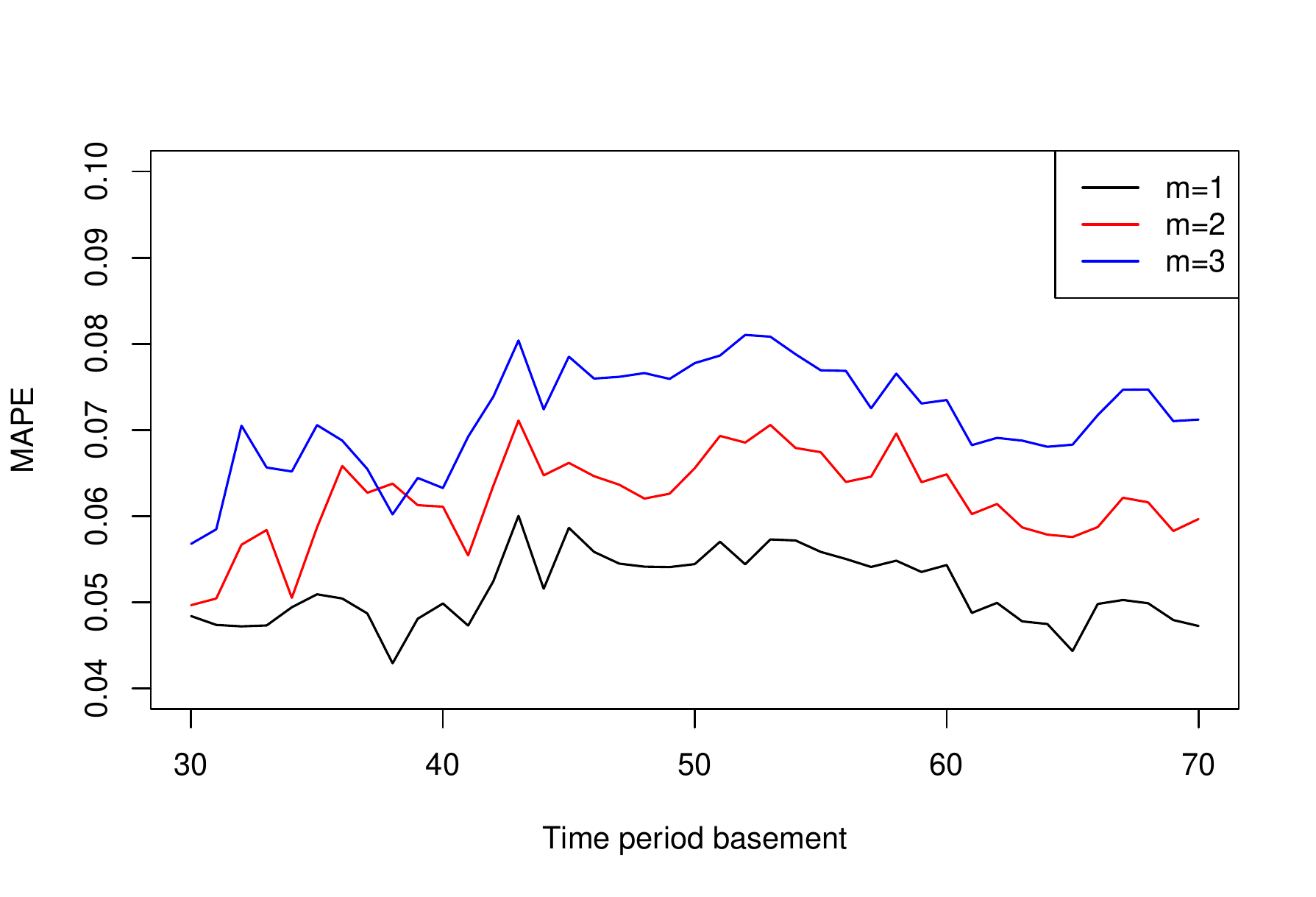}}}
  \hspace{0in}
  \subfigure[$x^{(l)}_t$]{
    \label{0.07_mape_l}
    \resizebox{3.5cm}{3.5cm}{\includegraphics{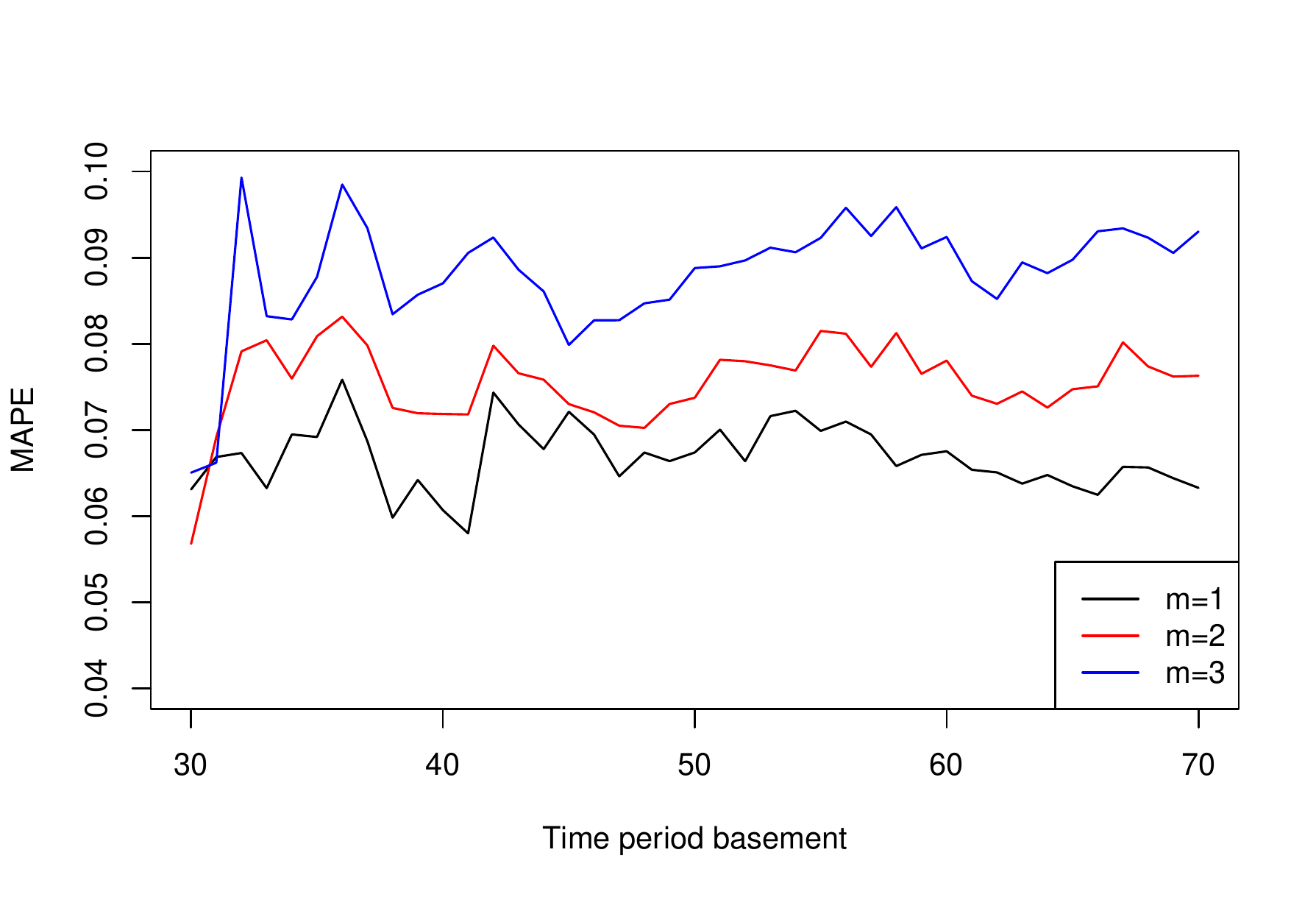}}}
  \hspace{0in}
  \subfigure[$x^{(c)}_t$]{
    \label{0.07_mape_c}
    \resizebox{3.5cm}{3.5cm}{\includegraphics{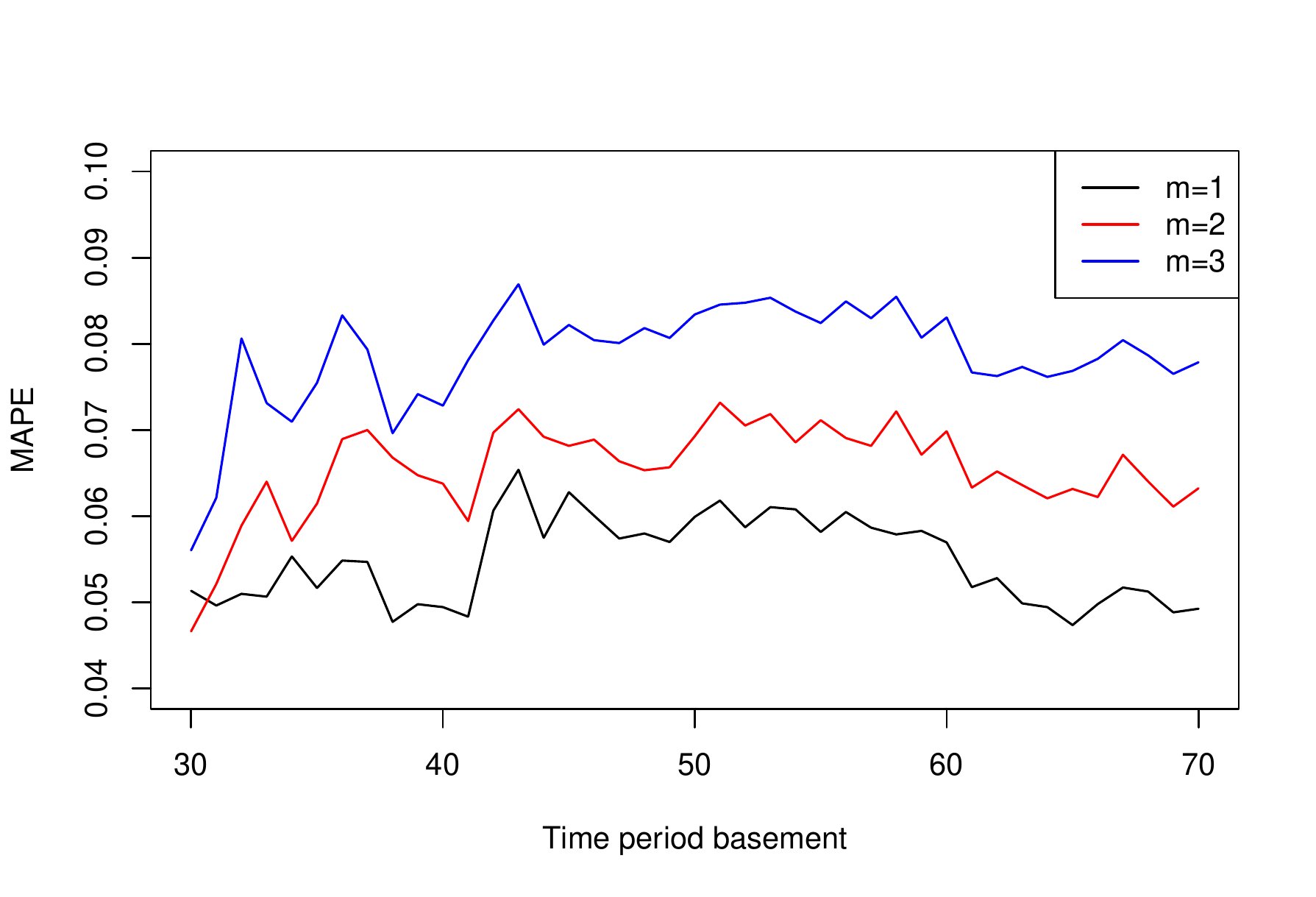}}}
  \hspace{0in}
  \subfigure[$x^{(o)}_t$]{
    \label{0.07_mape_o}
    \resizebox{3.5cm}{3.5cm}{\includegraphics{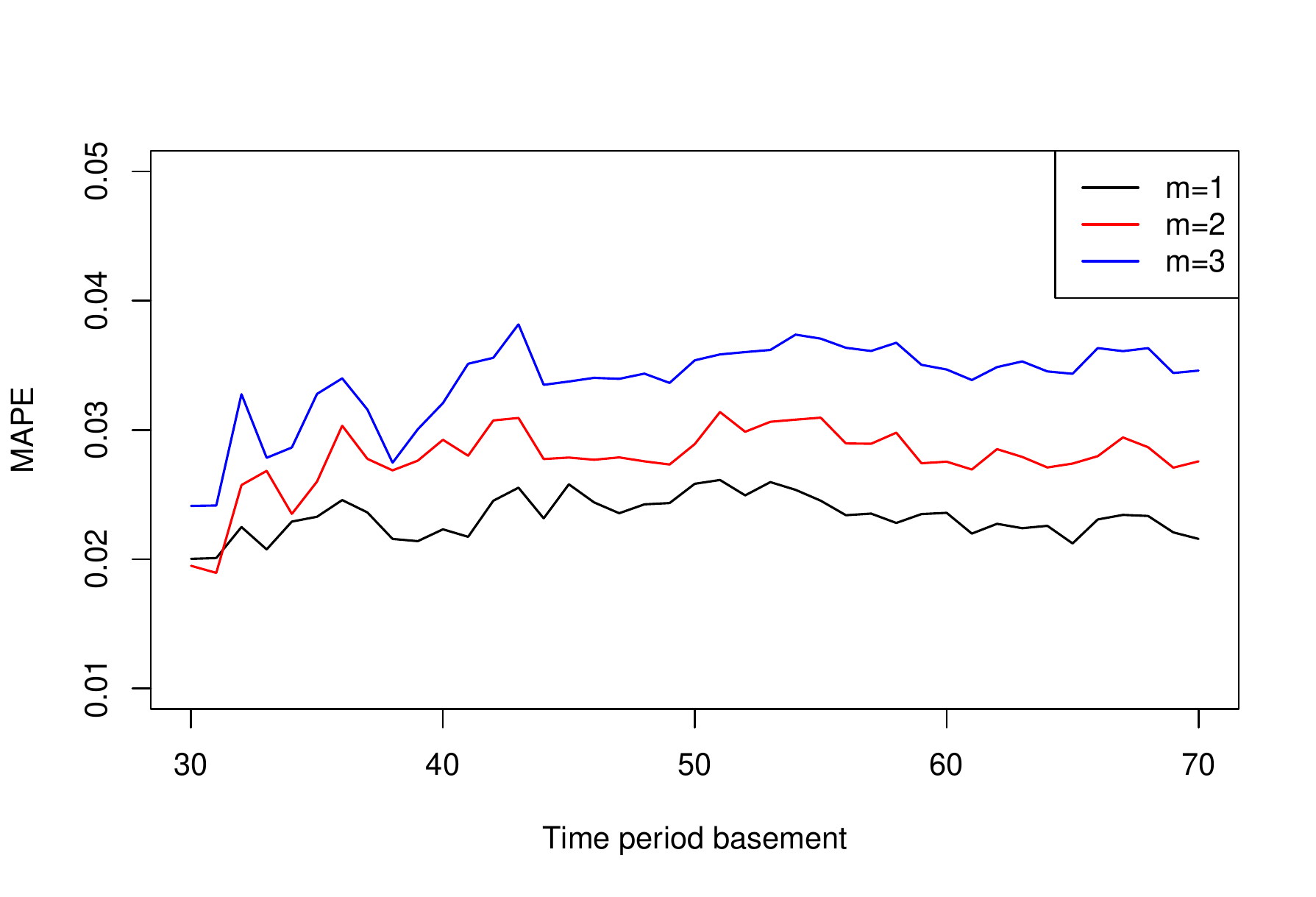}}}
  \hspace{0in}
  \subfigure[$x^{(h)}_t$]{
    \label{0.07_mape_h}
    \resizebox{3.5cm}{3.5cm}{\includegraphics{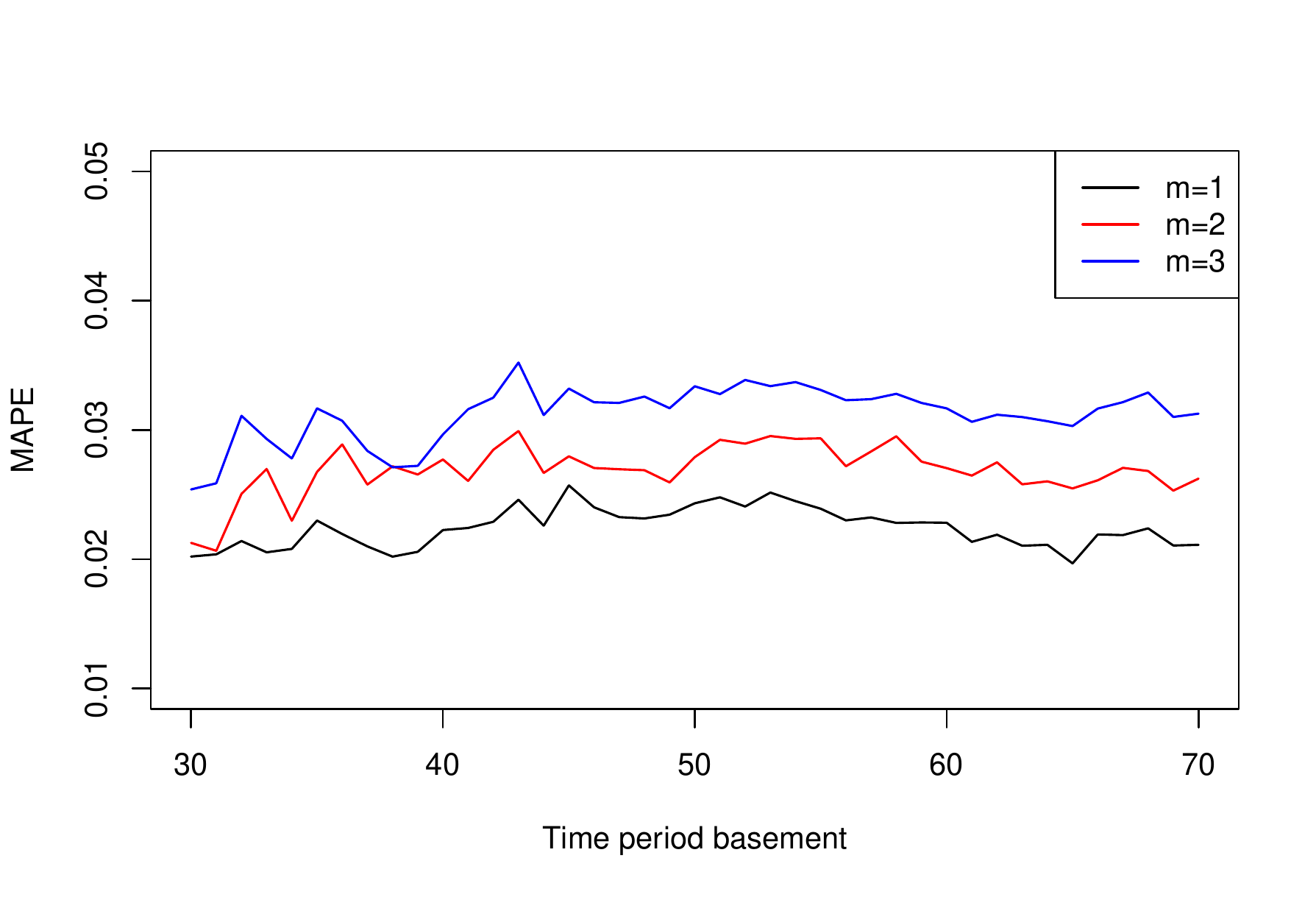}}}
  \hspace{0in}
  \subfigure[$x^{(l)}_t$]{
    \label{0.07_mape_l}
    \resizebox{3.5cm}{3.5cm}{\includegraphics{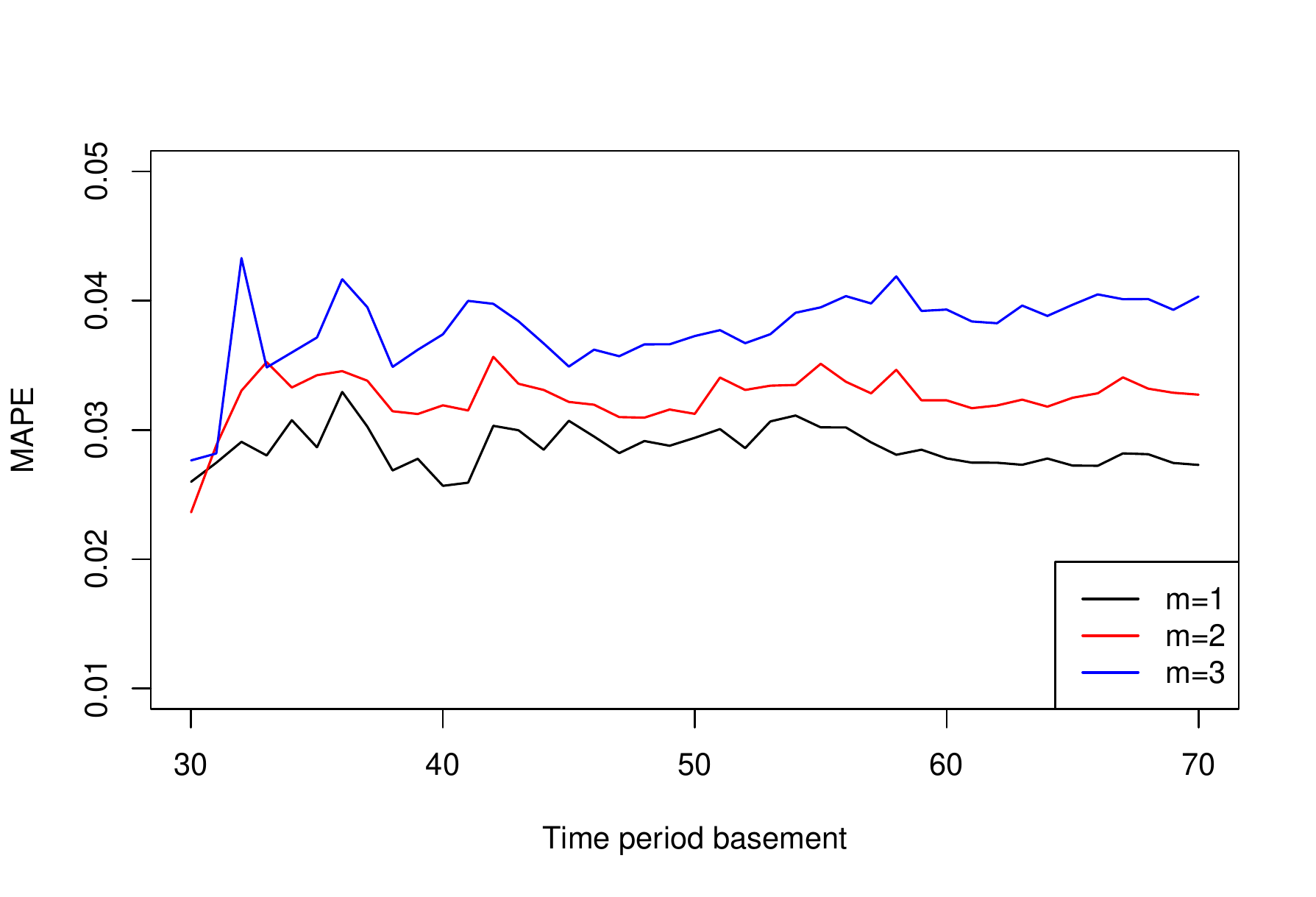}}}
  \hspace{0in}
  \subfigure[$x^{(c)}_t$]{
    \label{0.07_mape_c}
    \resizebox{3.5cm}{3.5cm}{\includegraphics{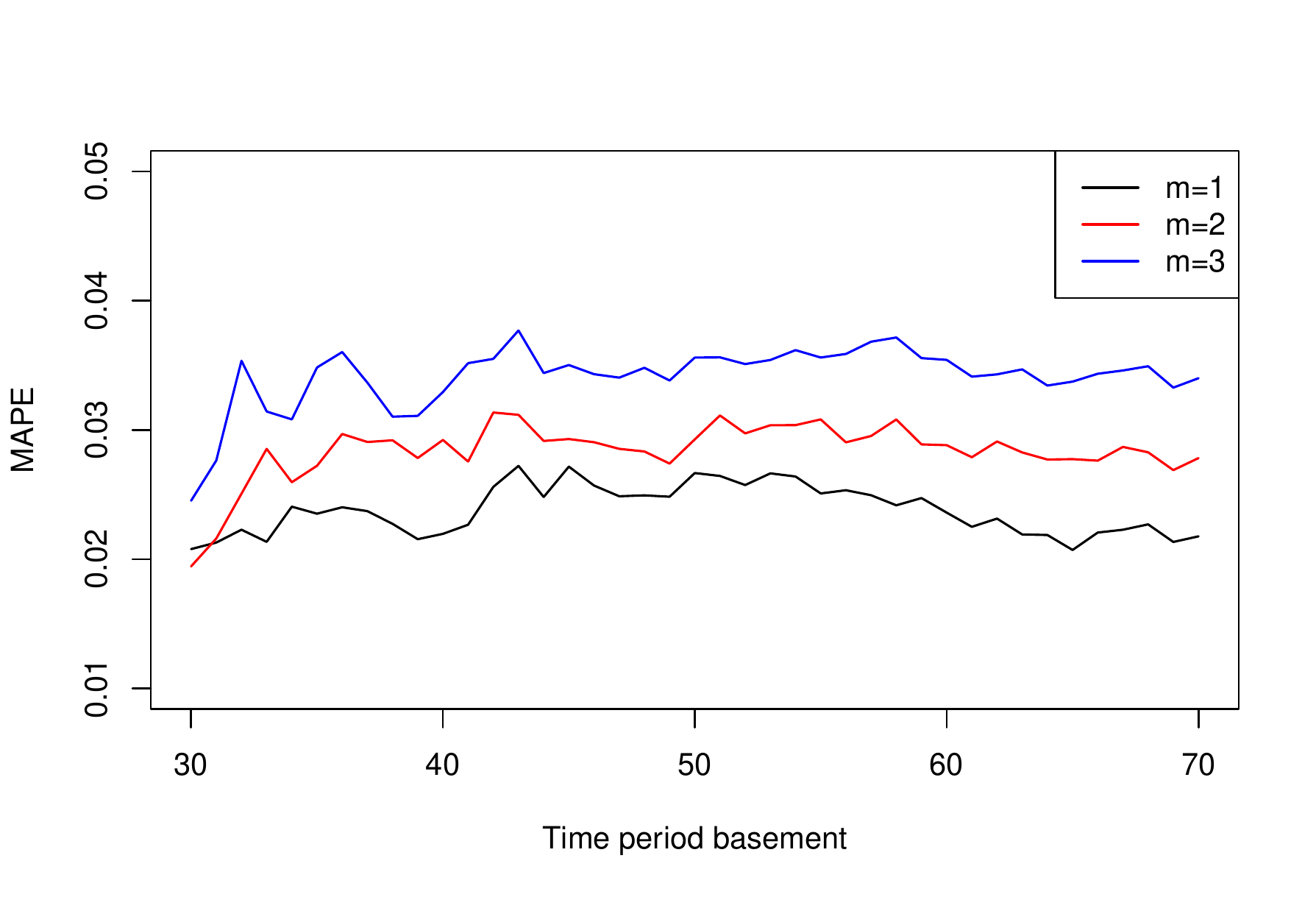}}}

  \caption{\rm MAPE of forecasted values with different $q$ and $m$ for Scenario 2 (The first row) and 3 (The second row), respectively}\label{Scenarios2&3}
\end{figure}

For Scenario $2$ and $3$, we also conduct simulations along the line with Scenario $1$, and the results exhibit the same trend. For the sake of space, we only show the forecasted results in terms of MAPE in Fig.\ref{Scenarios2&3} for Scenario $2$ with low signal-noise ratio (the first row) and Scenario $3$ with high signal-noise ratio (the second row), respectively. The MAPE in the first row of Fig.\ref{Scenarios2&3} is between $4.29\%$ and $9.93\%$, while the corresponding MAPE in the second row of Fig.\ref{Scenarios2&3} is between $1.89\%$ and $4.33\%$. The left panel of Fig.\ref{Scenario1} corresponds to the MAPE with medium signal-noise ratio, whose MAPE (ranges form $3.08\%$ to $7.13\%$) is between those in the second and the first row of Fig.\ref{Scenarios2&3}, which indicates the accuracy of the predicted results decreases as the signal-noise ratio decreases.

\section{Empirical analysis} \label{Sec 4}

We illustrated the practical utility of the proposed method via three different kinds of real data sets: the OHLC series of the Kweichow Moutai, CSI $100$ index, and $50$ ETF in the financial market of China. For each case, we first briefly describe the data set in Section \ref{sec4-sub1}, then apply the proposed method with different prediction basement period $q$ and prediction period $m$, and report the performance in terms of MAPE, RMSE, RMSEH and AR in Section \ref{sec4-sub2}.

\subsection{\textit{Raw OHLC data set description}}\label{sec4-sub1}

\begin{itemize}
  \item \textbf{OHLC series of the Kweichow Moutai}. The Kweichow Moutai is a well-known company in Chinese liquor industry, which has a long history and its stamp (SH: $600519$) is an important part of the China Securities Index (CSI $100$). Here we studies its OHLC series with the time ranging from $27/8/2001$ to $14/6/2019$, yielding $4243$ data in total.
  \item \textbf{OHLC series of the CSI $100$  index}. The CSI $100$ index is composed by the largest $100$ blue stocks selected from the CSI $300$ index stocks, reflecting the overall situation of the companies with the most market influence power in the Shanghai and Shenzhen stock markets. China Securities Index Co.,Ltd officially issued the CSI $100$ index on $30/12/2005$ and $1000$ being its base data and base point. We collected the OHLC series of the CSI $100$ index from $30/12/2005$ to $14/6/2019$, with a total of $3269$ periods.
  \item \textbf{OHLC series of the 50 ETF}. The SSE $50$ index (code: 510050) is China's first transactional exchange traded fund, compiled by the Shanghai Stock Exchange, whose base date and base point are $31/12/2003$ and $1000$, respectively.
    The investment objective of the $50$ ETF is to closely track the SSE $50$ index, minimizing tracking deviation and tracking error.
    This paper collected $3481$  OHLC data samples of the 50 ETF from $23/2/2005$ to $14/6/2019$.
\end{itemize}

\subsection{\textit{Results of empirical analysis}}\label{sec4-sub2}

For each raw data set $\{\bm{X}_t\}_{t=1}^T$, the very beginning operation is conducting the unconstrained transformation.
In terms of the unconstrained data set $\{\bm{Y}_t\}_{t=1}^T$, we first apply the ADF test together with the auto-correlation function (ACF) plot to examine the stability of each of the four variables.
If any variable is non-stationary, the Johansen test is further employed to examine the co-integration relationship between these four variables.
If no co-integration relationship exists, take one-order difference of the non-stationary variables and restart the ADF test.
In brief, we operate the proposed method shown in Fig.\ref{Fig process} with $q$ varying from $50$ to $90$ and $m=1,2,3$.

\begin{figure}[!h]
  \centering
  \subfigure[$\{\bm{y}_t^{(1)}\}_{t=1}^{90}$]{
    \label{maotai_raw}
    \resizebox{7cm}{4.8cm}{\includegraphics{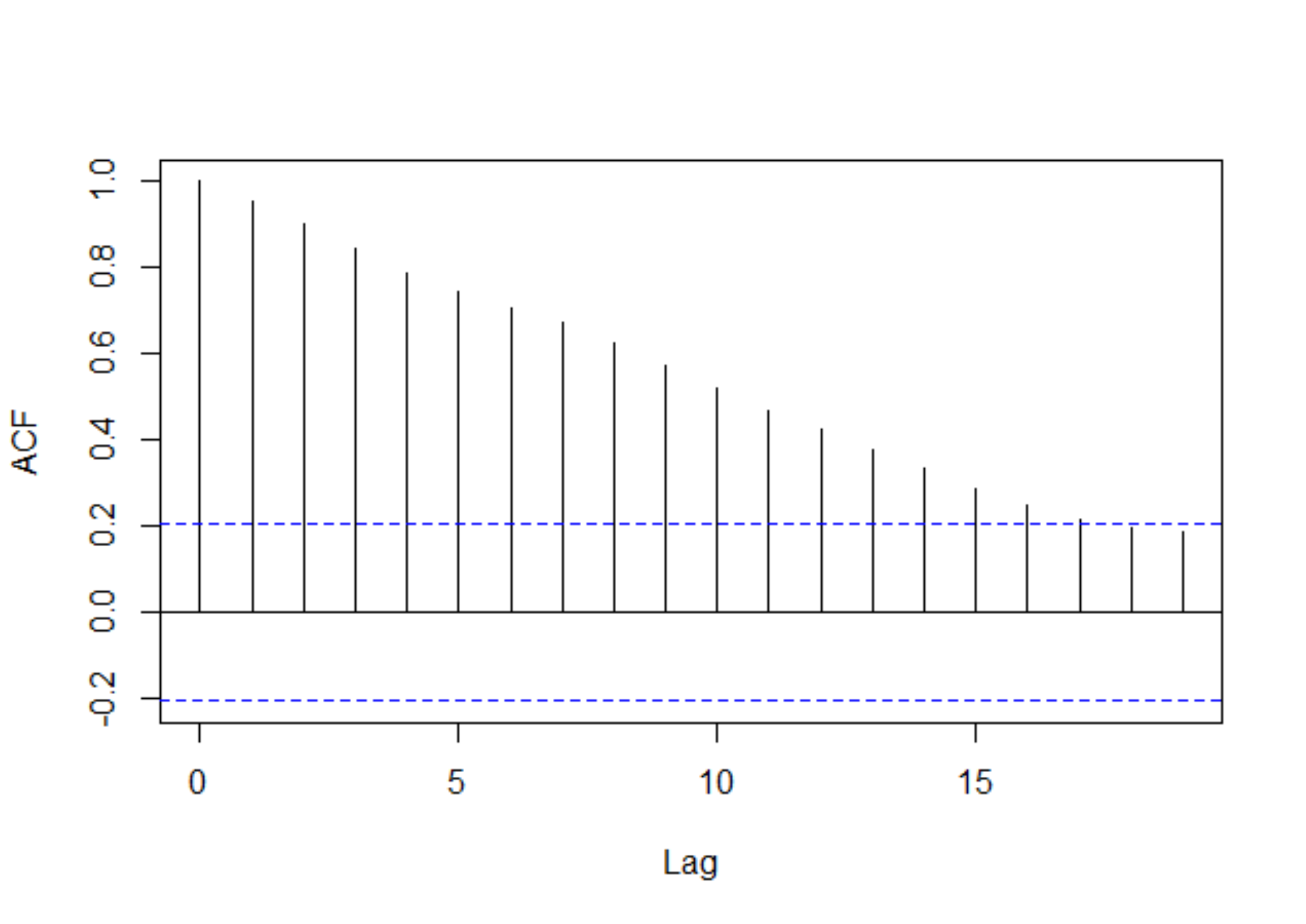}}}
  \hspace{0in}
   \subfigure[$\{\bm{y}_t^{(2)}\}_{t=1}^{90}$]{
    \label{CSI_raw}
   \resizebox{7cm}{4.8cm}{\includegraphics{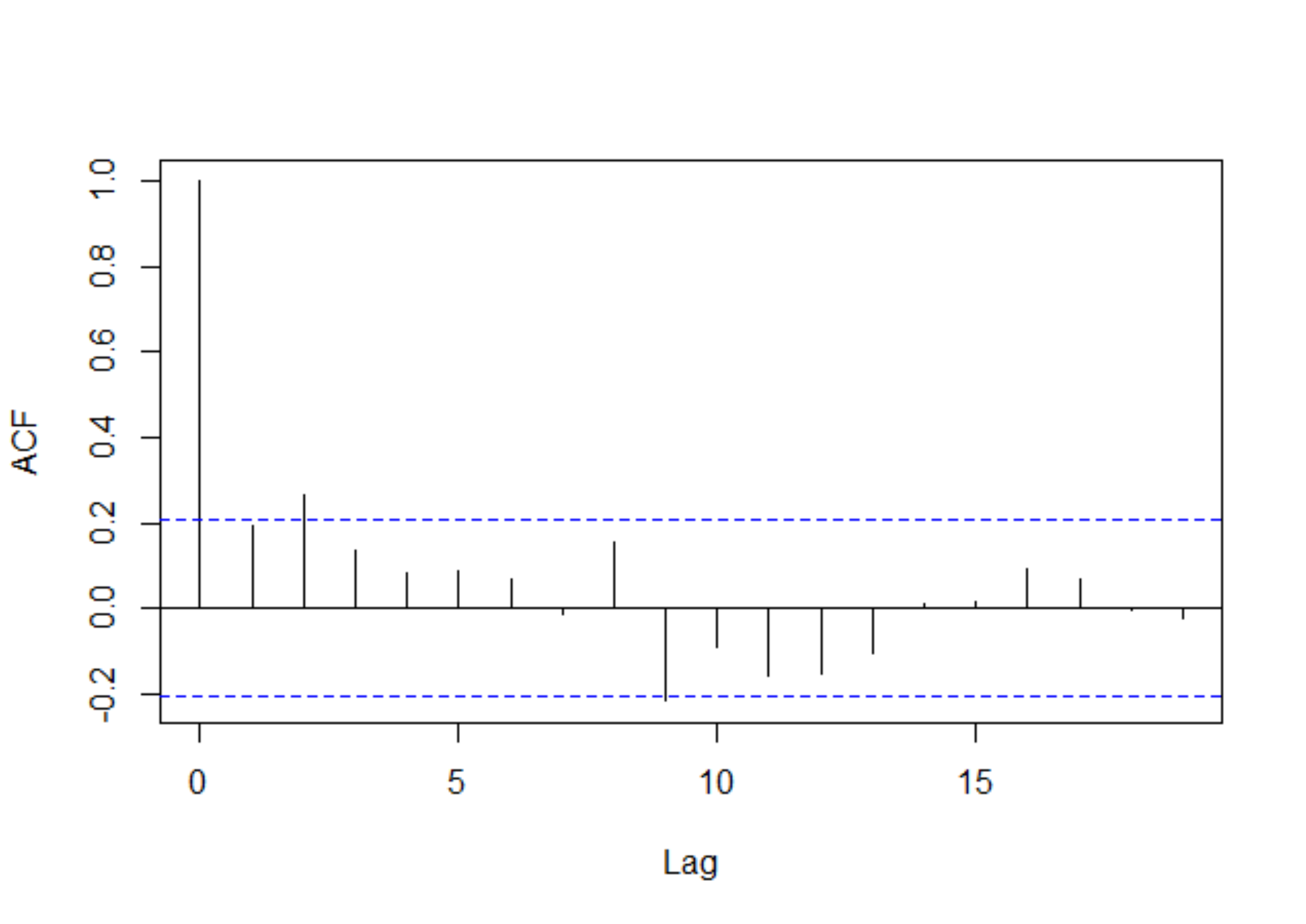}}}
  \hspace{0in}
  \subfigure[$\{\bm{y}_t^{(3)}\}_{t=1}^{90}$]{
    \label{50ETF_raw}
   \resizebox{7cm}{4.8cm}{\includegraphics{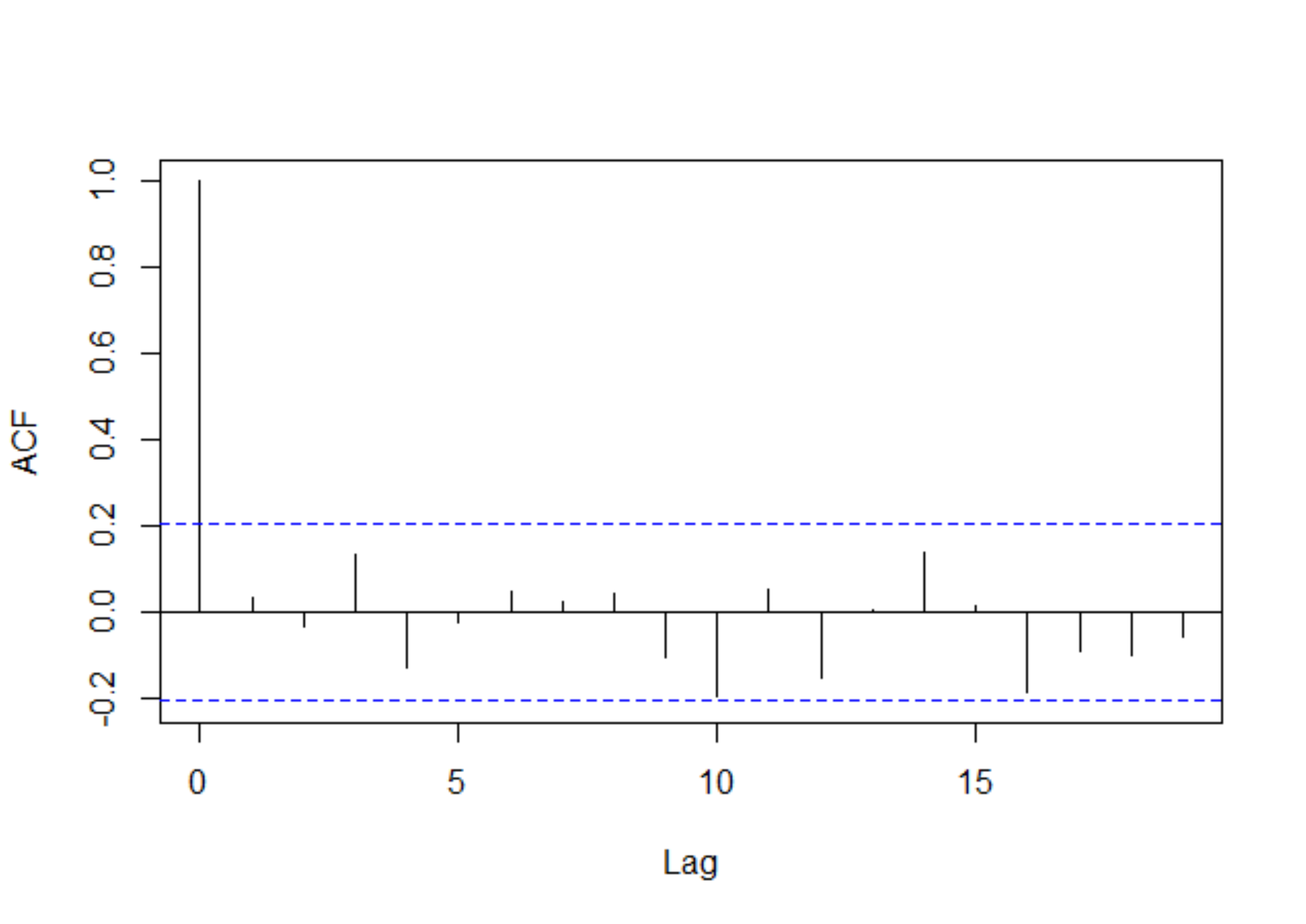}}}
  \hspace{0in}
  \subfigure[$\{\bm{y}_t^{(4)}\}_{t=1}^{90}$]{
    \label{maotai_predict}
   \resizebox{7cm}{4.8cm}{\includegraphics{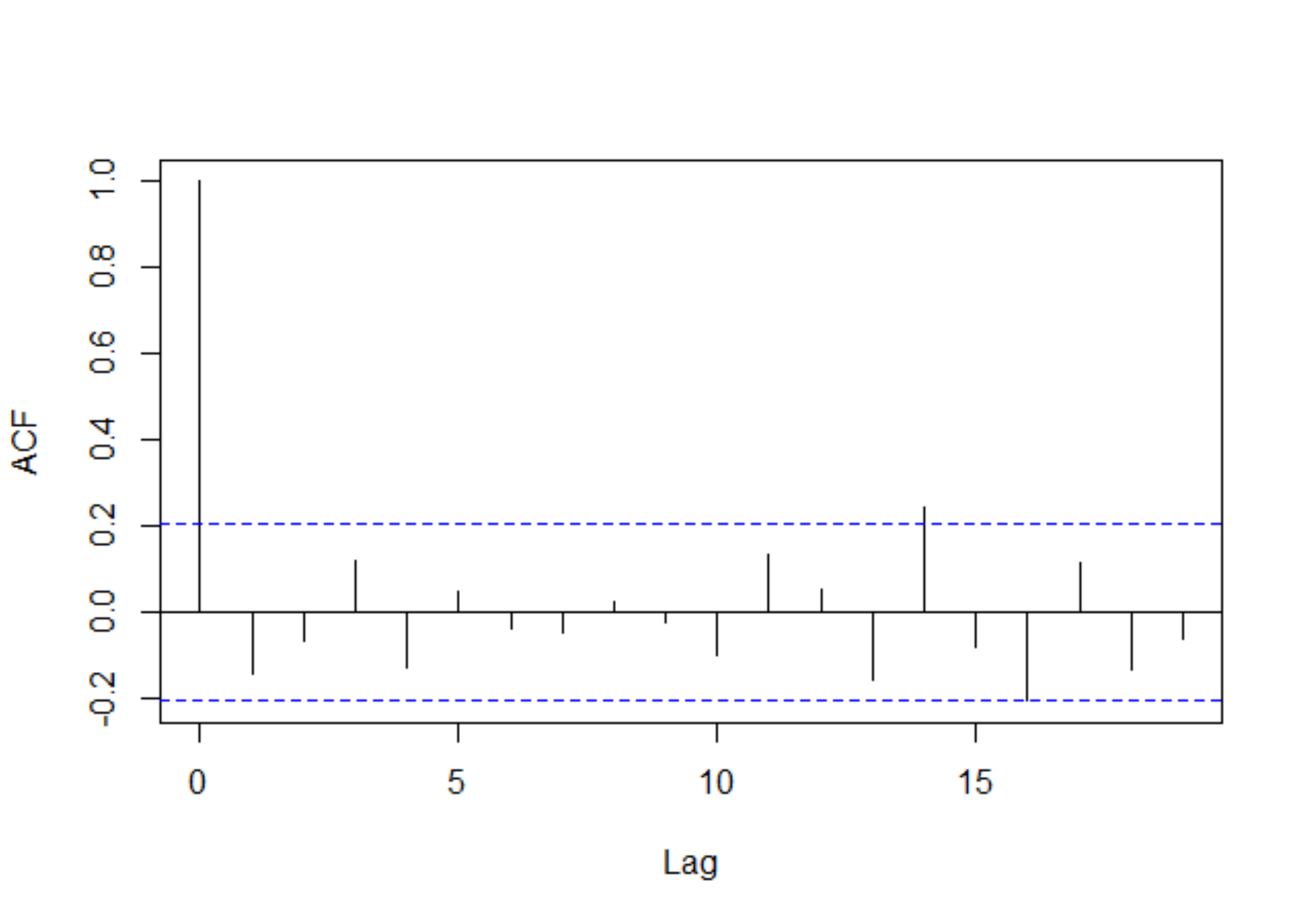}}}
  \caption{\rm ACF plots of the four time-series variables in $\{\bm{Y}_t\}_{t=1}^{90}$  of the 50 ETF}\label{ACF}
\end{figure}

Specifically, we take the first vector times series $\{\bm{Y}_t\}_{t=1}^{90}$ (i.e., $\bm{Y}_1^{(90)}$ in Algorithm \ref{algorithm2}) of the 50 ETF as an example to illustrate our modelling process.
At the significance level of $\alpha=0.1$, the four time series are stationary except $\{\bm{y}_t^{(1)}\}_{t=1}^{90}$ with the p-value of ADF test being $0.628$, indicating that $\{\bm{Y}_t\}_{t=1}^{90}$ cannot be modelled by VAR model.
The ACF plots in Fig.\ref{ACF} further demonstrate the distinct auto-correlation and non-stationary of $\{\bm{y}_t^{(1)}\}_{t=1}^{90}$.
Then, the Johansen test is applied to examine the co-integration relationship between the four variables in $\{\bm{Y}_t\}_{t=1}^{90}$, and the essence of Johansen test based on ``Trace" is investigating the number of co-integration vectors, which is recorded as $r$.
The results show that  the possible of $r\leq2$ is less than $1\%$ and the possible of $r\leq3$ is over than $10\%$, thus the $r$ in VEC model is determined as $3$. Finally, the VEC model  with the order of co-integration $r=3$ is established and the prediction values $\widehat{\bm{Y}_1}^{(90)}$ are obtained by the regression function. Through inverse transformation method, we can obtain predicted $\widehat{\bm{X}_1}^{(90)}$.
Traverse the following vector time series $\{\widehat{\bm{Y}}_{t+l}\}_{t=1}^{90} (l=0,1,\cdots, T-q-m)$, and the prediction accuracy can be evaluated.

\begin{figure}[!h]
  \centering
  \subfigure[$x^{(o)}_t$]{
    \label{maotai_o}
    \resizebox{4.5cm}{4cm}{\includegraphics{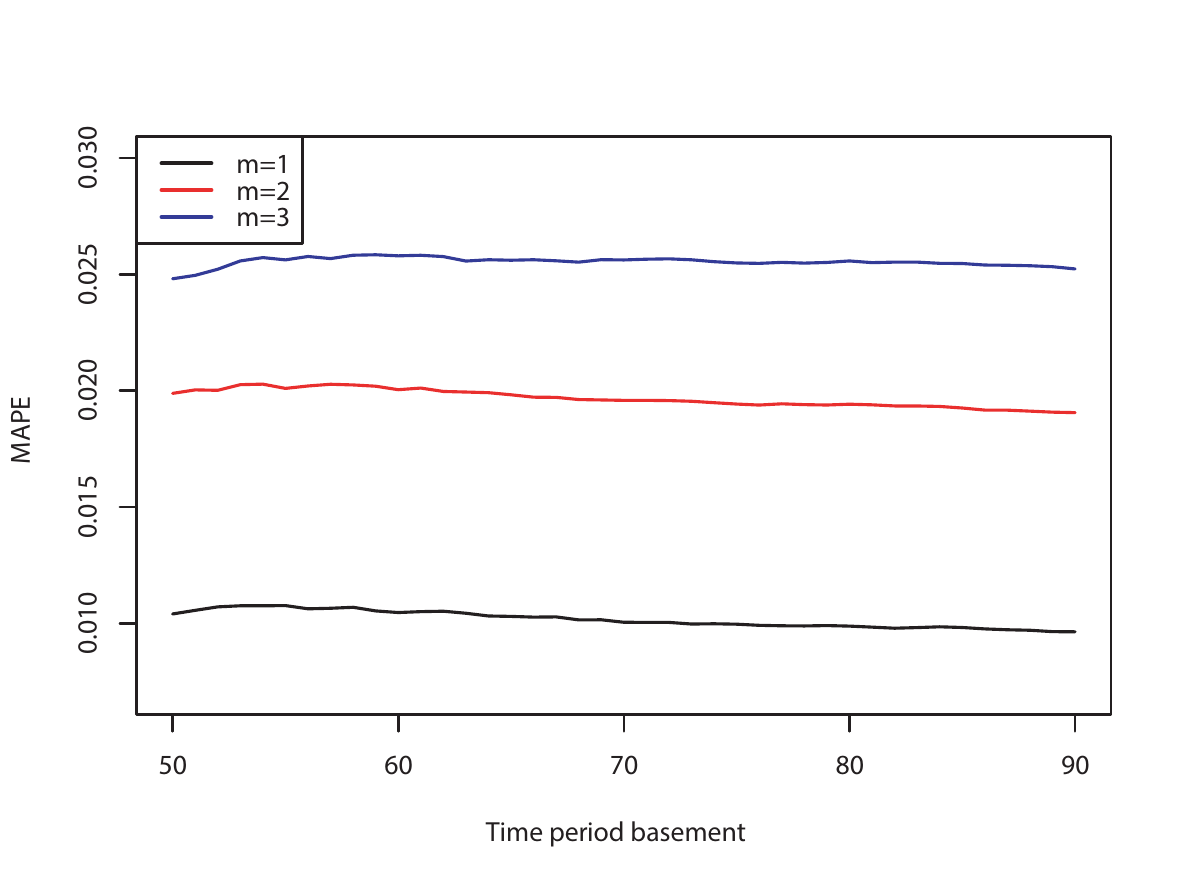}}}
  \hspace{0in}
   \subfigure[$x^{(o)}_t$]{
    \label{CSI_o}
    \resizebox{4.5cm}{4cm}{\includegraphics{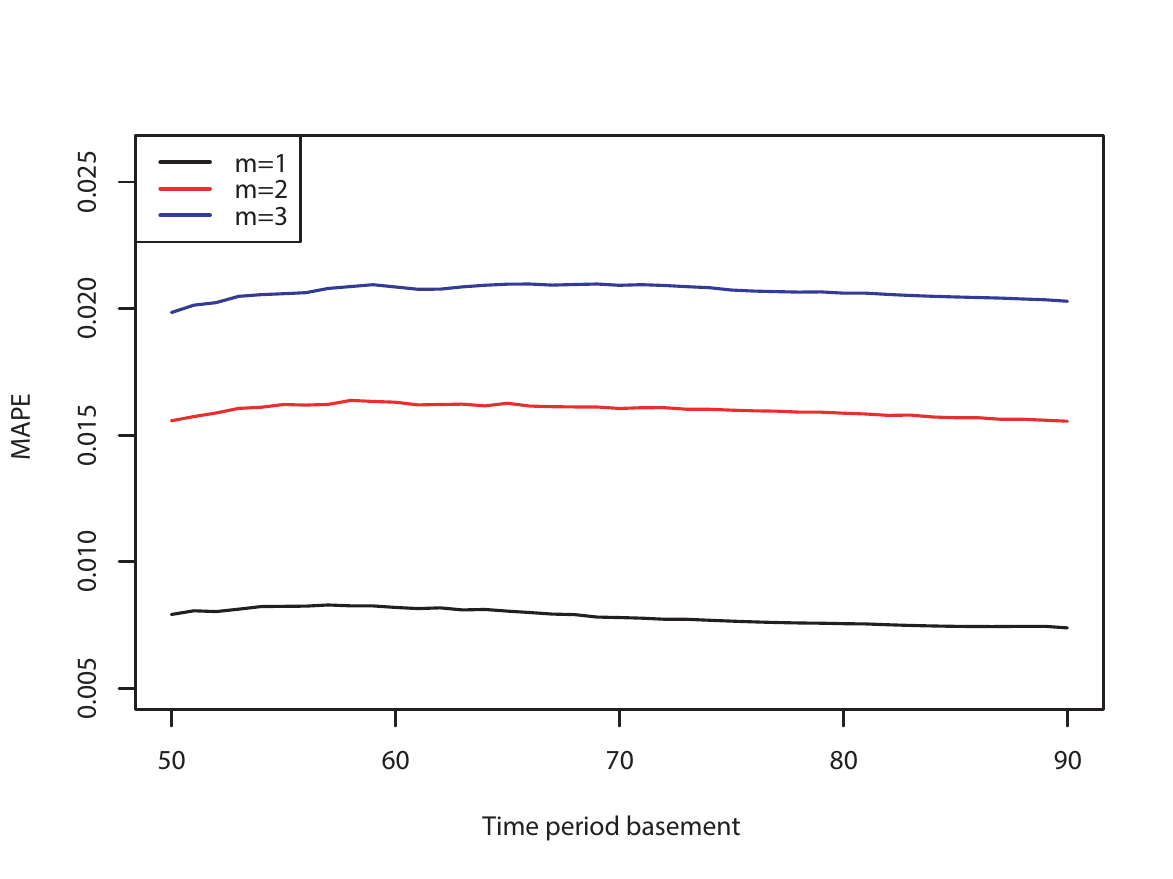}}}
  \hspace{0in}
    \subfigure[$x^{(o)}_t$]{
    \label{50ETF_o}
    \resizebox{4.5cm}{4cm}{\includegraphics{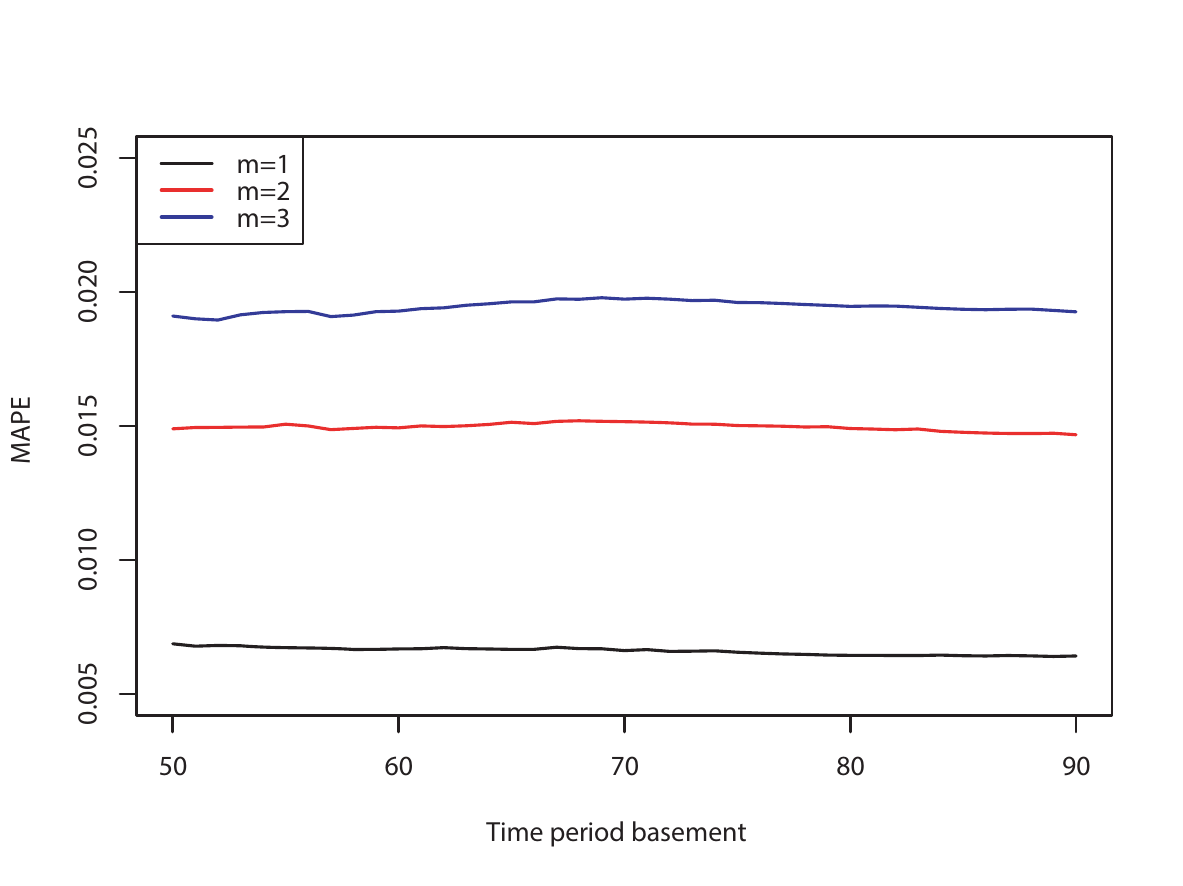}}}
  \hspace{0in}
  \subfigure[$x^{(h)}_t$]{
    \label{maotai_h}
    \resizebox{4.5cm}{4cm}{\includegraphics{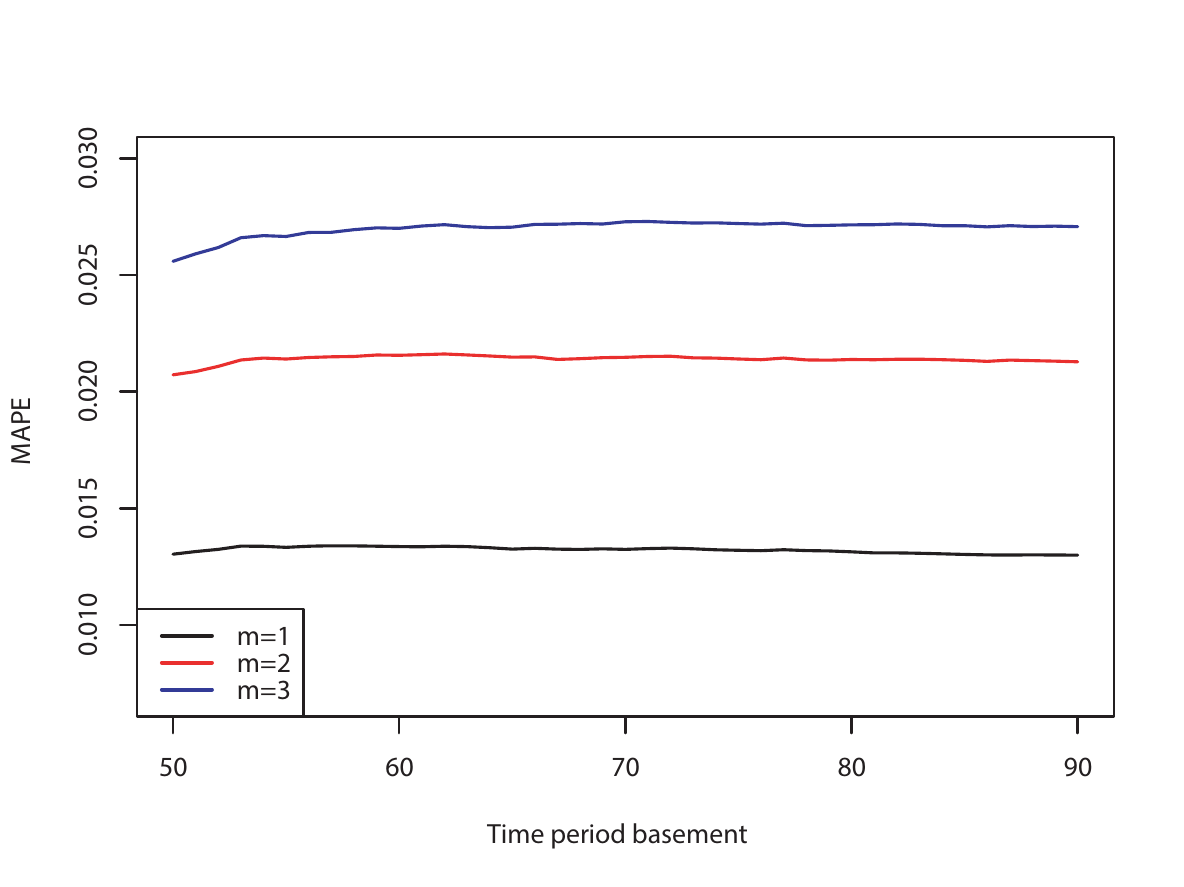}}}
  \hspace{0in}
   \subfigure[$x^{(h)}_t$]{
    \label{CSI_h}
    \resizebox{4.5cm}{4cm}{\includegraphics{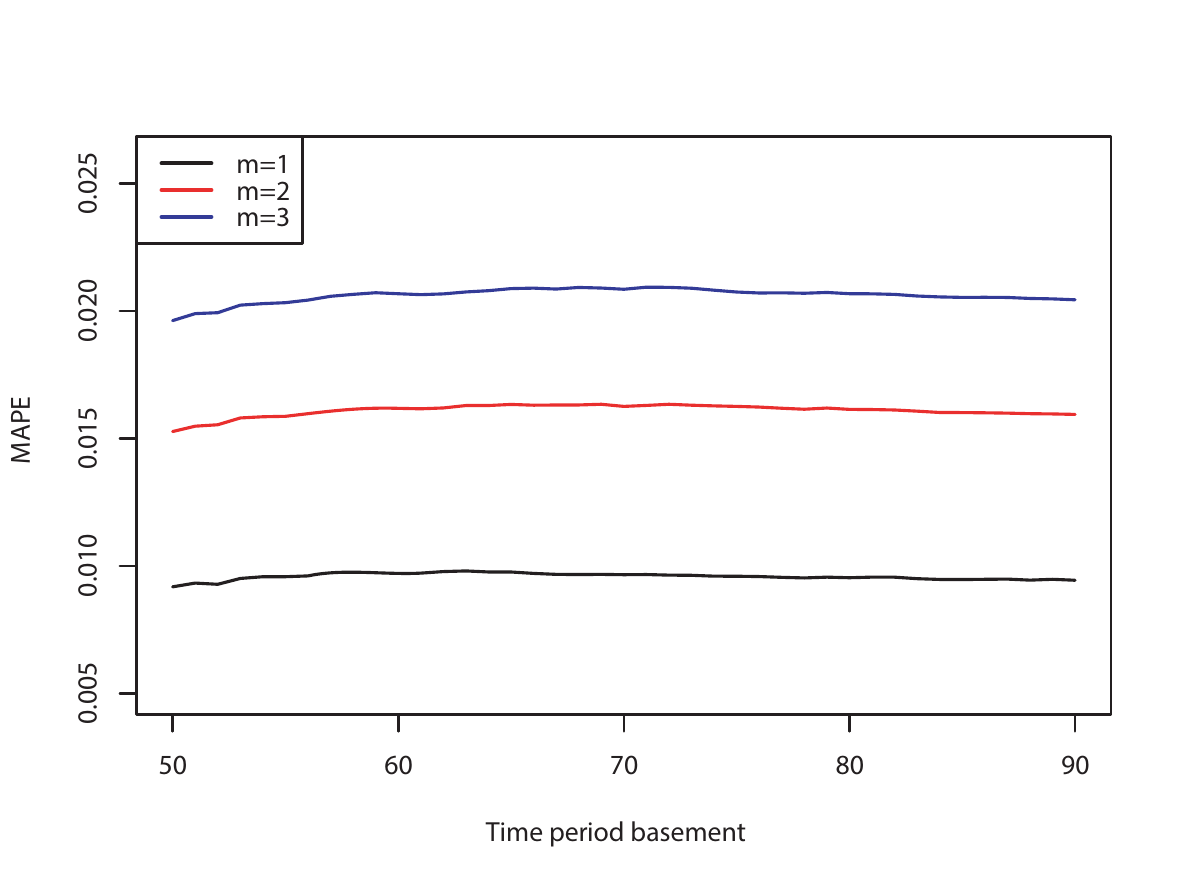}}}
  \hspace{0in}
   \subfigure[$x^{(h)}_t$]{
    \label{50ETF_h}
    \resizebox{4.5cm}{4cm}{\includegraphics{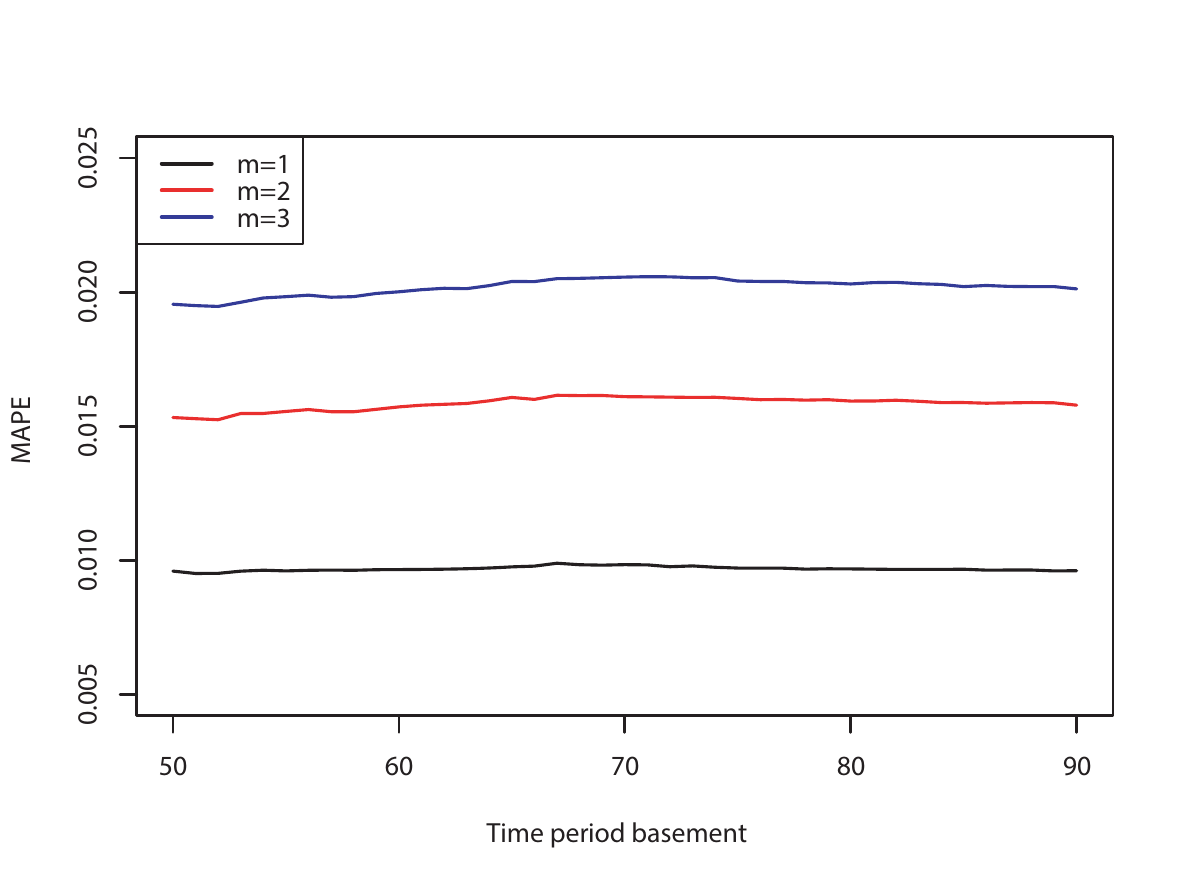}}}
  \hspace{0in}
  \subfigure[$x^{(l)}_t$]{
    \label{maotai_l}
    \resizebox{4.5cm}{4cm}{\includegraphics{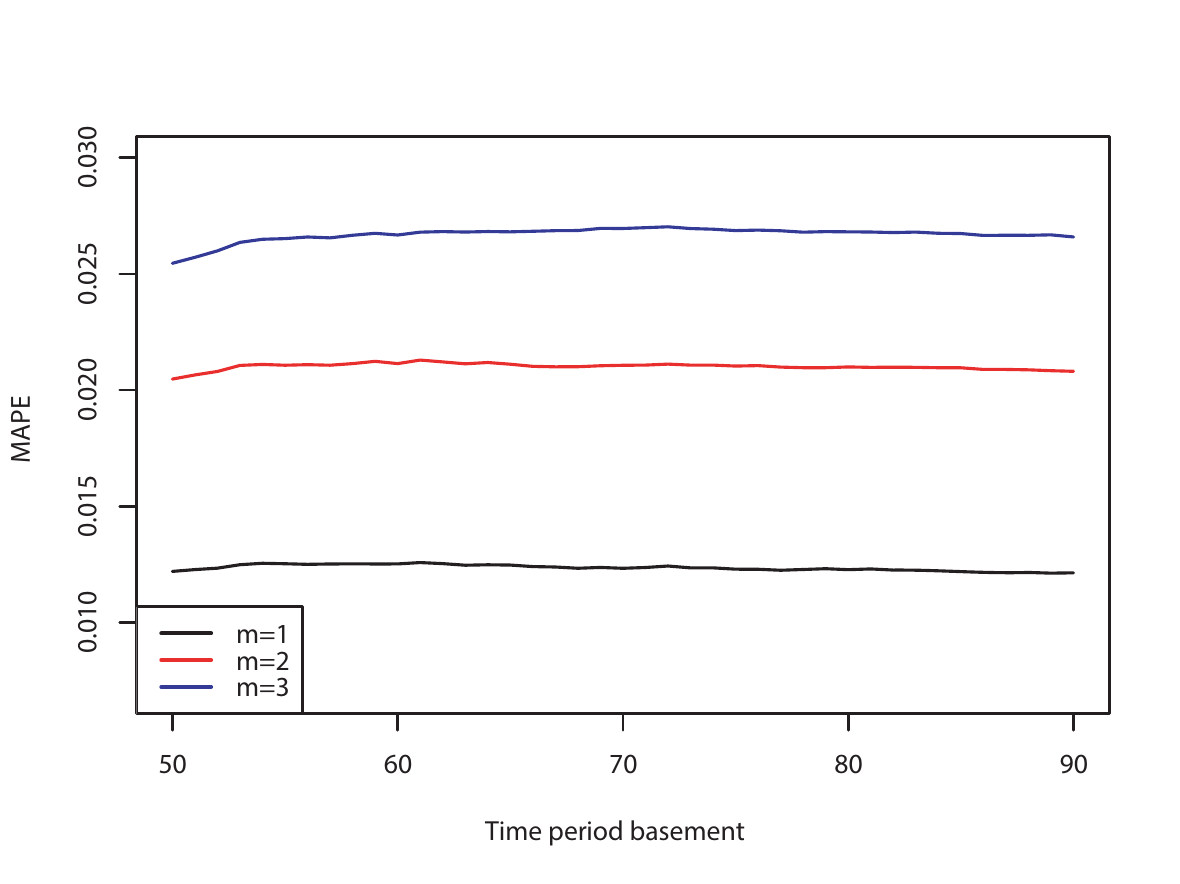}}}
  \hspace{0in}
   \subfigure[$x^{(l)}_t$]{
    \label{CSI_l}
    \resizebox{4.5cm}{4cm}{\includegraphics{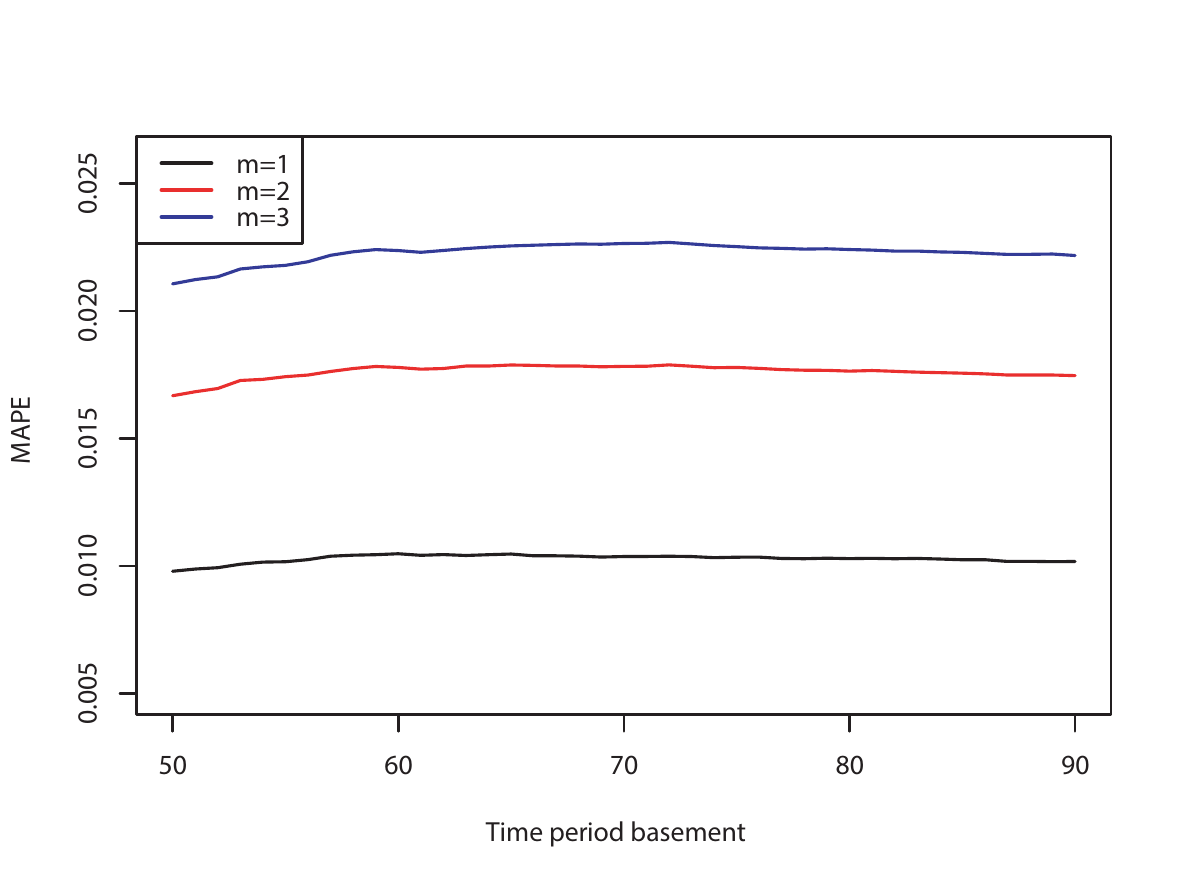}}}
    \hspace{0in}
     \subfigure[$x^{(l)}_t$]{
    \label{50ETF_l}
    \resizebox{4.5cm}{4cm}{\includegraphics{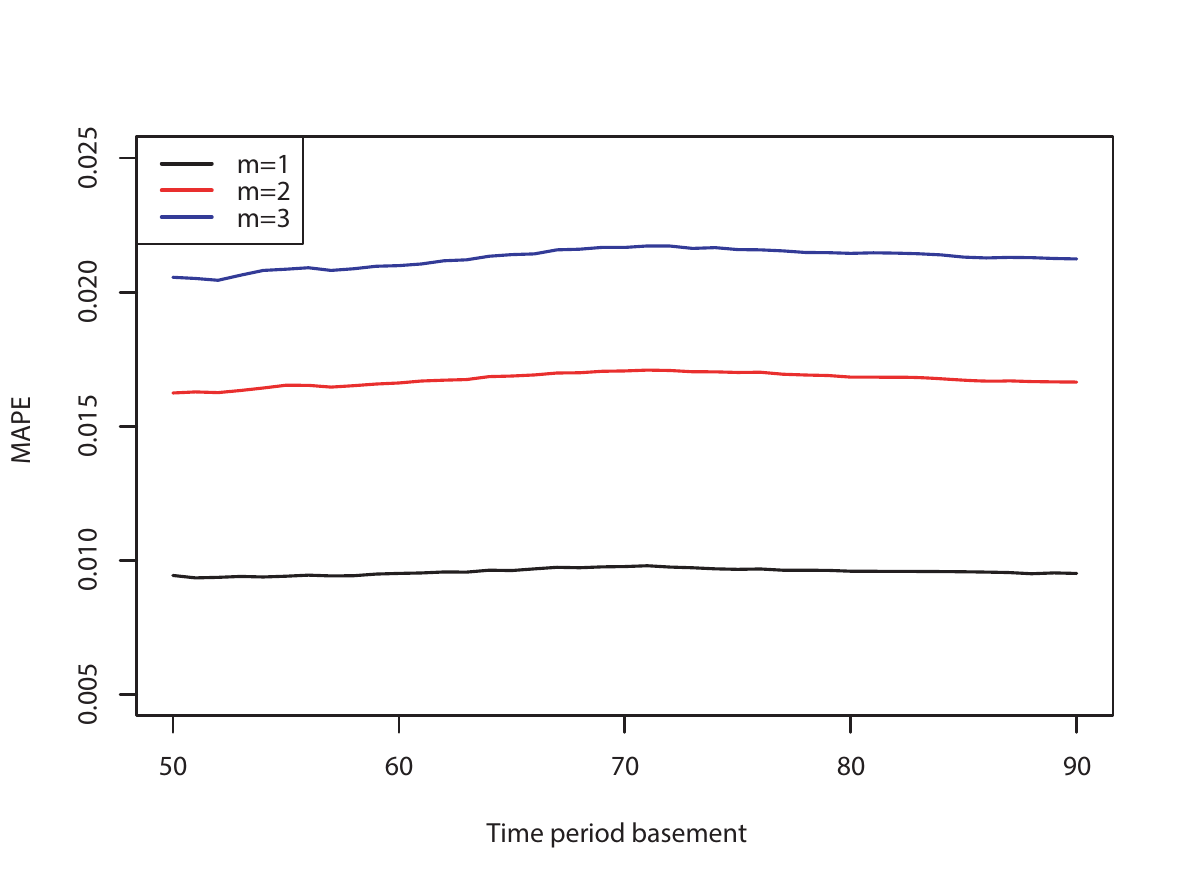}}}
  \hspace{0in}
  \subfigure[$x^{(c)}_t$]{
    \label{maotai_c}
    \resizebox{4.5cm}{4cm}{\includegraphics{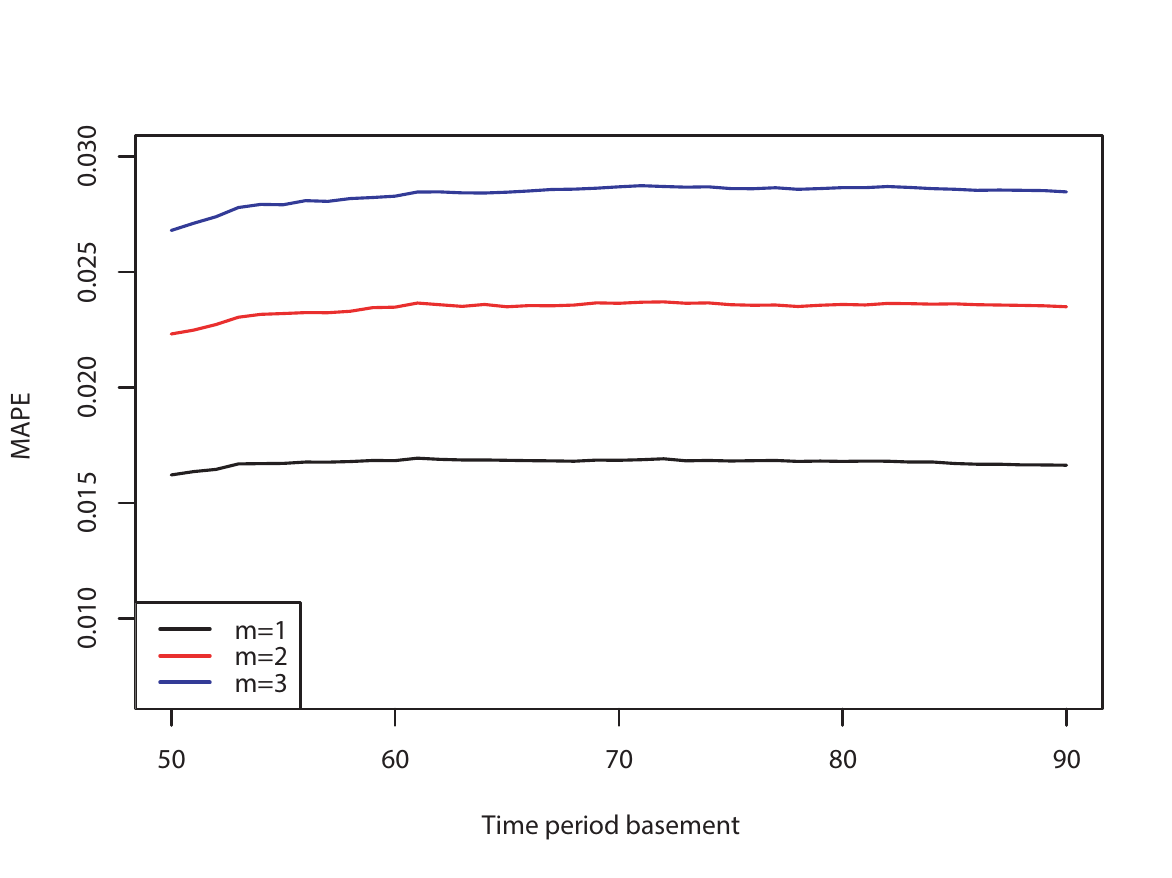}}}
  \hspace{0in}
  \subfigure[$x^{(c)}_t$]{
    \label{CSI_c}
    \resizebox{4.5cm}{4cm}{\includegraphics{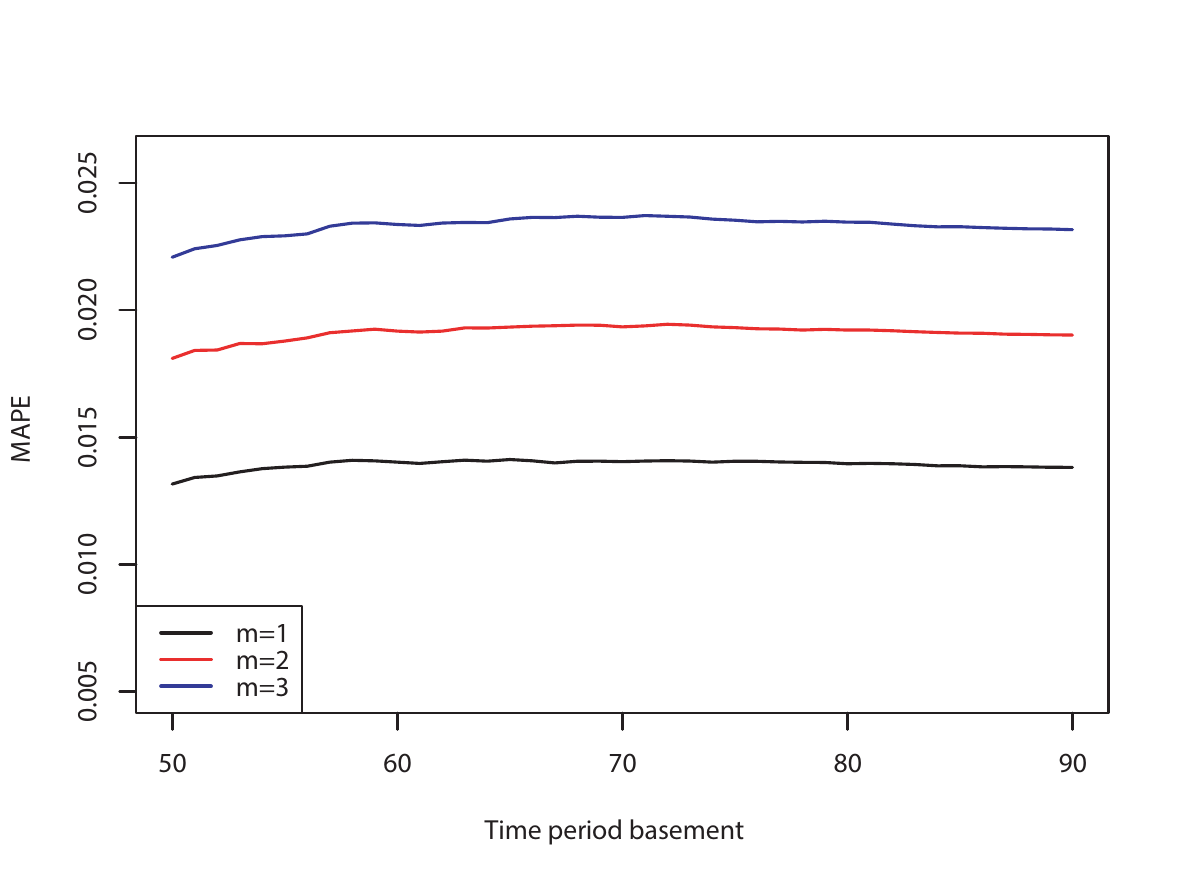}}}
  \hspace{0in}
   \subfigure[$x^{(c)}_t$]{
    \label{50ETF_c}
    \resizebox{4.5cm}{4cm}{\includegraphics{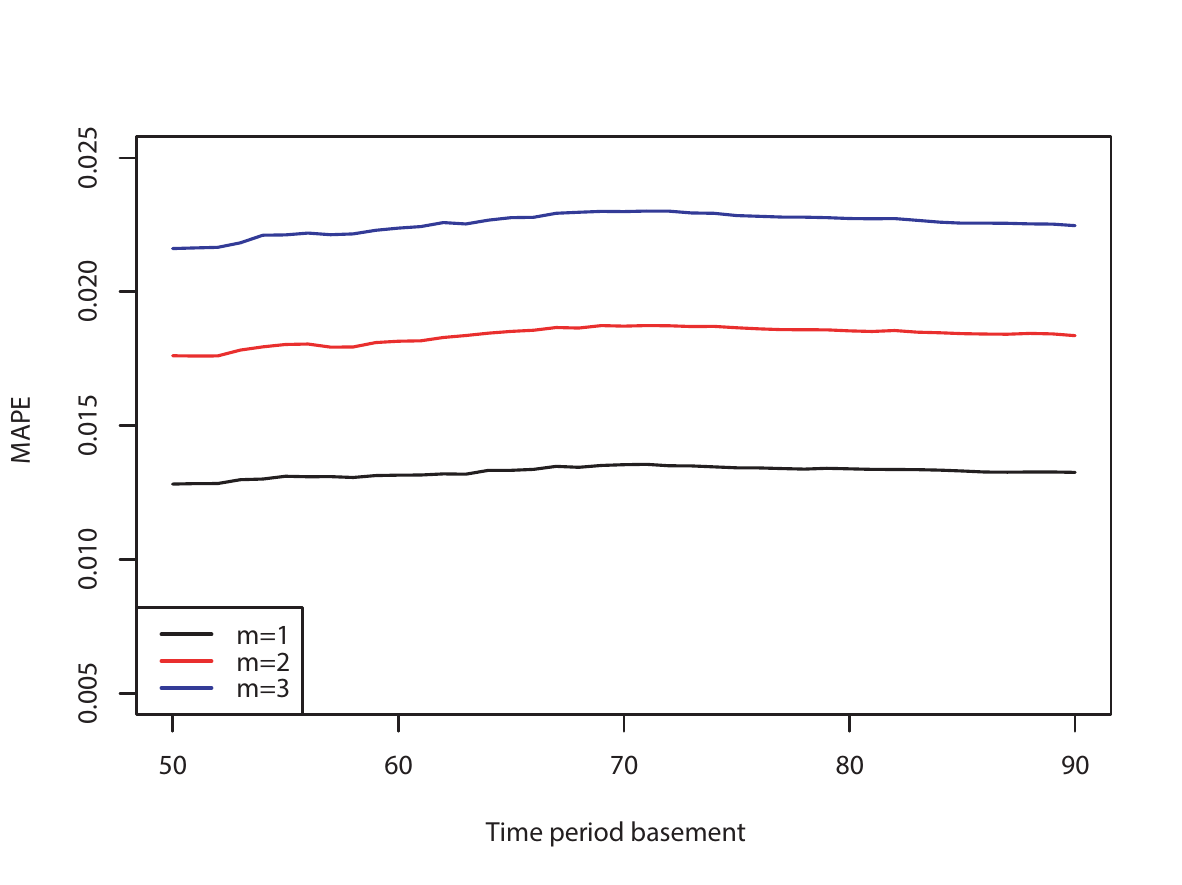}}}
  \caption{\rm MAPE of $x^{(o)}_t$ (The first row), $x^{(h)}_t$ (The second row), $x^{(l)}_t$ (The third row) and $x^{(c)}_t$ (The fourth row) for the Kweichow Moutai (Left panel), CSI 100 index (Middle panel), 50 ETF (Right panel) with different $q$ and $m$, respectively}\label{realexample}
\end{figure}

The performance in terms of MAPE for these three data sets of the Kweichow Moutai, CSI 100 index and 50 ETF is reported in the left, middle and right panels of Fig.\ref{realexample}, respectively.
From Fig.\ref{realexample}, the pattern of the performance exhibits similarly. It is obvious that under different setting $(q,m)$,  the prediction results are quite stable with all the values of MAPE below $2.87\%$ for the Kweichow Moutai,  $2.30\%$ for  the CSI $100$  index, and $2.37\%$ for the the 50 ETF, respectively. Moreover, the fewer the $m$ is, the more accurate the prediction results are.

\begin{table}[h]
  \setlength{\abovecaptionskip}{-0.2cm}
  \caption {Comparison of forecasted results for the Kweichow Moutai, CSI $100$  index, and 50 ETF with $q=90$ and $m=1$}
  \label{reaexample-comparison}
  \begin{center}
  \setlength\tabcolsep{4 pt}
  \renewcommand{\arraystretch}{0.8}
    \begin{tabular}{cccccccc}
    \toprule
  \multirow{2}{*}{Criterion}&& \multicolumn{2}{c}{the Kweichow Moutai}& \multicolumn{2}{c}{the CSI $100$  index}& \multicolumn{2}{c}{the 50 ETF}\\
  \cline{3-8}
  && Proposed & Naive& Proposed & Naive& Proposed & Naive\\
     \hline
    \multirow{4}{*}{MAPE} & $x^{(o)}_t$ & 0.821\%$^{*}$ & 1.602\% & 0.663\%$^{*}$ & 1.306\%& 0.729\%$^{*}$ & 1.254\%\\
    & $x^{(h)}_t$ & 1.307\%$^{*}$ & 1.401\%& 0.923\%$^{*}$ & 1.051\% & 0.977\%$^{*}$ & 1.050\%\\
    & $x^{(l)}_t$ & 1.208\%$^{*}$ & 1.407\%& 0.993\%$^{*}$ & 1.173\%& 0.980\%$^{*}$ & 1.114\%\\
    & $x^{(c)}_t$ & 1.654\% & 1.488\%& 1.354\% & 1.220\%& 1.349\% & 1.205\% \\
    \hline
    \multirow{4}{*}{RMSE} & $x^{(o)}_t$ & 22.689$^{*}$& 42.446 & 34.111$^{*}$ & 61.138& 0.030$^{*}$ & 0.048\\
    & $x^{(h)}_t$ & 34.544 & 35.799& 42.342$^{*}$ & 48.576 & 0.037$^{*}$ & 0.040\\
    & $x^{(l)}_t$ & 31.637$^{*}$& 35.922& 47.709$^{*}$ & 55.413& 0.039$^{*}$ & 0.044\\
    & $x^{(c)}_t$ & 44.589 & 40.634& 63.889 & 57.527& 0.052 & 0.047 \\
    \hline
    \multicolumn{2}{c}{RMSEH} & 40.663$^{*}$& 44.159& 54.990$^{*}$ & 63.760& 0.046$^{*}$ & 0.052\\
    \multicolumn{2}{c}{AR} & 0.447$^{*}$& 0.414& 0.424$^{*}$ & 0.371& 0.419$^{*}$ & 0.386\\
    \hline
    \multicolumn{2}{c}{Count of VAR} & \multicolumn{2}{c}{93}  & \multicolumn{2}{c}{40} & \multicolumn{2}{c}{33} \\
    \multicolumn{2}{c}{Count of VEC} & \multicolumn{2}{c}{4037} & \multicolumn{2}{c}{3129}& \multicolumn{2}{c}{3339}\\
    \bottomrule
      \end{tabular}
  \end{center}
\end{table}

Specifically, take the prediction basement period of $q=90$, and proceed one-step forecast ($m=1$) as an example, we summarized the results in term of MAPE, RMSE, RMSEH and AR in Table \ref{reaexample-comparison}, and compared them with the Naive method proposed by \cite{arroyo2011different}, whose indicators are calculated by taking the price of the previous day as the price of the day.
We also give the results of unilateral t-test in Table \ref{reaexample-comparison}.
The null hypothesis of the t-test of all indicators except AR is that the value of the proposed method is no less than that of the naive method.
If the p-value is less than 0.01, it indicates that we can reject the null hypothesis at 99\% confidence level (remarked by a superscript $*$ ), and choose the alternative hypothesis that the value of the proposed method is less than that of the naive method.
The t-test of AR is just the opposite, if the p-value is less than 0.01, it represents that the value of the proposed method is greater than that of the naive method.
From Table \ref{reaexample-comparison}, it can be seen that the forecasting effect of the proposed method is significantly accurate and most indicators pass the t-test. More specifically,
\begin{itemize}
  \item[(1)] for the Kweichow Moutai, the MAPE of the open price is only $0.821\%$, and the MAPE of the close price, high price, and low price of are also controlled within $1.7\%$. There is no evidence that the MAPE of the predicted close price from the proposed method are smaller than that of the naive method. Whereas, comparing to the naive method, the MAPE of the open price is improved by $46.55\%$, $6.71\%$ for the high price and $14.14\%$ for the low price; the RMSE of the open price is improved by $46.55\%$, $3.51\%$ for the high price, and $11.93\%$ for the low price. As for the RMSEH and the AR, the results of the proposed method are improved by $7.92\%$ and $7.97\%$ respectively.
  \item[(2)] for the CSI 100 index, the MAPE for the forecasted open price is only $0.663\%$, which is $49.23\%$ optimizer than that of the naive method. Meanwhile, the MAPE for the predicted high price and the low price are $12.18\%$ and $15.35\%$ optimizer than that of the naive method, respectively. The RMSE of the open price is improved by $44.21\%$, $12.83\%$ for the high price, and $13.90\%$ for the low price. As for the RMSEH and the AR, the results of the proposed method are improved by $13.75\%$ and $14.29\%$ respectively.
  \item[(3)] for the 50 ETF, the MAPE for the forecasted open price is $0.729\%$, which improved by $41.87\%$ than that of the naive method. At the same time, $6.95\%$ and $12.03\%$ improvement on the MAPE are made for the high price and the low price, respectively. The RMSE of the open price is improved by $37.50\%$, $7.50\%$ for the high, and $11.36\%$ for the low price.
      In terms of the RMSEH and AR, the results of the proposed method are improved by $11.54\%$ and $8.55\%$ respectively.
\end{itemize}
Table \ref{reaexample-comparison} also gives the counts of establishing VAR model and VEC model when employing the proposed framework.
We can conclude that the VEC model is far more frequently adopt comparing to the VAR model, which provides a proof to the view of \cite{cheung2007empirical} that the daily highs and lows of stocks follow a co-integration relationship.

\begin{figure}[!h]
  \centering
  \subfigure[$\bm{X}_t$]{
    \label{maotai_raw}
    \resizebox{5cm}{3cm}{\includegraphics{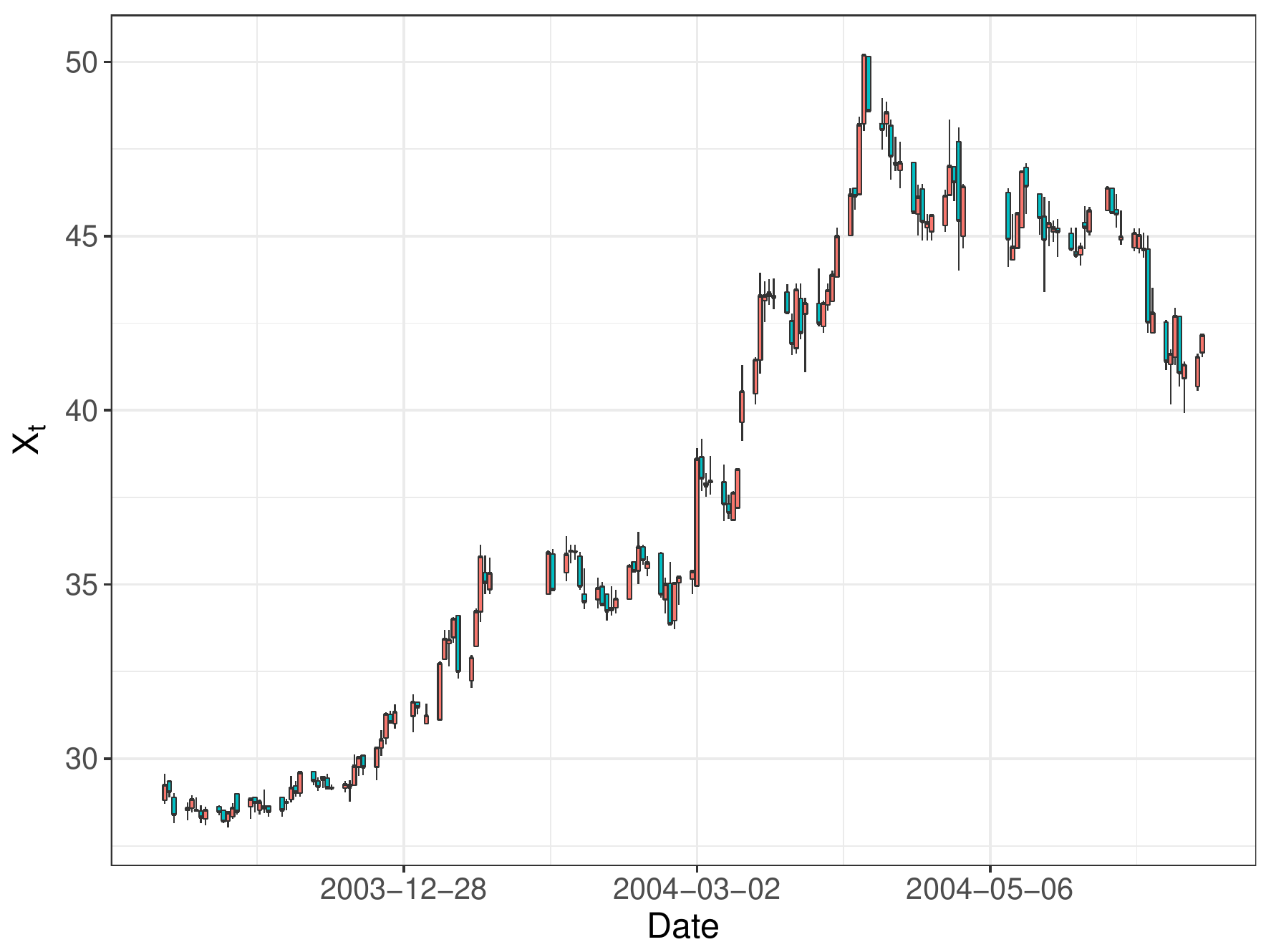}}}
  \hspace{0in}
   \subfigure[$\bm{X}_t$]{
    \label{CSI_raw}
   \resizebox{5cm}{3cm}{\includegraphics{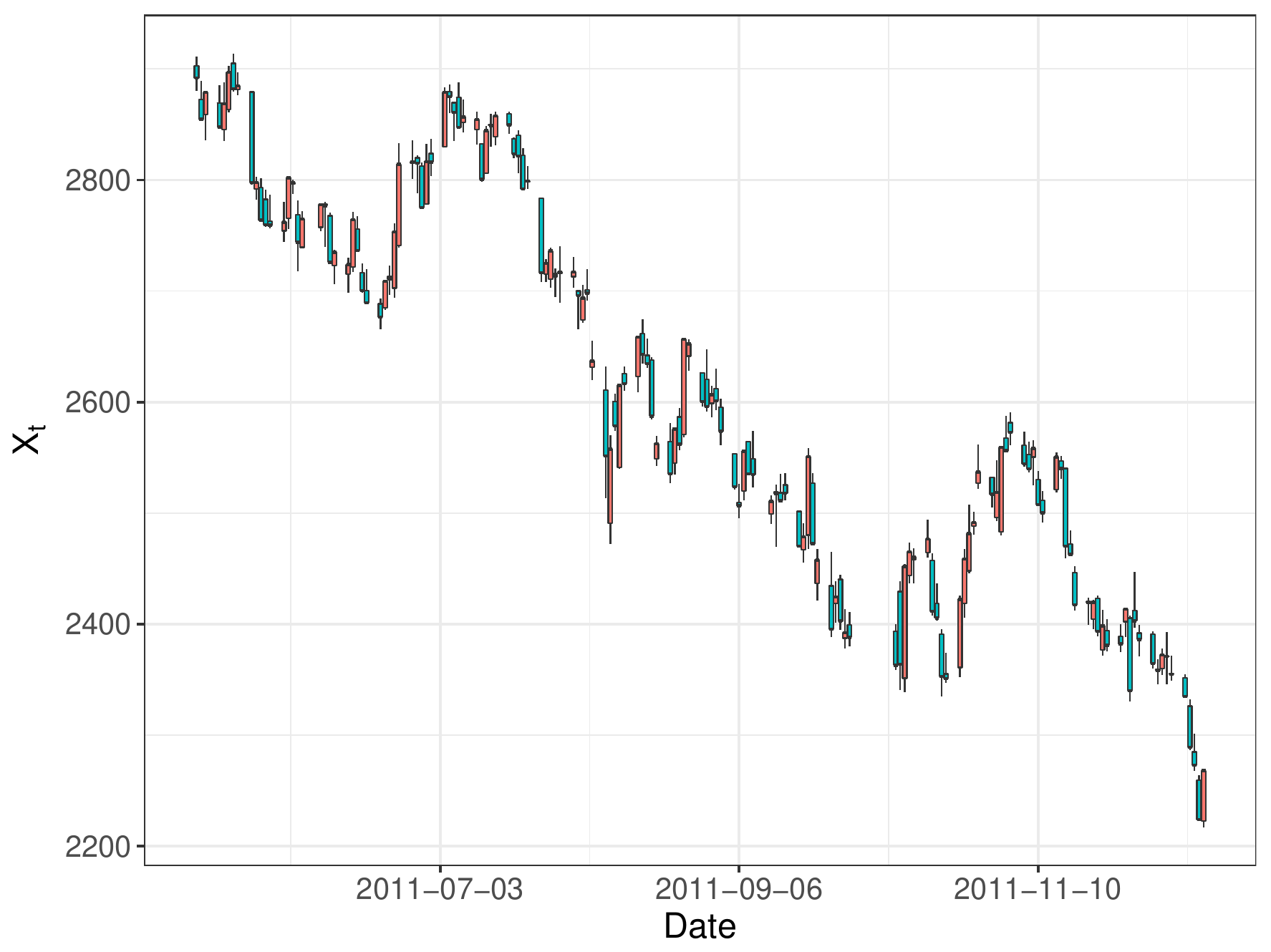}}}
  \hspace{0in}
  \subfigure[$\bm{X}_t$]{
    \label{50ETF_raw}
   \resizebox{5cm}{3cm}{\includegraphics{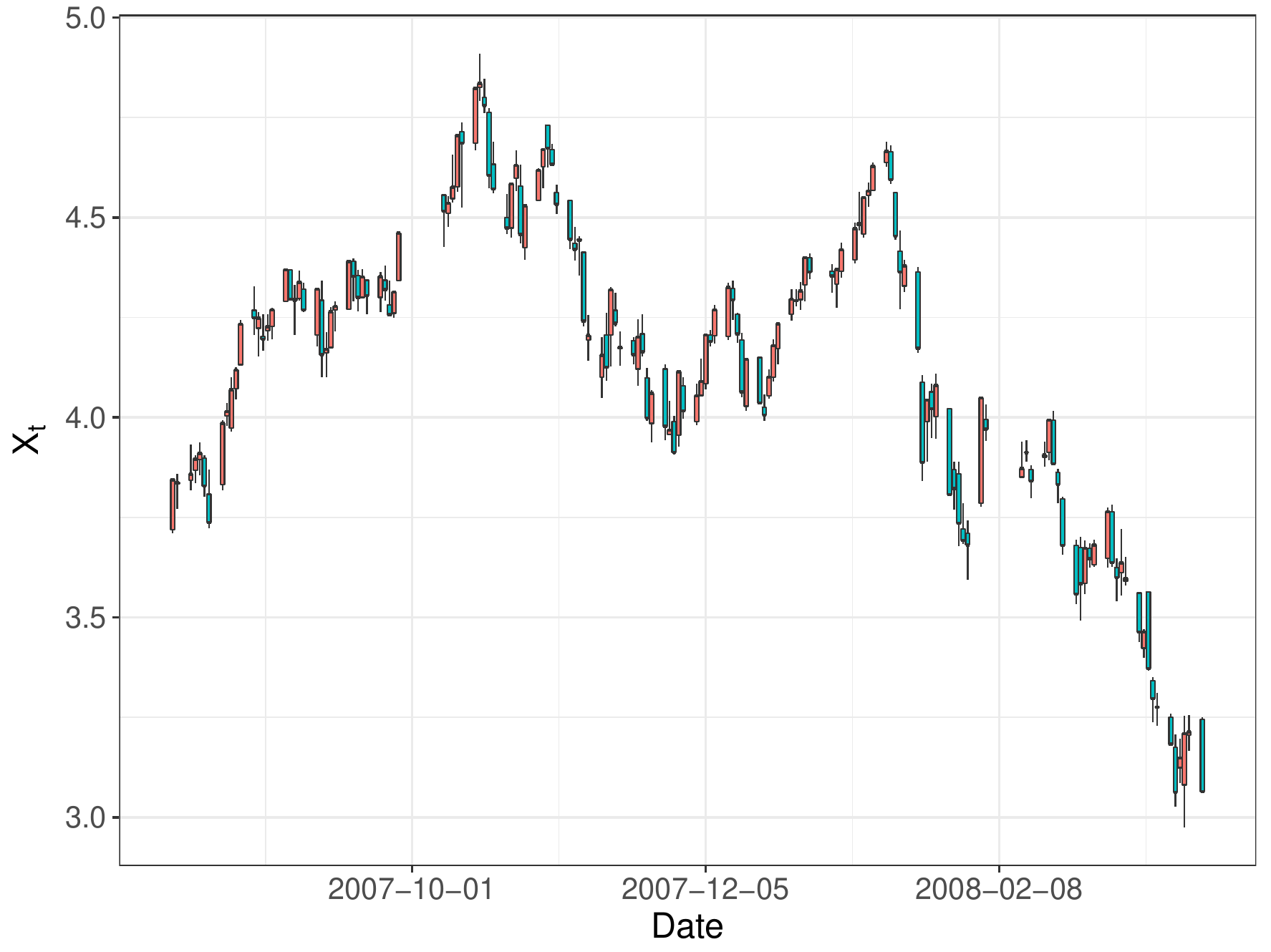}}}
  \hspace{0in}
  \subfigure[$\hat{\bm{X}}_t$]{
    \label{maotai_predict}
   \resizebox{5cm}{3cm}{\includegraphics{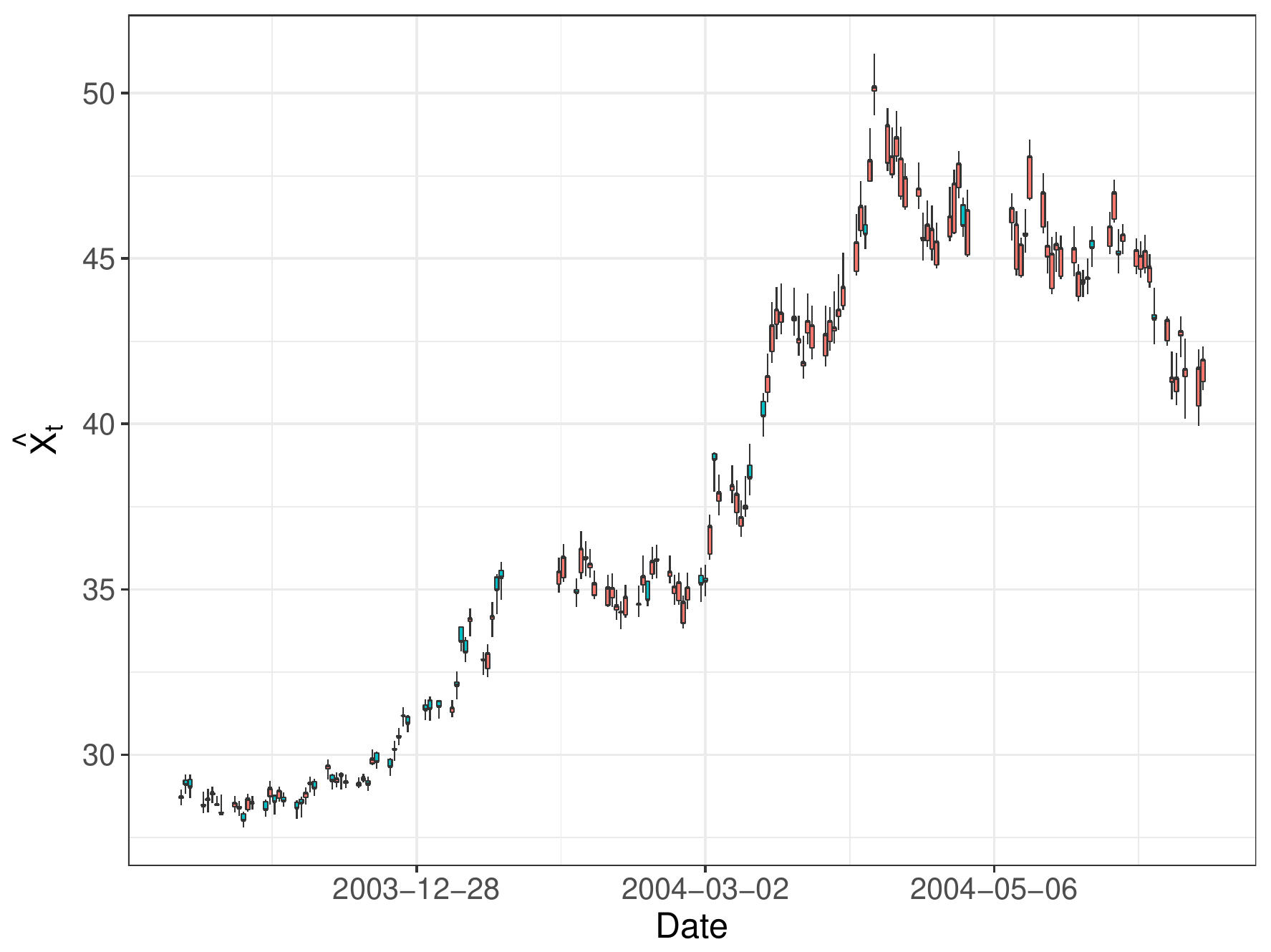}}}
  \hspace{0in}
  \subfigure[$\hat{\bm{X}}_t$]{
    \label{CSI_predict}
  \resizebox{5cm}{3cm}{\includegraphics{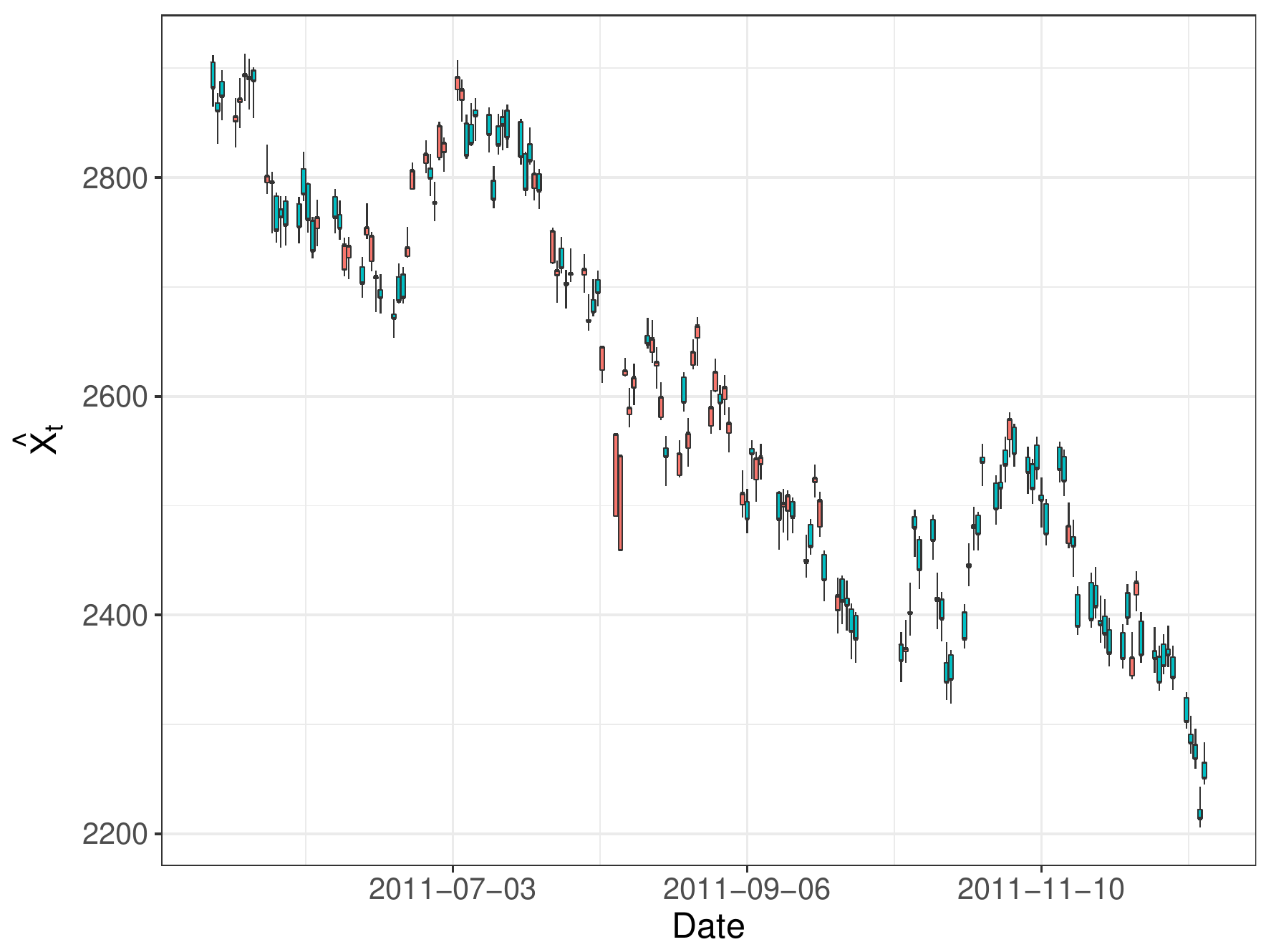}}}
    \hspace{0in}
  \subfigure[$\hat{\bm{X}}_t$]{
    \label{50ETF_predict}
   \resizebox{5cm}{3cm}{\includegraphics{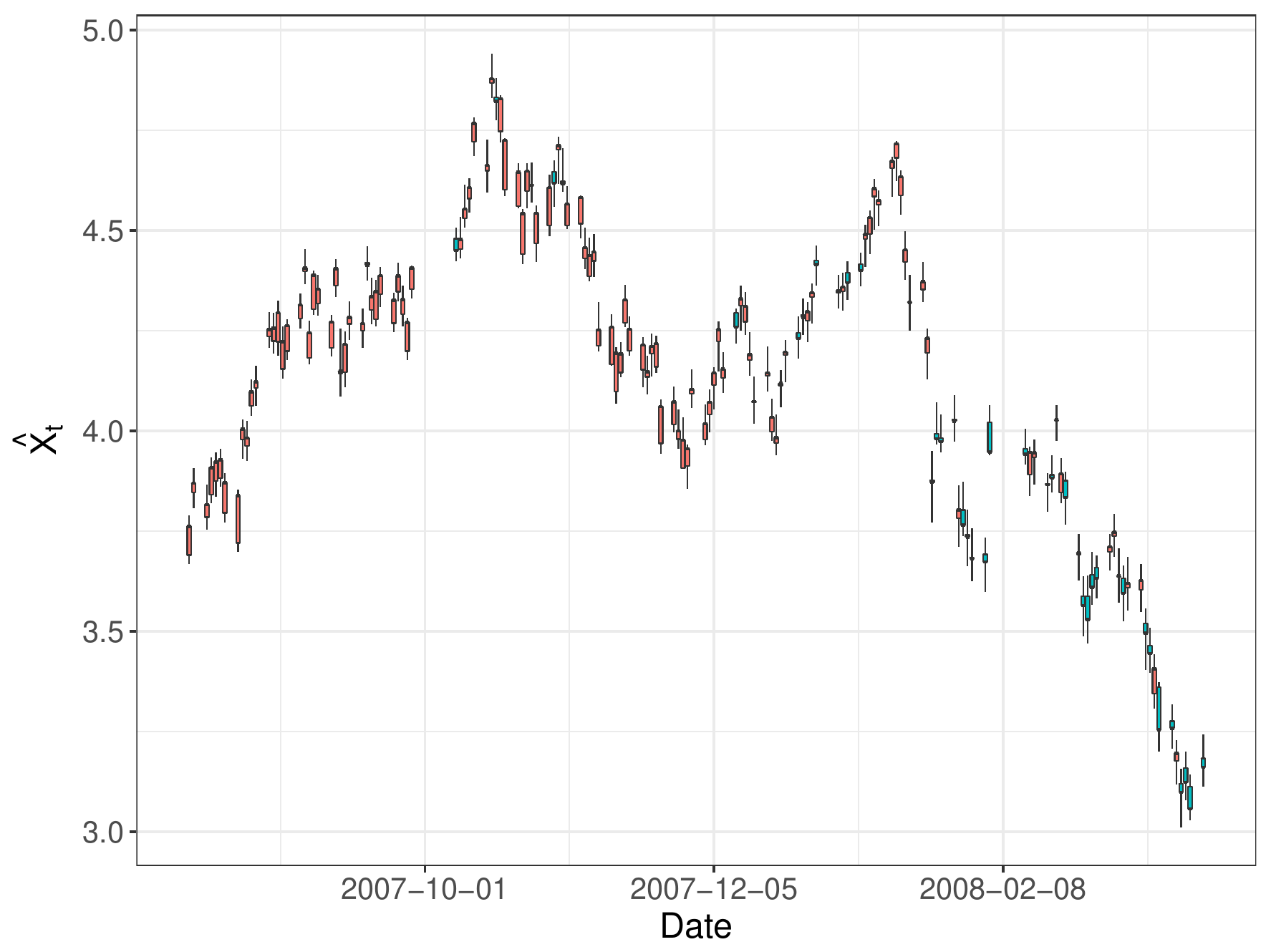}}}
  \caption{\rm Comparison of the real values (Top row) and predicted values (Bottom row) of the Kweichow Moutai (Left panel), CSI $100$ index (Middle panel), and $50$ ETF (Right panel)}\label{Figrealexample}
\end{figure}

Finally, to get a clear picture of the forecasted performance, we also compared the realistic and the predicted stock values of the Kweichow Moutai during the period from $5/11/2003$ to $22/6/2004$ (Left panel), the CSI $100$ index from $11/5/2011$ to $16/12/2011$ (Middle panel), and the $50$ ETF during the period from $9/8/2007$ to $24/3/2008$ (Right panel) in Fig.\ref{Figrealexample}. From it, we can see that the data predicted by the proposed method in this paper fits the reality well. Furthermore, for the Kweichow Moutai, the continuous rising that exists before $9/4/2004$ and the subsequent falling are perfectly forecasted; for the CSI $100$ index, the overall downward trend and the two rebounds around $4/7/2011$ and $9/11/2011$ are also fully reflected; for the $50$ ETF, two spikes around $16/10/2007$ and $15/1/2008$ coincided accurately.

\section{Conclusions} \label{Sec 5}

We investigated the forecasting problem of the OHLC data contained in candlestick chart, which plays an important role in the financial market. To address it, we proposed a novel transformation approach to relax the inherent constraints of the OHLC data along with its explicit inverse transformation, which facilitates the subsequent establishment of various prediction models and guarantee a meaningful predicted open-high-low-close prices. The transformation approach not only extends the range of variables to $\left(-\infty, +\infty\right)$, but also shares the flexibility of the well known log- and logit- transformation. Based on the unconstrained transformation, we establish a flexible and efficient framework for modelling the OHLC data with full use of its information. For illustration, we thereby show the detailed procedure via the VAR and VEC modelling.

The new approach has high practical utility because of its flexibility, simple implementation and straightforward interpretation. For example, it is applicable to a variety of positive interval data, and the selected model can be generalized to other types of statistical models and machine learning models. Investigations along this direction may merit further research but beyond the scope of this paper. From this perspective, the proposed method provides a completely new and useful alternative for OHLC data analysis, enriching the existing literatures.

Finally, we documented the finite performance of the proposed method via extensive simulation studies in terms of various measurements. And the analysis of the OHLC data of three different kinds of financial subjects in Chinese financial market: the Kweichow Moutai, CSI $100$ index, and $50$ ETF also illustrated the utility of the new methodology. All of these results fully demonstrate the stability and reasonability of the proposed method.

\section*{Compliance with Ethical Standards}

\section*{Funding}

This study was funded by the National Natural Science Foundation of China (grant Numbers. $71420107025$, $11701023$).

\section*{Conflict of interest}

Author Huiwen Wang declares that she has no conflict of interest. Author Wenyang Huang declares that he has no conflict of interest. Author Shanshan Wang declares that she has no conflict of interest.

\section*{Ethical approval}

This article does not contain any studies with human participants or animals performed by any of the authors.

\section*{Availability of data and material}

The data of the Kweichow Moutai, 50 ETF, CSI 100 is downloaded form a financial server called Wind.
The data can be uploaded as required.

\section*{Code availability}

This paper applies custom R code and can be uploaded as required.

\nolinenumbers
\small
\bibliography{KLineVAR}
\bibliographystyle{plainnat}

\end{document}